\documentclass[%
 reprint,longbibliography,
showpacs,preprintnumbers,
bibnotes,
 amsmath,amssymb,
prb,
]{revtex4-2}

\usepackage{graphicx}%
\usepackage[english]{babel}
\usepackage{microtype}          
\usepackage[breaklinks=true,colorlinks=true,linkcolor=blue,urlcolor=blue,citecolor=blue]{hyperref}

\usepackage{xcolor}     %
\usepackage{mathtools}
\mathtoolsset{showonlyrefs=true}
\usepackage{braket}
\usepackage{tikz}

 \usepackage{longtable}
  \usepackage{booktabs}
 \usepackage{siunitx}
 \usepackage{csquotes}
 
 \usepackage{multirow}
\usepackage{amssymb}

\newcommand{\refeqn}[1]{Eq.~\eqref{#1}}
\newcommand{\refeqns}[2]{Eqs.~\eqref{#1} and \eqref{#2}}

\newcommand{\eqrefs}[2]{Eqs.~\eqref{#1} and \eqref{#2}}
\newcommand{\eqrefss}[3]{Eqs.~\eqref{#1}, \eqref{#2} and \eqref{#3}}
\newcommand{\eqrefr}[2]{Eqs.~\eqref{#1} - \eqref{#2}}
\newcommand{\eqrefrno}[3]{Eqs.~\eqref{#1} - \eqref{#2}\noeqref{#3}}

\newcommand{\tref}[1]{Tab.~\ref{#1}}

\newcommand{\GKBA}{{}}
\newcommand{\GKBAd}{\Delta}

\newcommand{\lowerbossphantom}{\vphantom{\bar{\bar{x}}}}
\newcommand{\upperbossphantom}{\vphantom{\dagger}}
\newcommand{\tempop}[3][\textstyle]{\settowidth{\dimen1}{$#1\hat{#2}$}\makebox[\dimen1][l]{$#1\hat{#2\mspace{#3}}$}}
\newcommand{\xop}[1]{{\mathchoice{\tempop[\displaystyle]{#1}{3.5mu}}{\tempop{#1}{3.5mu}}{\tempop[\scriptstyle]{#1}{3.5mu}}{\tempop[\scriptscriptstyle]{#1}{3mu}}}}
\newcommand{\chat}[1]{\ensuremath{\xop{#1}}}
\newcommand{\aop}[2]{\ensuremath{\chat{c}_{#1#2\lowerbossphantom}^{\upperbossphantom}}}
\newcommand{\cop}[2]{\ensuremath{\chat{c}_{#1#2\lowerbossphantom}^{\dagger\upperbossphantom}}}
\newcommand{\cbar}[1]{\ensuremath{\xbar{#1}}}

\newcommand{\tempbar}[3][\textstyle]{\settowidth{\dimen1}{$#1\bar{#2}$}\makebox[\dimen1][l]{$#1\bar{#2\mspace{#3}}$}}
 
\newcommand{\xbar}[1]{{\mathchoice{\tempbar[\displaystyle]{#1}{3.5mu}}{\tempbar{#1}{3.5mu}}{\tempbar[\scriptstyle]{#1}{3.5mu}}{\tempbar[\scriptscriptstyle]{#1}{3mu}}}} 
\renewcommand{\i}{\mathrm{i}}
\renewcommand{\d}{\mathrm{d}}
\newcommand{\tn}[1]{\textnormal{#1}}

\usepackage{dcolumn}%
\usepackage{bm}%

\makeatletter
\g@addto@macro\bfseries{\boldmath}
\newcommand*{\balancecolsandclearpage}{%
   \close@column@grid
   \clearpage
   \twocolumngrid
 }
\makeatother

\begin{document}
\preprint{APS/123-QED}

\title{The G1--G2 Scheme: Dramatic Acceleration of Nonequilibrium Green Functions Simulations Within the Hartree--Fock-GKBA}

\author{Jan-Philip Joost, Niclas Schl\"unzen, and
Michael Bonitz
 \email{bonitz@theo-physik.uni-kiel.de}}
\affiliation{
Institut f\"ur Theoretische Physik und Astrophysik, 
Christian-Albrechts-Universit\"{a}t zu Kiel, D-24098 Kiel, Germany
}

\date{\today}%

\begin{abstract}
The time evolution in quantum many-body systems after external excitations is attracting high interest in many fields, including dense plasmas, correlated solids, laser excited materials or fermionic and bosonic atoms in optical lattices. The theoretical modeling of these processes is challenging, and the only rigorous quantum-dynamics approach that can treat correlated fermions 
 in two and three dimensions is nonequilibrium Green functions (NEGF). 
 However, NEGF simulations are computationally expensive due to their $T^3$-scaling with the simulation duration $T$. Recently, $T^2$-scaling was achieved with the generalized Kadanoff--Baym ansatz (GKBA), for the second-order Born (SOA) selfenergy, which has substantially extended the scope of NEGF simulations.
In a recent Letter [Schlünzen \textit{et al.}, Phys. Rev. Lett. \textbf{124}, 076601 (2020)] we 
demonstrated that
GKBA-NEGF simulations can be efficiently mapped onto coupled time-local equations for the single-particle and two-particle Green functions on the time diagonal, hence the method has been called G1--G2 scheme. This allows one to perform the same simulations with order $T^1$-scaling, both for SOA and $GW$ selfenergies giving rise to a dramatic speedup. Here we present more details on the G1--G2 scheme, including derivations of the basic equations including results for a general basis, for Hubbard systems and for jellium. Also, we demonstrate how to incorporate initial correlations into the G1--G2 scheme. Further, the derivations are extended to a broader class of selfenergies, including the $T$ matrix in the particle--particle and particle--hole channels, and the dynamically screened-ladder approximation. Finally, we demonstrate that, for all selfenergies, the CPU time scaling of the G1--G2 scheme with the basis dimension, $N_\tn{b}$, can be improved compared to our first report: the overhead compared to the original GKBA, is not more than an additional factor $N_\tn{b}$, 
even for Hubbard systems.
\end{abstract}

\maketitle

\section{Introduction}\label{s:intro}
Nonequibrium Green functions (NEGF) \cite{keldysh64,bonitz_pss_19_keldysh, kadanoff-baym-book}, have proven highly successful in simulations of the dynamics of correlated many-body systems. This is due to a number of attractive properties that  include conservation laws and the existence of systematic approximations schemes that are based on Feynman diagrams. Moreover, NEGF allow for a rigorous derivation of quantum kinetic equations and for their systematic improvement; for recent overviews, see the text books \cite{haug_2008_quantum, bonitz_qkt,stefanucci_cambridge_2013}. 

While early computational applications focused on spatially homogeneous systems such as nuclear matter \cite{DANIELEWICZ_84_ap2, koehler_prc_95},
optically excited semiconductors \cite{schaefer_wegener, haug_2008_quantum}, and dense plasmas \cite{semkat_99_pre,kremp_99_pre}, during the recent 15 years the scope of applications has substantially broadened. This includes the excitation and ionization dynamics of small atoms and molecules \cite{dahlen_solving_2007,balzer_pra_10,balzer_pra_10_2}, the correlated-electron dynamics in the Hubbard model \cite{von_friesen_successes_2009,von_friesen_kadanoff-baym_2010-1, hermanns_prb14}, the dynamics of fermionic atoms \cite{schluenzen_prb16, schluenzen_cpp16}, and the stopping of ions in correlated materials \cite{balzer_prl_18,bonitz_pss_18,schluenzen_cpp_18}.
This success was caused, among others, by progress in the numerical solution of the basic equations of NEGF---the Keldysh--Kadanoff--Baym equations \cite{koehler-kwong-code,bonitz_rinton,marini_2009_yambo,dahlen_solving_2007}. Furthermore, improved time propagation and integration schemes have allowed to increase the 
efficiency and accuracy of the simulations \cite{schluenzen_prb17_comment,schluenzen_prb17}. Moreover, the implementation of more advanced selfenergies, such as the $T$-matrix selfenergy, have allowed to increase the accuracy and predictive capability; for a recent review, see Ref.~\cite{schluenzen_jpcm_19}. In particular, very good agreement with cold-atom experiments \cite{schluenzen_prb16} and with \textit{ab initio} density-matrix-renormalization-group (DMRG) simulations were reported \cite{schluenzen_prb17}.
A particular advantage of NEGF simulation is that they are well capable to treat electronic correlations, in contrast to density-functional theory (DFT), and that they are neither restricted to 1D systems, such as DMRG, nor to short times, such as continuous-time quantum Monte Carlo \cite{gull_continous_2011}.

The main disadvantage of NEGF is their high numerical effort. The majority of many-body methods, including time-dependent DFT (TDDFT), Boltzmann-type quantum kinetic equations, hydrodynamics or semiclassical molecular dynamic---and even the exact solution of the time-dependent Schrödinger equation---require a simulation time that grows linearly with the physical time. In contrast, for NEGF, the propagation in the two-time plane, together with the memory integration in the scattering contributions, gives rise to a $N_\tn{t}^3$-scaling, where $N_\tn{t}$ is the propagation time (number of time steps). A substantial acceleration is possible when the 
generalized Kadanoff--Baym ansatz (GKBA) is applied  \cite{lipavsky_generalized_1986} which restricts the propagation to a time-stepping along the time diagonal. If combined with Hartree--Fock propagators (HF-GKBA) 
\cite{bonitz-etal.96jpcm,bonitz-etal.96pla,balzer_nonequilibrium_2012}
the CPU time scaling can be reduced to $N_\tn{t}^2$, which has given rise to a drastic increase of the number of HF-GKBA simulations in recent years, e.g. Refs.~\cite{hermanns_prb14,latini_charge_2014,schluenzen_prb17,marini_competition_2013,karlsson_gkba18,perfetto_pra_15,bostrom_charge_2018}. However, this improved scaling is achieved only for the simplest selfenergy---for the second-Born approximation (SOA). If the HF-GKBA is applied to improved selfenergies, such as the $T$-matrix selfenergy \cite{schluenzen_cpp16,schluenzen_prb17}, which is required for strongly correlated systems \cite{schluenzen_prb16}, or the $GW$ selfenergy \cite{schluenzen_jpcm_19} which is required to capture dynamical screening effects, the CPU time scaling is again increased to $N_\tn{t}^3$.

In a recent Letter we reported a breakthrough for NEGF simulations within the HF-GKBA scheme: we demonstrated that \textit{time-linear scaling}, i.e. a CPU time that is of order $N_\tn{t}^1$, can be achieved if the equations of motion are reformulated, without any approximations. The alternative approach solves the time-local equations for the time-diagonal single- and two-particle Green functions and was called G1--G2 scheme \cite{schluenzen_19_prl}. While the equivalence of the HF-GKBA to time-local equations was pointed out before \cite{bonitz_qkt,hermanns_jpcs13}, a comparison of the numerical behavior of both approaches was performed only in Ref.~\cite{schluenzen_19_prl}. There we predicted $N_\tn{t}^1$-scaling for SOA and $GW$ selfenergies and any type of single-particle basis. The scaling was demonstrated for small Hubbard clusters which turned out to be the most unfavorable case because the CPU time of the G1--G2 scheme was found to grow by a factor $N_\tn{b}^2$ faster than for the standard HF-GKBA approach.

In this article we present extensive additional results for the G1--G2 scheme. First, we present all necessary details for the derivation of the equation of motion for the time-diagonal two-particle Green function. The results are derived for a general basis, for the Hubbard model and for jellium. Second, we discuss how initial correlations can be
incorporated. Third, we extend the analysis to other selfenergies: the $T$-matrix approximation in the particle--particle (TPP) and particle--hole (TPH) channels and the dynamically-screened-ladder (DSL) approximation. 
Fourth, numerical results are demonstrated for all selfenergy approximations which clearly confirm the $N_\tn{t}^1$-scaling. Finally, we re-evaluate the $N_\tn{b}$-dependence of the CPU time and report an additional optimization that reduces the overhead of the new scheme from  
$N_\tn{b}^2$ to only $N_\tn{b}^1$, for the Hubbard model, for all selfenergies.

This paper is organized as follows: In Sec.~\ref{s:theory} we summarize the main required formulas of NEGF theory and the properties of the two-particle Green function. In Sec.~\ref{s:soa} we present the basic formulas for the G1--G2 scheme for the case of SOA selfenergy---separately, for a general basis, the Hubbard basis and for jellium. The same analysis is then extended to $GW$ and $T$-matrix selfenergies in Secs.~\ref{s:gw} and \ref{s:t-matrix}, and to the screened-ladder approximation in Sec.~\ref{s:dyn-scr-ladder}. Finally, the analysis of the scaling behavior with $N_\tn{t}$ and $N_\tn{b}$ for all selfenergies and numerical results are presented in Sec.~\ref{s:numerics}.

\section{Theoretical framework}\label{s:theory}
\subsection{Keldysh--Kadanoff--Baym Equations and two-particle Green function}
\label{ss:framework}
We consider a nonequilibrium quantum many-particle system with the generic Hamiltonian
\begin{align}
    \chat{H}(t) &= \sum_{ij} h^{(0)}_{ij} \cop{i}{} \aop{j}{} + \frac{1}{2} \sum_{ijkl} w_{ijkl} \cop{i}{} \cop{j}{} \aop{l}{} \aop{k}{} \, ,
    \label{eq:h}
\end{align}
containing a single-particle contribution $h^{(0)}$ and a pair interaction $w$. The matrix elements are computed with an orthonormal system of single-particle orbitals $|i\rangle$.
The creation ($\chat c^\dagger_i$) and annihilation ($\chat c_i$) operators of particles in state $|i\rangle$ define the one-body nonequilibrium Green function (correlation function) for contour-time arguments $z$ on the Keldysh contour $\mathcal{C}$~\cite{schluenzen_jpcm_19},
\begin{align}
    G_{ij}(z,z')=\frac{1}{\i\hbar}\left\langle \mathcal{T}_\mathcal{C} \left\{\chat{c}_i(z)\chat{c}^\dagger_j(z')\right\} \right\rangle\,.
\end{align}
Here, $\mathcal{T}_\mathcal{C}$ is the time-ordering operator on the contour, and the averaging is performed with the correlated unperturbed density operator of the system.
The equations of motion (EOMs) for the NEGF are the Keldysh--Kadanoff--Baym equations (KBE)~\footnote{Throughout this work, ``$\pm$'' refers to bosons/fermions.}
\begin{align}\label{eq:KBE1}
    \sum_k & \left[ \i\hbar \frac{\d}{\d z} \delta_{ik} - h^{(0)}_{ik}(z) \right] G_{kj}(z,z') = \delta_{ij} \delta_\mathcal{C}(z,z') \\ 
    &\pm \i\hbar \sum_{klp} \int_\mathcal{C} \d \cbar{z}\, w_{iklp}(z,\cbar{z})G^{(2)}_{lpjk}(z,\cbar{z},z',\cbar{z}^+)\,,
    \nonumber \\
   \sum_k & G_{ik}(z,z') \left[- \i\hbar \frac{\overset{\leftarrow}{\d}}{\d z'} \delta_{kj} - h^{(0)}_{kj}(z') \right] = \delta_{ij} \delta_\mathcal{C}(z,z') \label{eq:KBE2}\quad \\ 
   &\pm \i\hbar \sum_{klp} \int_\mathcal{C} \d \cbar{z}\, G^{(2)}_{iklp}(z,\cbar{z}^-,z',\cbar{z}) w_{lpjk}(\cbar{z},z')
   \, ,\nonumber
    \label{eq:kbe}
\end{align} 
(the times ${z^\pm \coloneqq z \pm \epsilon, \epsilon \ll 1}$ are slightly shifted to disambiguate time ordering)
that couple to the two-particle Green function
\begin{align}
    &G^{(2)}_{ijkl}(z_1,z_2,z_3,z_4)\nonumber \\
    &\qquad=\frac{1}{\left(\i\hbar\right)^2}\left\langle \mathcal{T}_\mathcal{C} \left\{\chat{c}_i(z_1)\chat{c}_j(z_2)\chat{c}^\dagger_l(z_4)\chat{c}^\dagger_k(z_3)\right\} \right\rangle\,,
\end{align}
which contains a mean-field (Hartree--Fock) and a correlation contribution
\begin{align}\label{eq:G2_parts}
    G^{(2)}_{ijkl}(z_1,z&_2,z_3,z_4) \\ &=G^{(2),\tn{H}}_{ijkl}(z_1,z_2,z_3,z_4) \pm G^{(2),\tn{F}}_{ijkl}(z_1,z_2,z_3,z_4) \nonumber\\
    &\quad+ G^{(2),\tn{corr}}_{ijkl}(z_1,z_2,z_3,z_4)\,.\nonumber
    \end{align}
Our scheme involves the special case of two-particle functions that depend either on one or two times and their real-time components~\cite{schluenzen_jpcm_19}, that we define as follows
    \begin{align}
    \mathcal{G}^{\tn{H}}_{ijkl}(z,z') &\coloneqq G^{(2),\tn{H}}_{ijkl}(z,z,z',z') = G_{ik}(z,z') G_{jl}(z,z')\,,\nonumber\\
    \mathcal{G}^{\tn{F}}_{ijkl}(z,z') &\coloneqq G^{(2),\tn{F}}_{ijkl}(z,z',z,z') = G_{il}(z,z') G_{jk}(z',z)\,,\nonumber\\
    \mathcal{G}^{\tn{corr}}_{ijkl}(z,z') &\coloneqq G^{(2),\tn{corr}}_{ijkl}(z,z,z',z^+)\,,\nonumber\\
    \mathcal{G}^{\tn{H},\gtrless}_{ijkl}(t,t') &\coloneqq G^\gtrless_{ik}(t,t') G^\gtrless_{jl}(t,t')\,,
    \label{eq:g2h-def}\\
    \mathcal{G}^{\tn{F},\gtrless}_{ijkl}(t,t') &\coloneqq G^\gtrless_{il}(t,t') G^\lessgtr_{jk}(t',t)\,,
    \label{eq:g2f-def}\\
    \mathcal{G}^{\tn{H},\gtrless}_{ijkl}(t) &\coloneqq \mathcal{G}^{\tn{H},\gtrless}_{ijkl}(t,t)\,,\nonumber\\
    \mathcal{G}^{\tn{F},\gtrless}_{ijkl}(t) &\coloneqq \mathcal{G}^{\tn{F},\gtrless}_{ijkl}(t,t)\,,\nonumber\\
    \mathcal{G}_{ijkl}(t) &\coloneqq \mathcal{G}^{\tn{corr},<}_{ijkl}(t,t)\, .
    \label{eq:gscor-def}
\end{align}
The time-diagonal two-particle Green function, $\mathcal{G}(t)$, defined by \refeqn{eq:gscor-def}, 
is the central quantity of the G1--G2 scheme. In general, and for all considered selfenergy approximations in this work, it obeys the following (pair-) exchange symmetries,
\begin{align}
    \mathcal{G}_{ijkl}(t) &= \mathcal{G}_{jilk}(t)\, , \label{eq:Gsymm_a}\\
    \mathcal{G}_{ijkl}(t) &= \Big[\mathcal{G}_{klij}(t)\Big]^*\, . \label{eq:Gsymm_b}
\end{align}

\subsection{Time-diagonal KBE}
\label{ss:diagonal-kbe}
Following \eqrefs{eq:KBE1}{eq:KBE2} the EOM for the less component of the NEGF on the real-time diagonal, ${G_{ij}^\gtrless(t) \coloneqq G_{ij}^\gtrless(t,t)}$, can be rewritten as~\cite{schlunzen_nonequilibrium_2016}
\begin{align}
 \i\hbar \frac{\d}{\d t}G_{ij}^<(t) - \left[ h^\mathrm{HF}(t),G^<(t) \right]_{ij} &= \big[I + I^\dagger\big]_{ij}(t)\,, \label{eq:eom_gone}
\end{align}
where the right-hand side contains the collision integral
\begin{align}
 I_{ij}(t) &= \sum_{k} \int_{t_0}^t \mathrm{d}\cbar{t} \left[ \Sigma_{ik}^>(t,\cbar{t}) G_{kj}^<(\cbar{t},t) - \Sigma_{ik}^<(t,\cbar{t}) G_{kj}^>(\cbar{t},t) \right]
 \nonumber\\
\label{eq:definition_D}
  &= \pm\i\hbar \sum_{klp} w_{iklp}(t) \mathcal{G}_{lpjk}(t) \,.
\end{align}
Here, $t_0$ marks the time, at which the system's initial state is prepared; the treatment of initial correlations is discussed in Sec.~\ref{ss:inicor}. 
On the time diagonal the less component of the NEGF can be written as 
\begin{align}
    G_{ij}^<(t) &= G_{ij}^>(t) - \frac{1}{\i\hbar}\delta_{ij} = \pm \frac{1}{\i\hbar} n_{ij}(t)\,,
\label{eq:g-glsymm}
\end{align}
where $n_{ij}$ is the single-particle density matrix. In \refeqn{eq:eom_gone} the mean-field part of the two-particle Green function, cf. \refeqn{eq:G2_parts}, is included in an effective single-particle Hartree--Fock Hamiltonian which is defined as~\cite{schluenzen_jpcm_19}
\begin{align}
 h^\mathrm{HF}_{ij}(t) = h^{(0)}_{ij}(t) \pm \mathrm{i}\hbar \sum_{kl} w^\pm_{ikjl}(t) G^<_{lk}(t,t)\,, \label{eq:h_HF}
\end{align}
where we introduced the (anti-)symmetrized four-dimensional interaction matrix elements
\begin{align}
 w^\pm_{ijkl}(t) \coloneqq w_{ijkl}(t) \pm w_{ijlk}(t)\,.
\end{align}
The interaction tensor obeys the same symmetries as $\mathcal{G}(t)$ [cf. \refeqns{eq:Gsymm_a}{eq:Gsymm_b}],
\begin{align}
    w_{ijkl}(t) &= w_{jilk}(t)\, , \label{eq:wsymm_a}\\
    w_{ijkl}(t) &= \Big[w_{klij}(t)\Big]^*\, , \label{eq:wsymm_b}
\end{align}
which also leads to
\begin{align}
    w^\pm_{ijkl}\left( t\right) = \pm w^\pm_{ijlk}\left( t\right) \, . \label{eq:wsymm_c}
\end{align}
The selfenergy $\Sigma$ in the collision integral \eqref{eq:definition_D} contains only the remaining correlation contribution of the two-particle Green function.
\section{Second-Order Born Selfenergy}\label{s:soa}
In the following
we introduce the G1--G2 scheme for the simplest case of choosing the selfenergy in the second-Born approximation~\cite{schlunzen_nonequilibrium_2016},
\begin{align}
 \Sigma^\gtrless_{ij}\left(t,t'\right) 
 &= \left(\i\hbar\right)^2 \sum_{klpqrs} \, w_{iklp}\left(t\right) w^\pm_{qrsj}\left(t'\right) 
 \label{eq:sigma-soa} \\ \nonumber
 &\qquad \qquad \times G^\gtrless_{lq}\left(t,t'\right) G^\gtrless_{pr}\left(t,t'\right) G^\lessgtr_{sk}\left(t',t\right)\, .
\end{align}
With that, the collision integral of the time-diagonal equation \eqref{eq:definition_D} transforms into:
\begin{align}
 I_{ij}(t) =
 & \left(\i\hbar\right)^2 \sum_{klpqrsu} \, w_{iklp}\left(t\right) \int_{t_0}^t \d\cbar t\, w^\pm_{qrsu}\left(\cbar t\right) \times \\
 & \times\Big[ G^>_{lq}\left(t,\cbar t\right) G^>_{pr}\left(t,\cbar t\right) G^<_{sk}\left(\cbar t,t\right) G^<_{uj}\left(\cbar t,t\right) \nonumber \\
 &\quad - G^<_{lq}\left(t,\cbar t\right) G^<_{pr}\left(t,\cbar t\right) G^>_{sk}\left(\cbar t,t\right) G^>_{uj}\left(\cbar t,t\right) \Big] \nonumber \\
 = & \left(\i\hbar\right)^2 \sum_{klpqrsu} \, w_{iklp}\left(t\right) \int_{t_0}^t \d\cbar t\, w^\pm_{qrsu}\left(\cbar t\right) \times \\
 & \times \Big[ \mathcal{G}^{\tn{H},>}_{lpqr}(t,\cbar t) \mathcal{G}^{\tn{H},<}_{usjk}(\cbar t, t) - \mathcal{G}^{\tn{H},<}_{lpqr}(t,\cbar t) \mathcal{G}^{\tn{H},>}_{usjk}(\cbar t, t) \Big] \nonumber \\
 = & \left(\i\hbar\right)^2 \sum_{klpqrsu} \, w_{iklp}\left(t\right) \int_{t_0}^t \d\cbar t\, w^\pm_{qrsu}\left(\cbar t\right) \times \\
 & \times \Big[ \mathcal{G}^{\tn{F},>}_{ljqu}(t,\cbar t) \mathcal{G}^{\tn{F},<}_{srkp}(\cbar t, t) - \mathcal{G}^{\tn{F},<}_{ljqu}(t,\cbar t) \mathcal{G}^{\tn{F},>}_{srkp}(\cbar t, t) \Big] \, ,
\end{align}
where we presented several equivalent formulations that will be used below.
At this point, it is possible to identify $\mathcal{G}$ [cf. \refeqn{eq:definition_D}] in SOA,
\begin{align}
 \mathcal{G}_{ijkl}(t) =& \pm \i\hbar\sum_{pqrs} \int_{t_0}^t \d\cbar t\, w^\pm_{pqrs}\left(\cbar t\right) \times \\
 & \times\Big[ \mathcal{G}^{\tn{H},>}_{ijpq}(t,\cbar t) \mathcal{G}^{\tn{H},<}_{srkl}(\cbar t, t) - \mathcal{G}^{\tn{H},<}_{ijpq}(t,\cbar t) \mathcal{G}^{\tn{H},>}_{srkl}(\cbar t, t) \Big] \nonumber \\
  \overset{(*)}{=}& \i\hbar\sum_{pqrs} \int_{t_0}^t \d\cbar t\, w^\pm_{pqrs}\left(\cbar t\right) \times \label{eq:g2-integral}
  \\
 & \times\Big[ \mathcal{G}^{\tn{H},>}_{ijpq}(t,\cbar t) \mathcal{G}^{\tn{H},<}_{rskl}(\cbar t, t) - \mathcal{G}^{\tn{H},<}_{ijpq}(t,\cbar t) \mathcal{G}^{\tn{H},>}_{rskl}(\cbar t, t) \Big] \nonumber \, ,
\end{align}
where, in the transformation (*), the symmetry of \refeqn{eq:wsymm_c} is used and two summation indices are switched, $r\leftrightarrow s$. 

\subsection{%
$\mathcal{G}$ within the GKBA}\label{ss:g2-hf-gkba}
The G1--G2 scheme is a reformulation of the ordinary solution of the time-diagonal KBE in the HF-GKBA. When applying the GKBA the time off-diagonal elements of the less and greater NEGF are reconstructed from the time diagonal value via~\cite{schlunzen_nonequilibrium_2016}
\begin{align}
 G_{ij}^\gtrless(t, t') = \i \hbar \sum_k \left[G_{ik}^\mathrm{R}(t,t') G_{kj}^\gtrless(t')-G_{ik}^\gtrless(t) G_{kj}^\mathrm{A}(t,t')\right]\, . \label{eq:GKBA}
\end{align}
We now rewrite this reconstruction, taking into account that in the collision integral \eqref{eq:definition_D} only $G^>(t\geq \cbar{t})$ and $G^<(\cbar{t}\leq t)$ appear. Replacing them with the GKBA and taking into account the special case of equal times results in
\begin{align}
 G_{ij}^\gtrless(t'\leq t) &= -\i \hbar \sum_k G_{ik}^\gtrless(t') \Big[ G_{kj}^\mathrm{A}(t',t) - \delta_{\cbar t,t} G_{kj}^\mathrm{R}(t,t) \Big]\nonumber\\
 &= -\i \hbar \sum_k G_{ik}^\gtrless(t') \Big[ G_{kj}^\mathrm{A}(t',t) - G_{kj}^\mathrm{R}(t',t) \Big]\nonumber\\
 &= \phantom{-}\i \hbar \sum_k G_{ik}^\gtrless(t')  \mathcal{U}_{kj}(t',t)\,, \label{eq:GKBA_propa} \\
 G_{ij}^\gtrless(t\geq t') &= \phantom{-}\i \hbar \sum_k \Big[G_{ik}^\mathrm{R}(t,t') - \delta_{\cbar t,t} G_{ik}^\mathrm{A}(t,t) \Big] G_{kj}^\gtrless(t')
 \nonumber \\
 &= \phantom{-}\i \hbar \sum_k \Big[G_{ik}^\mathrm{R}(t,t') - G_{ik}^\mathrm{A}(t,t') \Big] G_{kj}^\gtrless(t')\nonumber \\
&= \phantom{-}\i \hbar \sum_k \mathcal{U}_{ik}(t,t') G_{kj}^\gtrless(t')\,, \label{eq:GKBA_propb}
\end{align}
where we introduce the modified propagator (time-evolution operator)
\begin{align}
 \mathcal{U}_{ik}(t,t') &= G_{ik}^\mathrm{R}(t,t') - G_{ik}^\mathrm{A}(t,t')\,, \label{eq:prop_b}
\end{align}
which on the time diagonal reduces to~\cite{schlunzen_nonequilibrium_2016},
\begin{align}
 \mathcal{U}_{ij}(t,t) &= G_{ij}^\mathrm{R}(t,t) - G_{ij}^\mathrm{A}(t,t) = G_{ij}^>(t) - G_{ij}^<(t) \nonumber \\ &= \frac{1}{\i\hbar} \delta_{ij}\, . \label{eq:unity_prop}
\end{align}
By inserting the GKBA, \eqrefs{eq:GKBA_propa}{eq:GKBA_propb}, for the Green functions into \refeqn{eq:g2-integral}, we find (cf. Appendix~\ref{app:Usymm})
\begin{align}
 \mathcal{G}^\GKBA_{ijkl}(t) &= \i\hbar \sum_{pqrsuvxy}\int_{t_0}^t \mathrm{d}\cbar{t}\, w^\pm_{pqrs}(\cbar{t}) \times
\label{eq:dtilde-solution-short-phi}
 \\ & \quad\nonumber \times
  \mathcal{U}_{ijuv}^{(2)}(t,\cbar{t}) 
  \Phi^{uvrs}_{pqxy}(\cbar{t})
  \mathcal{U}_{xykl}^{(2)}(\cbar{t},t)\,,
\end{align}
where we now introduce short notations for the two-particle propagator $\mathcal{U}^{(2)}$ and define the occupation factors $\Phi^\gtrless$,
\begin{align}
  \mathcal{U}_{ijkl}^{(2)}(t,t') &= \mathcal{U}_{ik}(t,t') \mathcal{U}_{jl}(t,t') = \mathcal{U}_{jilk}^{(2)}(t,t')\,, \label{prop_two}
  \\
\Phi^{ijkl}_{pqrs}({t})&=\Phi^{ijkl,>}_{pqrs}({t}) - \Phi^{ijkl,<}_{pqrs}({t})\,,
  \\
\Phi^{ijkl,\gtrless}_{pqrs}({t}) &= (\i\hbar)^4 \mathcal{G}^{\tn{H},\gtrless}_{ijpq}({t}) \mathcal{G}^{\tn{H},\lessgtr}_{klrs}({t}) \\
&= (\i\hbar)^4 \mathcal{G}^{\tn{F},\gtrless}_{ikrp}({t}) \mathcal{G}^{\tn{F},\gtrless}_{jlsq}({t}) = (\i\hbar)^4 \mathcal{G}^{\tn{F},\gtrless}_{ilsp}({t}) \mathcal{G}^{\tn{F},\lessgtr}_{kjqr}({t})\,. \nonumber
\end{align}
An even more compact notation can be achieved by introducing the two-particle source term
\begin{align}
    \Psi^\pm_{ijkl}(t) = \frac{1}{\left(\i\hbar\right)^2}\sum_{pqrs} w^\pm_{pqrs}(t)\Phi^{ijrs}_{pqkl}(t)\,,
    \label{eq:psi-pm-def}
\end{align}
which results in
\begin{align}\label{eq:dtilde-solution-short}
 \mathcal{G}^\GKBA_{ijkl}(t) &= \left(\i\hbar\right)^3 \sum_{pqrs}\int_{t_0}^t \mathrm{d}\cbar{t}\, 
  \mathcal{U}_{ijpq}^{(2)}(t,\cbar{t}) 
  \Psi^\pm_{pqrs}(\cbar{t})
  \mathcal{U}_{rskl}^{(2)}(\cbar{t},t)\,.\phantom{..}
\end{align}
The function $\Psi^\pm$ has the meaning of pair correlations produced in the system via two-particle scattering per unit time.

\subsection{Time-linear integral solution for $\mathcal{G}^\GKBA$} \label{ss:integral_sol}
When applying the HF-GKBA the retarded and advanced propagators, $G^\mathrm{R}(t,t')$ and $G^\mathrm{A}(t,t')$, are described in HF approximation and, thus, obey the group property~\cite{karlsson_gkba18} for $t> \cbar{t}> t'$:
\begin{align}
 G_{ij}^\mathrm{A}(t',t) &= -\i \hbar \sum_k G_{ik}^\mathrm{A}(t',\cbar{t}) G_{kj}^\mathrm{A}(\cbar{t},t)\,, \label{group_ret}\\
 G_{ij}^\mathrm{R}(t,t') &= \phantom{-}\i \hbar \sum_k G_{ik}^\mathrm{R}(t,\cbar{t}) G_{kj}^\mathrm{R}(\cbar{t},t')\,. \label{group_adv}
\end{align}
As we show in Appendix~\ref{app:Ugroup} it directly follows that the two-particle propagator also possesses the group property,
\begin{align}\label{eq:prop_two_group}
 \mathcal{U}^{(2)}_{ijkl}(t,t') = \left(\i\hbar\right)^2 \sum_{pq} \mathcal{U}^{(2)}_{ijpq}(t,\cbar{t}) \mathcal{U}^{(2)}_{pqkl}(\cbar{t},t')\,.
\end{align}
To reveal the time-linear core of \refeqn{eq:dtilde-solution-short} for Hartree--Fock propagators, we consider a time $T+\Delta$ for which the time integral can be split into two intervals $[t_0,T]$ and $[T,T+\Delta]$, resulting in
\begin{align}
 &\mathcal{G}^\GKBA_{ijkl}(T+\Delta) %
 = \mathcal{G}^\GKBAd_{ijkl}(T) \\
  &+\left(\i\hbar\right)^3 \sum_{pqrs}\int_{t_0}^{T} \mathrm{d}\cbar{t}\, \mathcal{U}_{ijpq}^{(2)}(T+\Delta,\cbar{t}) 
  \Psi^\pm_{pqrs}(\cbar{t})
  \mathcal{U}_{rskl}^{(2)}(\cbar{t},T+\Delta)\,,\nonumber
  \end{align}
with
\begin{align}
 \mathcal{G}^\GKBAd_{ijkl}(T) &\coloneqq \left(\i\hbar\right)^3 \sum_{pqrs}\int_{T}^{T+\Delta} \mathrm{d}\cbar{t}\, \times
 \\ & \nonumber \times
  \mathcal{U}_{ijpq}^{(2)}(T+\Delta,\cbar{t}) 
  \Psi^\pm_{pqrs}(\cbar{t})
  \mathcal{U}_{rskl}^{(2)}(\cbar{t},T+\Delta)\,.
\end{align}
Applying the group property of the two-particle propagator, \refeqn{eq:prop_two_group}, leads to
\begin{align}
 &\mathcal{G}^\GKBA_{ijkl}(T+\Delta) = \mathcal{G}^\GKBAd_{ijkl}(T) \\&+ \left(\i\hbar\right)^7 \sum_{pqrsuvxy}\int_{t_0}^{T} \mathrm{d}\cbar{t}\, 
  \mathcal{U}_{ijpq}^{(2)}(T+\Delta,T) \mathcal{U}_{pqrs}^{(2)}(T,\cbar{t})\times\nonumber\\ 
  &\quad\times \Psi^\pm_{rsuv}(\cbar{t}) \mathcal{U}_{uvxy}^{(2)}(\cbar{t},T)
  \mathcal{U}_{xyjm}^{(2)}(T,T+\Delta)\nonumber\, ,
\end{align}
where we identify the two-particle Green function at time $T$,
\begin{align} \label{eq:linintG}
 &\mathcal{G}^\GKBA_{ijkl}(T+\Delta) = \mathcal{G}^\GKBAd_{ijkl}(T)\\
  &+\left(\i\hbar\right)^4 \sum_{pqrs} \mathcal{U}_{ijpq}^{(2)}(T+\Delta,T)
  \mathcal{G}^\GKBA_{pqrs}(T)
  \mathcal{U}_{rskl}^{(2)}(T,T+\Delta) \nonumber \,.
\end{align}
The above expression only contains a time integral of fixed length $\Delta$. Thus, provided that the solution $\mathcal{G}^\GKBA(T)$ is known, the propagation to $T+\Delta$ can be done in a constant amount of time, independent of $T$. While \refeqn{eq:linintG}, in principle, provides the basis for a time-linear propagation scheme, its integral form proves to be unfavorable for numerical implementation. Therefore, in the following, an alternative approach (G1--G2 scheme \cite{schluenzen_19_prl}), that is based on the solution of a (time-local) differential equation, is derived which will be analyzed throughout this paper.

\subsection{Time-linear differential solution for $\mathcal{G}^\GKBA$: SOA-G1--G2 equations for a general basis}
Here we consider a general single-particle basis where spin degrees of freedom are included in the basis index. Below we will separately consider the special cases of a Hubbard basis and the jellium model for electrons where the two spin projections will be indicated explicitly.
In order to find the differential equation for $\mathcal{G}^\GKBA$, the EOMs for the retarded/advanced Green functions in HF-GKBA along both time-directions and the diagonal are repeated~\cite{schlunzen_nonequilibrium_2016}:
\begin{align}
\i\hbar \frac{\d}{\d t} G^\tn{R/A}_{ij}(t,t') &= \sum_k h^\tn{HF}_{ik}(t) G^\tn{R/A}_{kj} (t,t') + \delta_{ij} \delta(t,t') 
\nonumber\\
\i\hbar \frac{\d}{\d t} G^\tn{R/A}_{ij}(t',t) &=- \sum_k G^\tn{R/A}_{ik} (t',t) h^\tn{HF}_{kj}(t) - \delta_{ij} \delta(t,t') 
\nonumber\\
\i\hbar \frac{\d}{\d t} G^\tn{R/A}_{ij}(t'=t) 
&= \left[h^\tn{HF}(t), G^\tn{R/A} (t,t)\right]_{ij} \, .
\label{eq:eom_twotime}
\end{align}
For the two-particle propagators similar Schrödinger-type EOMs hold as shown in Appendix~\ref{app:Ueom},
\begin{align}
 \frac{\d}{\d t} \mathcal{U}^{(2)}_{ijkl} (t,t')  \label{prod_ret}
 &= \frac{1}{\i\hbar} \sum_{pq} h^{(2),\tn{HF}}_{ijpq} (t) \mathcal{U}^{(2)}_{pqkl} (t,t')\, ,\\
 \frac{\d}{\d t} \mathcal{U}^{(2)}_{ijkl} (t',t) &= -\frac{1}{\i\hbar} \sum_{pq} \mathcal{U}^{(2)}_{ijpq}(t',t) h^{(2),\tn{HF}}_{pqkl} (t)\label{prod_adv} \, , 
\end{align}
where we define the two-particle Hartree--Fock Hamiltonian as the sum of two single-particle parts:
\begin{align}
 h^{(2),\tn{HF}}_{ijkl}(t) = \delta_{jl} h^\tn{HF}_{ik}(t) + \delta_{ik}h^\tn{HF}_{jl}(t)\, . \label{twopart_hamiltonian}
\end{align}
With that we now compute the time derivative of the time-diagonal two-particle Green function within the HF-GKBA, $\mathcal{G}^\GKBA$, which contains two parts,
\begin{align}\label{eq:eom_parts_soa}
  \frac{\d}{\d t} \mathcal{G}^\GKBA_{ijkl} (t) &= \left[\frac{\d}{\d t} \mathcal{G}^\GKBA_{ijkl} (t)\right]_{\int^{}_{}} + \left[\frac{\d}{\d t} \mathcal{G}^\GKBA_{ijkl} (t)\right]_{\mathcal{U}^{(2)}}\,.
\end{align}
The first contribution ($\int^{}_{}$) originates from the integration boundaries,
\begin{align}
 \left[\frac{\d}{\d t} \mathcal{G}^\GKBA_{ijkl} (t) \right]_{\int^{}_{}}&= \left(\i\hbar\right)^3 \sum_{pqrs} \mathcal{U}_{ijpq}^{(2)}(t,t) \Psi^\pm_{pqrs}(t)
  \mathcal{U}_{rskl}^{(2)}(t,t)\nonumber \\
  &= \frac{1}{\i\hbar} \Psi^\pm_{ijkl}(t)\, ,
\label{eq:dgdt-a}
\end{align}
where the latter equation holds due to the identity [cf. \refeqns{eq:unity_prop}{prop_two}]
\begin{align}
\mathcal{U}_{ijkl}^{(2)}(t,t) = \frac{1}{\left(\i\hbar\right)^2}\delta_{ik} \delta_{jl} \, . \label{eq:unity_prop_two}
\end{align}
The second contribution to the derivative results from the time dependence of the integrand, i.e. of $\mathcal{U}^{(2)}$,
\begin{align}
  &\left[\frac{\d}{\d t} \mathcal{G}^\GKBA_{ijkl} (t) \right]_{\mathcal{U}^{(2)}}= \left(\i\hbar\right)^3 \sum_{pqrs}\int_{t_0}^t \mathrm{d}\cbar{t}\, \Psi^\pm_{pqrs}(\cbar{t}) \times
 \\ & \nonumber \times \Bigg\{\left[\frac{\d}{\d t} \mathcal{U}_{ijpq}^{(2)}(t,\cbar{t}) \right]
  \mathcal{U}_{rskl}^{(2)}(\cbar{t},t) +\mathcal{U}_{ijpq}^{(2)}(t,\cbar{t})
  \left[\frac{\d}{\d t}\mathcal{U}_{rskl}^{(2)}(\cbar{t},t)\right]\Bigg\}\,,
\end{align}
and, using the results from \eqrefs{prod_ret}{prod_adv}, we obtain
 \begin{align}
  &\left[\frac{\d}{\d t} \mathcal{G}^\GKBA_{ijkl} (t) \right]_{\mathcal{U}^{(2)}}= \left(\i\hbar\right)^3 \sum_{pqrs}\int_{t_0}^t \mathrm{d}\cbar{t}\, \times
 \nonumber\\ & \nonumber \times
  \Bigg\{ \bigg[\frac{1}{\i\hbar} \sum_{uv} h^{(2),\tn{HF}}_{ijuv} (t) \mathcal{U}^{(2)}_{uvpq} (t,\cbar t) \bigg]
  \Psi^\pm_{pqrs}(\cbar{t}) \mathcal{U}_{rskl}^{(2)}(\cbar{t},t) \\ \nonumber
  &\quad +\mathcal{U}_{ijpq}^{(2)}(t,\cbar{t})
  \Psi^\pm_{pqrs}(\cbar{t}) \bigg[-\frac{1}{\i\hbar} \sum_{uv} \mathcal{U}^{(2)}_{rsuv}(\cbar t,t) h^{(2),\tn{HF}}_{uvkl} (t) \bigg]\Bigg\}\,,
\end{align}
where we identify $\mathcal{G}^\GKBA$ again, to get
  \begin{align}
  \left[\frac{\d}{\d t} \mathcal{G}^\GKBA_{ijkl} (t) \right]_{\mathcal{U}^{(2)}}&= \frac{1}{\i\hbar} \sum_{pq} h^{(2),\tn{HF}}_{ijpq} (t)\, \mathcal{G}^\GKBA_{pqkl} (t) 
 \label{eq:dgdt-b}
  \\ &\quad -\frac{1}{\i\hbar} \sum_{lf} \mathcal{G}^\GKBA_{ijpq}(t)\, h^{(2),\tn{HF}}_{pqkl} (t)\,. \nonumber
\end{align}
With that, the full derivative of the time-diagonal two-particle Green function is obtained by adding up the results of \eqrefs{eq:dgdt-a}{eq:dgdt-b},
\begin{align}
\i\hbar \frac{\d}{\d t} \mathcal{G}^\GKBA_{ijkl} (t) - %
  \left[h^{(2),\tn{HF}}, \mathcal{G}^\GKBA\right]_{ijkl} (t) 
 \label{eq:result_eom_soa} = \Psi^\pm_{ijkl}(t) \, . 
\end{align}

We now summarize the equations of the G1--G2 scheme for second order Born selfenergies, for a general basis. The scheme consists of the equation for the time-diagonal element of the single-particle Green function, cf. \refeqn{eq:eom_gone},
\begin{align}
 \i\hbar \frac{\d}{\d t}G_{ij}^<(t) &= \left[ h^\mathrm{HF}(t),G^<(t) \right]_{ij} + \left[I + I^\dagger\right]_{ij}(t)\,, \quad 
 \label{eq:eom_gone-2}
\\
 I_{ij}(t) &= \pm\i\hbar \sum_{klp} w_{iklp}(t) \mathcal{G}^\GKBA_{lpjk}(t) \,,
 \label{eq:eom_gone-I}
\end{align}
coupled to \refeqn{eq:result_eom_soa}---the EOM of the time-diagonal element of the two-particle Green function.
Equations \eqref{eq:result_eom_soa}, \eqref{eq:eom_gone-2}, and \eqref{eq:eom_gone-I} constitute 
a closed system of time-local differential equations, for which the computational effort for a numerical implementation scales linearly with time. This was achieved by eliminating the non-Markovian (memory) structure of the collision integral. All transformations so far are exact and reproduce the standard HF-GKBA result, as was demonstrated in Ref.~\cite{schluenzen_19_prl}. The linear scaling with $N_t$, as opposed to the quadratic scaling of the standard HF-GKBA in SOA, is the basis for a potentially dramatic speedup of NEGF simulations. The price to pay is the need to compute the entire matrix of the time-diagonal two-particle Green function the effort for which only depends on the basis dimension $N_\tn{b}$. This will be analyzed in detail in Sec.~\ref{s:numerics}.

In similar manner as for the SOA selfenergy, a time-local equation for $\mathcal{G}^\GKBA$ corresponding to more advanced selfenergies can be derived for which the speedup of the G1--G2 scheme is even larger. This will be demonstrated in the subsequent sections. But before that, we consider the G1--G2 scheme in SOA for two important special cases of basis sets---the Hubbard basis and the spatially uniform jellium model (plane-wave basis). 

\subsection{SOA-G1--G2 equations for the Hubbard model}\label{ss:hubbard-soa}
The Hubbard model \cite{hubbard_1963} is among the fundamental models in condensed matter physics, in particular, for the analysis of strong electronic correlations. More recently it has been widely used to study the behavior of fermionic and bosonic atoms in optical lattices \cite{bloch_probing_2014} and, in particular time-dependent correlation phenomena, see, e.g, Refs.~\cite{kajala_expansion_2011,schneider_fermionic_2012,von_friesen_kadanoff-baym_2010-1, hermanns_prb14}.
For the Fermi--Hubbard model, the general pair-interaction matrix element becomes
\begin{align}
 w_{ijkl}^{\alpha\beta\gamma\delta}(t) = U(t) \delta_{ij} \delta_{ik} \delta_{il} \delta_{\alpha \gamma} \delta_{\beta \delta} \cbar{\delta}_{\alpha \beta}\, ,
 \label{eq:wmatrix-hubbard}
\end{align}
with the on-site interaction $U$ and the spin projection labeled by greek indices. The kinetic energy matrix is replaced by a hopping Hamiltonian,
\begin{align}
    h^{(0)}_{ij} = - \delta_{\left<i,j\right>} J \, ,
\end{align}
which includes hopping processes between nearest-neighbor sites $\left<i,j\right>$ with amplitude $J$.
Thus, the total Hamiltonian is given by
\begin{align}
 \chat{H}(t) = -J\sum_{\left<i,j\right>} \sum_\alpha \cop{i}{\alpha}\aop{j}{\alpha} + U(t) \sum_i \chat{n}_{i}^{\uparrow} \chat{n}_{i}^{\downarrow}\, .
\label{eq:h-hubbard}
\end{align}
Extensions to more complicated models, going beyond the nearest neighbor single-band case are straightforward but will not be considered here.

The time-diagonal EOM for the single-particle Green function, \refeqn{eq:eom_gone}, takes the following form (from here we give all Hubbard equations for the spin-up component; the spin-down equations follow from the replacement $\uparrow\,\leftrightarrow\,\downarrow$.)
\begin{align}\label{eq:eom_g1_hubbard}
\i\hbar \frac{\d}{\d t} G^{<,\uparrow}_{ij}(t) &= \left[h^{\tn{HF}}, G^{<}\right]^{\uparrow}_{ij}(t) + \left[I+I^\dagger\right]^{\uparrow}_{ij}(t)\, ,\quad \\
 I^{\uparrow}_{ij}(t) &= -\i\hbar U(t) \mathcal{G}^{\uparrow\downarrow\uparrow\downarrow}_{iiji}(t) \,, \label{eq:colint_hubbard}
\end{align}
where for electrons there exist two collision integrals, $I^\uparrow$ and $I^\downarrow$,  that enter the single-particle EOMs. 
The Hartree--Fock Hamiltonian in \refeqn{eq:eom_g1_hubbard} in the Hubbard basis becomes [cf. \refeqn{eq:h_HF}]:
\begin{align}
    h^{\tn{HF},\uparrow}_{ij}(t) = h^{(0)}_{ij} -\i\hbar \delta_{ij} U(t) G^{<,\downarrow}_{ii}(t) \, .
\end{align}
The equation for the time-diagonal two-particle Green function, \refeqn{eq:result_eom_soa}, now reads
\begin{align}
 \i\hbar\frac{\d}{\d t} & \mathcal{G}^{\uparrow \downarrow \uparrow \downarrow}_{ijkl} (t) - \left[h^{(2),\tn{HF}}_{\uparrow\downarrow}, \mathcal{G}^{\uparrow \downarrow \uparrow \downarrow}\right]_{ijkl} (t) 
 \nonumber \\ 
 &= \frac{1}{\left(\i\hbar\right)^2} U(t)\,
\Phi^{\uparrow\downarrow\uparrow\downarrow}_{ijkl}(t) \eqqcolon \Psi_{ijkl}^{\uparrow\downarrow\uparrow\downarrow}(t) \,, \label{eq:g2-hubbard}
\end{align}
where
\begin{align}
 h^{(2),\tn{HF}}_{ijkl,\uparrow\downarrow}(t) = \delta_{jl} h^{\tn{HF},\uparrow}_{ik}(t) + \delta_{ik}h^{\tn{HF},\downarrow}_{jl}(t)\, , \label{eq:twopart_hamiltonian_hubbard}
\end{align}
and
\begin{align}
\Phi^{\uparrow\downarrow\uparrow\downarrow}_{ijkl}(t) &\coloneqq \left(\i\hbar\right)^4 \sum_{p}
  \left[ G^{>,\uparrow}_{ip}(t) G^{>,\downarrow}_{jp}(t) G^{<,\uparrow}_{pk}(t) G^{<,\downarrow}_{pl}(t)\right. \nonumber \\
  &\qquad \left.-G^{<,\uparrow}_{ip}(t) G^{<,\downarrow}_{jp}(t) G^{>,\uparrow}_{pk}(t) G^{>,\downarrow}_{pl}(t) \right] \, .
\label{eq:phi-hubbard}  
\end{align}
The \refeqns{eq:eom_g1_hubbard}{eq:g2-hubbard} form a coupled system of four equations. For SOA, no further spin combinations of $\mathcal{G}$ contribute. Numerical examples will be presented in Sec.~\ref{s:numerics}.

\subsection{SOA-G1--G2 equations for jellium}\label{ss:soa-jellium}
As the second example we consider the jellium Hamiltonian \cite{dornheim_physrep_18}, 
\begin{align}
 \chat{H}(t) &=& \sum_{\bm{p}\alpha} %
 \frac{\bm{p}^2}{2m}\cop{\bm{p}}{\alpha}\aop{\bm{p}}{\alpha} 
 + \sum_{\bm{p}\bm{p'}\bm{q}\alpha\beta} %
 v_{\left|\bm{q}\right|} \cop{\bm{p}+\bm{q}}{\alpha} \cop{\bm{p'}-\bm{q}}{\beta} \aop{\bm{p'}}{\beta} \aop{\bm{p}}{\alpha}\,,
\label{eq:h-jellium}
\end{align}
with the Coulomb matrix element $v_{\left|\bm{q}\right|} = \frac{4\pi e^2}{\left|\bm{q}\right|^2}$. 
This model is of relevance for the electron gas in metals \cite{ziman_electrons_1960, mahan-book}, for electron-hole plasmas in semiconductors~\cite{haug_2008_quantum}, and for dense quantum plasmas, e.g. \cite{bonitz_qkt, semkat_99_pre}, as well as for model development \cite{mahan-book, dornheim_physrep_18}.

The matrix element of the pair interaction in a plane wave basis is
\begin{align}
 w_{\bm{k_1}\bm{k_2}\bm{k_3}\bm{k_4}}^{\alpha\beta\gamma\delta}(t) = \delta_{\alpha \gamma} \delta_{\beta \delta} \delta(\bm{k_1}+\bm{k_2}-\bm{k_3}-\bm{k_4}) v_{\left|\bm{k_1}-\bm{k_3}\right|}(t) \, ,
\label{eq:jellium-w-matrix}
\end{align}
where $v_q$ denotes the spatial Fourier transform of the pair potential, and the delta function arises from momentum conservation (spatial homogeneity). 

The EOM for the single-particle Green function, \refeqn{eq:eom_gone}, is now %
\begin{align}
\i\hbar \frac{\d}{\d t} G^{<,\alpha}_{\bm{p}}(t) = I^\alpha_{\bm{p}}(t) = \pm\i\hbar \sum_{\bm{\cbar p},\bm{q}} \sum_\beta v_{\left|\bm{q}\right|}(t) \mathcal{G}_{\bm{p} \bm{\cbar p} \bm{q}}^{ \alpha \beta }(t) \,, \qquad\label{eq:jellium-g1}
\end{align}
where we defined 
\begin{align}
  \mathcal{G}_{\bm{p},\bm{\cbar p},\bm{q}}^{\alpha \beta}(t) &:= \mathcal{G}_{\bm{p}-\bm{q},\bm{\cbar p}+\bm{q},\bm{p},\bm{\cbar p}}^{\alpha \beta \alpha \beta}(t)\, ,
\label{eq:g2-def-jellium}
\end{align}
and the equation for the time-diagonal two-particle Green function becomes
\begin{align}
 \i\hbar\frac{\d}{\d t} & \mathcal{G}_{\bm{p} \bm{\cbar p} \bm{q}}^{ \alpha \beta }(t) - \mathcal{G}_{\bm{p} \bm{\cbar p} \bm{q}}^{ \alpha \beta }(t) 
 \left(h^{\tn{HF},\alpha}_{\bm{p} -\bm{q}} + h^{\tn{HF},\beta}_{\bm{\cbar{p}} +\bm{q}} - h^{\tn{HF},\alpha}_{\bm{p}} - h^{\tn{HF},\beta}_{\bm{\cbar{p}}} \right) \nonumber \\
 &= \frac{1}{\left(\i\hbar\right)^2} 
 \left[
 v_{|\bm{q}|} (t) \pm \delta_{\alpha\beta}  v_{|\bm{p}-\bm{q}-\bm{\cbar{p}}|} (t)
 \right] \Phi_{\bm{p} \bm{\cbar p} \bm{q}}^{ \alpha \beta }(t) \nonumber \\ 
 &\eqqcolon \Psi_{\bm{p} \bm{\cbar p} \bm{q}}^{ \pm, \alpha \beta }(t) \,,
 \label{eq:jellium-g2}
\end{align}
where
\begin{align}
    h^{\tn{HF},\alpha}_{\bm{p}} (t) = \frac{\bm{p}^2}{2m} + \i\hbar \sum_{\bm{\cbar{p}}} v_{|\bm{p}-\bm{\cbar{p}}|} (t) G^{<,\alpha}_{\bm{\cbar{p}}}(t) \,,
\end{align}
and
\begin{align}
  \Phi_{\bm{p} \bm{\cbar p} \bm{q}}^{ \alpha \beta}(t) &= \left(\i\hbar\right)^4\Big[
  G^{>,\alpha}_{\bm{p} -\bm{q}}(t)  G^{>,\beta}_{\bm{\cbar{p}} +\bm{q}}(t)  G^{<,\alpha}_{\bm{p}}(t)  G^{<,\beta}_{\bm{\cbar{p}}}(t) \nonumber \\
  &\quad -G^{<,\alpha}_{\bm{p} -\bm{q}}(t)  G^{<,\beta}_{\bm{\cbar{p}} +\bm{q}}(t)  G^{>,\alpha}_{\bm{p}}(t)  G^{>,\beta}_{\bm{\cbar{p}}}(t) \Big] \nonumber \,. 
\end{align}
This result agrees with the one derived in Refs.~\cite{bonitz-etal.96pla, bonitz_qkt}.

\subsection{Initial pair correlations in the G1--G2 scheme}\label{ss:inicor}
We conclude this section by considering the question of initial values in the G1--G2 scheme. Obviously, the solution of the differential equations \eqref{eq:eom_gone-2}, for $G^<(t)$, and \refeqn{eq:result_eom_soa}, for $\mathcal{G}(t)$, are defined only up to arbitrary constants which we can fix by choosing the initial values, $G^{0,<}=G^<(t=t_0)$ and $\mathcal{G}^0=\mathcal{G}(t=t_0)$. Recalling the definitions \eqref{eq:g-glsymm} and \eqref{eq:gscor-def}, the former is related to the initial value of the single-particle density matrix, and the latter to the initial value of the correlated part of the two-particle density matrix,
\begin{align}
    G^{0,<}_{ij} &= \pm \frac{1}{\i\hbar}n_{ij}(t_0)=\pm \frac{1}{\i\hbar}n_{ij}^0\,,
    \label{eq:ini-gless}
    \\
    \mathcal{G}^{0}_{ijkl} &=  \frac{1}{(\i\hbar)^2}\left\{
    n_{ijkl}^0 - n^0_{ik}n^0_{jl} \mp n_{il}^0n_{jk}^0
    \right\}\,,
    \label{eq:ini-g2}
\end{align}
i.e., pair correlations existing in the system at the initial time $t=t_0$.  While, mathematically, any initial value is compatible with the differential equation, physical considerations do impose restrictions, as was discussed e.g. in Refs.~\cite{semkat_cpp_03,bonitz_pss_18}. The result can be summarized as follows: only such pair correlations are physically relevant that can be produced by a dynamic evolution of the form
\begin{align}
\mathcal{G}^0_{ijkl} &= \left(\i\hbar\right)^3 \sum_{pqrs}\int_{-\infty}^{t_0} \mathrm{d}\cbar{t}\, 
\label{eq:g0-solution-short}
  \mathcal{U}_{ijpq}^{(2)}(t_0,\cbar{t}) 
  \Psi^\pm_{pqrs}(\cbar{t})
  \mathcal{U}_{rskl}^{(2)}(\cbar{t},t_0)\,,
\end{align} 
starting from an uncorrelated system at $\cbar{t}\to-\infty$.
We underline that the treatment of initial values in the G1--G2 scheme is not restricted to the second-Born approximation but can be generalized to more sophisticated selfenergies.
In the context of NEGF theory and the GKBA, the question of initial correlations has been extensively discussed before, see, e.g., Refs.~\cite{DANIELEWICZ_84_ap2,kremp_99_pre,semkat_00_jmp,stefanucci_cambridge_2013}, for more recent investigations, see Refs.~\cite{karlsson_gkba18,bonitz_pss_18,hopjan_initial_2019}.
Going back to our starting point---the integral representation of $\mathcal{G}$, cf. e.g. \refeqns{eq:g2-integral}{eq:dtilde-solution-short}---it is clear that these expressions vanish, in the limit $t\to t_0$, i.e., these expressions are valid only for the case of an initially uncorrelated system. These integral solutions are readily extended to the case of arbitrary initial correlations \cite{bonitz_qkt}: in that case, the previous solution,
\refeqn{eq:dtilde-solution-short}, has to be supplemented by a homogeneous solution of the differential equation for $\mathcal{G}$, which we denote $\mathcal{G}^{\rm IC}$,  
\begin{align}
    \mathcal{G}^\GKBA(t) &\to     \mathcal{G}^\GKBA(t) + \mathcal{G}^{\rm IC}(t),
\label{eq:g2-general}\\    
\mathcal{G}^{\rm IC}_{ijkl}(t) &= 
\left(\i\hbar\right)^4 \sum_{pqrs}\mathcal{U}^{(2)}_{ijpq}(t,t_0)\,\mathcal{G}^{0}_{pqrs}\,\mathcal{U}^{(2)}_{rskl}(t_0,t)\,,
\label{eq:g2-ic}
\end{align}
which recovers the structure of \refeqn{eq:linintG}.
Both terms \eqref{eq:g2-general} can be combined into a single expression according to 
\begin{align}
&\mathcal{G}^\GKBA_{ijkl}(t) = \label{eq:g2-ic_compact}
\left(\i\hbar\right)^4 \sum_{pqrs} \int_{t_0}^t \d \cbar{t}\,\times\\
&\quad\times \mathcal{U}^{(2)}_{ijpq}(t,\cbar{t})\,\Big[\delta(t_0,\cbar{t})\mathcal{G}^{0}_{pqrs} + \frac{1}{\i\hbar} \Psi^\pm_{pqrs}(\cbar{t})\Big]\,\mathcal{U}^{(2)}_{rskl}(\cbar{t},t)\,.\nonumber
\end{align}

While \refeqn{eq:g2-ic_compact} in its presented form 
holds for the second-Born approximation, this functional form 
is generally valid. The main difference, for more complicated selfenergies, is the explicit form 
of the two-particle propagators. For the additional approximations considered in this work [$GW$ (Sec.~\ref{s:gw}), $T$ matrix (Sec.~\ref{s:t-matrix})] the respective expressions are presented in Appendix~\ref{app:IC}. 
\section{$GW$ Selfenergy}\label{s:gw}
The static second-Born approximation that was considered above neglects screening effects and the dynamics of screening. These effects are captured by the $GW$ approximation
for which the selfenergy is given by,
\begin{align}
 \Sigma_{ij}^\gtrless(t,t') = \i \hbar \sum_{kl} W^\gtrless_{ilkj}(t,t') G^\gtrless_{kl}(t,t')\, .
\label{eq:sigma-gw}
\end{align}
Here, $W$ is the dynamically screened interaction, which can be expressed in terms of the bare interaction and the inverse dielectric function,
\begin{align}\label{eq:W-eps}
 W_{ijkl}^\gtrless(t,t') = \sum_{pq} w_{ipkq}(t) \varepsilon^{-1,\gtrless}_{pjql}(t,t')\, ,
\end{align}
which allows us to transform the selfenergy \eqref{eq:sigma-gw} into,
\begin{align}
 \Sigma_{ij}^\gtrless(t,t') = \i \hbar \sum_{klpq} w_{ipkq}(t) \varepsilon^{-1,\gtrless}_{plqj}(t,t') G^\gtrless_{kl}(t,t')\, .
\end{align}

The collision integral of the time-diagonal equation then becomes,
\begin{align}
 I_{ij} (t) &= \sum_k \int_{t_0}^t \d \cbar t \, \Big[ \Sigma^>_{ik}(t,\cbar{t}) G^<_{kj}(\cbar t, t) - \Sigma^<_{ik}(t,\cbar{t}) G^>_{kj}(\cbar t, t)  \Big] 
 \nonumber\\
&= \i\hbar \sum_{klpqr} w_{ipkq}(t) \int_{t_0}^t \d \cbar t\, \times \\ &\quad \times \Big[ \varepsilon^{-1,>}_{plqr}(t,\cbar t) \mathcal{G}^{\tn{F},>}_{krjl}(t,\cbar t) - \varepsilon^{-1,<}_{plqr}(t,\cbar t) \mathcal{G}^{\tn{F},<}_{krjl}(t, \cbar t)\Big] \nonumber \,.
 \end{align} 
Recalling the definition \eqref{eq:definition_D}, 
we identify the time-diagonal element of the two-particle Green function in $GW$ approximation,
\begin{align}
 \mathcal{G}_{ijkl} (t)=& \pm \sum_{pq} \int_{t_0}^t \d \cbar t \, \Big[ \varepsilon^{-1,>}_{lpjq}(t,\cbar t) \mathcal{G}^{\tn{F},>}_{iqkp}(t,\cbar t) \nonumber \\
 & \qquad \qquad - \varepsilon^{-1,<}_{lpjq}(t,\cbar t) \mathcal{G}^{\tn{F},<}_{iqkp}(t,\cbar t)\Big] \, .
 \end{align} 
 By construction, the screened interaction tensor obeys the following symmetry [cf. \refeqn{eq:wsymm_a}],
 \begin{align}
  W^\gtrless_{ijkl}(t,t') &= W^\lessgtr_{jilk}(t',t)\, . \label{eq:Wsymm}
 \end{align}
 From Hedin's equations \cite{hedin_new_1965} we derive the following relation for the dynamically screened interaction $W$ from which we subtract the singular part, i.e. $W^\gtrless_{ijkl}(t,t') \to W^\gtrless_{ijkl}(t,t') -w_{ijkl}\delta(t-t')$~\cite{schluenzen_jpcm_19},
\begin{widetext}
\begin{align}
\label{eq:W-def}
    & W^\gtrless_{ijkl}(t,t') = \pm \i \hbar \sum_{pqrs} w_{ipkq}(t) w_{rjsl}(t') \mathcal{G}^{\tn{F},\gtrless}_{qspr}(t, t') \\
    &\quad \pm \i\hbar \sum_{pqrs} w_{ipkq}(t) \Bigg[ \int_{t_0}^t \d\cbar{t} \, \Big( \mathcal{G}^{\tn{F},>}_{qspr}(t, \cbar t) - \mathcal{G}^{\tn{F},<}_{qspr}(t, \cbar t) \Big) W^\gtrless_{rjsl}(\cbar{t},t') \nonumber %
    +\int_{t_0}^{t'} \d \cbar{t} \, \mathcal{G}^{\tn{F},\gtrless}_{qspr}(t, \cbar t) \left(W^<_{rjsl}(\cbar{t},t') - W^>_{rjsl}(\cbar{t},t') \right)\Bigg] \nonumber
\end{align}
By comparison with \refeqn{eq:W-eps} and using the symmetry of \refeqn{eq:Wsymm} one can identify a recursive equation for $\varepsilon^{-1}$,
\begin{align}
 &  \varepsilon^{-1,\gtrless}_{ijkl}(t,t') = \pm \i \hbar \sum_{pq} w_{pjql}(t') \mathcal{G}^{\tn{F},\gtrless}_{kqip}(t, t') \\
  & \quad\pm \i\hbar \sum_{pqrs} w_{jrls} (t') \nonumber 
  \Bigg[ \int_{t_0}^t \d \cbar t\, \Big( \mathcal{G}^{\tn{F},>}_{kqip}(t, \cbar t) - \mathcal{G}^{\tn{F},<}_{kqip}(t, \cbar t) \Big) 
  \varepsilon^{-1,\lessgtr}_{rpsq}(t',\cbar t) \nonumber 
  + \int_{t_0}^{t'} \d \cbar t\, \mathcal{G}^{\tn{F},\gtrless}_{kqip}(t, \cbar t) \Big( \varepsilon^{-1,>}_{rpsq}(t',\cbar t) - \varepsilon^{-1,<}_{rpsq}(t',\cbar t) \Big) \Bigg] \, .
 \end{align}
 The time-diagonal equation for the inverse dielectric function can be further simplified,
   \begin{align}
  \varepsilon^{-1,\gtrless}_{ijkl}(t,t) =& \pm \i \hbar \sum_{pq} w_{pjql}(t) \mathcal{G}^{\tn{F},\gtrless}_{kqip}(t) %
  \pm \i\hbar \sum_{pqrs} w_{jrls} (t) \int_{t_0}^t \d \cbar t\, \Big( \mathcal{G}^{\tn{F},>}_{kqip}(t, \cbar t) \varepsilon^{-1,>}_{rpsq}(t,\cbar t) - \mathcal{G}^{\tn{F},<}_{kqip}(t, \cbar t) \varepsilon^{-1,<}_{rpsq}(t,\cbar t) \Big) \nonumber \\
  =& \pm \i \hbar \sum_{pq} w_{pjql}(t) \mathcal{G}^{\tn{F},\gtrless}_{kqip}(t) + \i \hbar \sum_{pq} w_{pjql}(t) \mathcal{G}_{kqip}(t)\, . \label{eq:eps_diag}
  \end{align}
\end{widetext}
\balancecolsandclearpage
\subsection{$GW$ approximation within the HF-GKBA}
We now apply the HF-GKBA [cf. \eqrefr{eq:GKBA}{eq:prop_b}] and obtain the following expressions for $\mathcal{G}^\GKBA$,
 \begin{align}
 \mathcal{G}^\GKBA_{ijkl} (t)=& \pm \left(\i\hbar\right)^2\sum_{pqrs} \int_{t_0}^t \d \cbar t \, \mathcal{U}_{ir}(t,\cbar t)\Big[ \varepsilon^{-1,>}_{lpjq}(t,\cbar t) \mathcal{G}^{\tn{F},>}_{rqsp}(\cbar t) \nonumber \\ &\quad - \varepsilon^{-1,<}_{lpjq}(t,\cbar t) \mathcal{G}^{\tn{F},<}_{rqsp}(\cbar t)\Big] \mathcal{U}_{sk}(\cbar t, t) \, , \label{eq:gtwogwgkba1}
 \end{align} 
 as well as for $\varepsilon^{-1}_\GKBA$,
  \begin{align}
  \varepsilon^{-1,\gtrless}_{ijkl}&(t\ge t') \label{eq:inv_di_func} \\
  =& \pm \left(\i \hbar\right)^3 \sum_{pqrs} w_{pjql}(t') \mathcal{U}_{kr}(t,t') \mathcal{G}^{\tn{F},\gtrless}_{rqsp}(t') \mathcal{U}_{si}(t',t) \nonumber \\
  & \pm \left(\i \hbar\right)^3 \sum_{pqrsuv} w_{jrls} (t') \nonumber \Bigg[ \int_{t_0}^t \d \cbar t\, \mathcal{U}_{ku}(t,\cbar t) \times 
  \\ &\times  \Big( \mathcal{G}^{\tn{F},>}_{uqvp}(\cbar t) - \mathcal{G}^{\tn{F},<}_{uqvp}(\cbar t) \Big) \nonumber \mathcal{U}_{vi}(\cbar t,t) \varepsilon^{-1,\lessgtr}_{rpsq}(t',\cbar t) \nonumber \\
  &\nonumber \quad + \int_{t_0}^{t'} \d \cbar t\, \mathcal{U}_{ku}(t,\cbar t) \mathcal{G}^{\tn{F},\gtrless}_{uqvp}(\cbar t) \mathcal{U}_{vi}(\cbar t, t) \nonumber \times \\ &\qquad \times \Big( \varepsilon^{-1,>}_{rpsq}(t',\cbar t) - \varepsilon^{-1,<}_{rpsq}(t',\cbar t) \Big) \Bigg] \, ,
\nonumber
 \end{align}
 where $\mathcal{U}$ is given by \refeqns{eq:eom-ua}{eq:eom-ub}.
 By using the symmetry relation of \refeqn{eq:Wsymm} we easily find an expression for the time derivative of the off-diagonal inverse dielectric function,
\begin{align}
 &\frac{\d}{\d t} \varepsilon^{-1,\gtrless}_{ijkl}(t\ge t') \\ \nonumber 
 &= \frac{1}{\i\hbar} \sum_p
 \left\{ h^\textnormal{HF}_{kp}(t) \varepsilon^{-1,\gtrless}_{ijpl}(t\ge t') 
 \nonumber 
 -\varepsilon^{-1,\gtrless}_{pjkl}(t\ge t') h^\textnormal{HF}_{pi}(t) \right\}
 \nonumber \\
 &\quad \pm \i\hbar \sum_{pqrs} w_{prqs}(t) \Big[ \mathcal{G}^{\tn{F},>}_{kqip}(t) - \mathcal{G}^{\tn{F},<}_{kqip}(t) \Big] \varepsilon^{-1,\gtrless}_{rjsl} (t \ge t')\nonumber\\
 &= \frac{1}{\i\hbar} \sum_{pq} \Big[ \mathfrak{h}^{\varepsilon,\tn{HF}}_{pkqi}(t) + \mathfrak{h}^{\varepsilon,\tn{corr}}_{pkqi}(t) \Big] \varepsilon^{-1,\gtrless}_{pjql}(t\ge t')
 \, , \nonumber
\end{align}
where we introduced the modified two-particle Hartree--Fock Hamiltonian
\begin{align}\label{eq:HF-gw}
    \mathfrak{h}^{\varepsilon,\tn{HF}}_{ijkl}(t) = \delta_{il}h^\tn{HF}_{jk}(t)-\delta_{jk}h^\tn{HF}_{il}(t)\,,
\end{align}
which matches the index structure of the effective quasi-Hamiltonian, defined as
   \begin{align}
       \mathfrak{h}^{\varepsilon,\tn{corr}}_{ijkl}(t) = \pm \left(\i \hbar\right)^2 \sum_{pq} w_{qipk}(t) \Big[ \mathcal{G}^{\tn{F},>}_{jplq}(t) - \mathcal{G}^{\tn{F},<}_{jplq}(t) \Big]\,.\quad \label{eq:gwhcorr} \phantom{..}
   \end{align}
Combining these Hamiltonians into a single one, 
\begin{align}
    \mathfrak{h}^{\varepsilon}_{ijkl}(t) = \mathfrak{h}^{\varepsilon,\tn{HF}}_{ijkl}(t) + 
    \mathfrak{h}^{\varepsilon,\tn{corr}}_{ijkl}(t)\,,
  \label{eq:h-epsilon-def}
\end{align}
we observe that the inverse dielectric function, within the $GW$-HF-GKBA, obeys a time-dependent two-particle Schrödinger equation,
\begin{align}
 & \i\hbar\frac{\d}{\d t} \varepsilon^{-1,\gtrless}_{ijkl}(t\ge t') 
 =  \sum_{pq}  \mathfrak{h}^{\varepsilon}_{pkqi}(t) \varepsilon^{-1,\gtrless}_{pjql}(t\ge t')
 \, , 
\label{eq:tdse-eps}
\end{align}
with the Hamiltonian \eqref{eq:h-epsilon-def},
that is equivalent to the rather complicated integral equation \eqref{eq:inv_di_func}.

In the following, we demonstrate that, for the $GW$-HF-GKBA, again, a time-local G1--G2 scheme can be derived which retains time-linear scaling \cite{schluenzen_19_prl}.
\subsection{$GW$-G1--G2 equations for a general basis}
To derive the G1--G2 scheme, we compute the time derivative of $\mathcal{G}$
, yielding,
  \begin{align}\label{eq:eom_parts_epsilon}
   &\frac{\d}{\d t} \mathcal{G}_{ijkl} (t) \\ &= \left[\frac{\d}{\d t} \mathcal{G}_{ijkl} (t)\right]_{\int^{}_{}} + \left[\frac{\d}{\d t} \mathcal{G}_{ijkl} (t)\right]_{\varepsilon} + \left[\frac{\d}{\d t} \mathcal{G}_{ijkl} (t)\right]_{\mathcal{U}}\nonumber\,,
   \end{align}
where the first contribution, which originates from the derivative of the integration boundaries, is given by
  \begin{align}
   &\left[\frac{\d}{\d t} \mathcal{G}_{ijkl} (t)\right]_{\int^{}_{}}\nonumber\\
   &\,=\pm \sum_{pq} \Big[ \varepsilon^{-1,>}_{lpjq}(t,t) \mathcal{G}^{\tn{F},>}_{iqkp}(t) - \varepsilon^{-1,<}_{lpjq}(t, t) \mathcal{G}^{\tn{F},<}_{iqkp}(t)\Big]  \nonumber \\
   &\,=  \i \hbar \sum_{pqrs} w_{rpsq}(t) \Big[ \mathcal{G}^{\tn{F},>}_{jslr}(t)\mathcal{G}^{\tn{F},>}_{iqkp}(t) - \mathcal{G}^{\tn{F},<}_{jslr}(t) \mathcal{G}^{\tn{F},<}_{iqkp}(t) \Big] \nonumber \\ &\,\quad
   \pm \i \hbar \sum_{pqrs} w_{rpsq}(t) \mathcal{G}_{jslr}(t) \Big[ \mathcal{G}^{\tn{F},>}_{iqkp}(t) - \mathcal{G}^{\tn{F},<}_{iqkp}(t) \Big]\nonumber\\
   &\,= \frac{1}{\i\hbar} \Psi_{ijkl}(t)
  - \frac{1}{\i\hbar} \sum_{pq}  \mathcal{G}_{qjpl}(t) \Big[\mathfrak{h}^{\varepsilon,\tn{corr}}_{qkpi}(t)\Big]^*\,\nonumber\,.
   \end{align}
   Here, the two-particle source term is defined as
   \begin{align}
       \Psi_{ijkl}(t) = \frac{1}{\left(\i\hbar\right)^2}\sum_{pqrs} w_{pqrs}(t)
  \Phi^{ijrs}_{pqkl}(t)\, .
   \end{align}
The second contribution to \refeqn{eq:eom_parts_epsilon}, resulting from the time derivative of $\varepsilon^{-1}$, is given by
\begin{align}
    &\left[\frac{\d}{\d t} \mathcal{G}_{ijkl} (t)\right]_{\varepsilon}\nonumber\\
   &\,= \pm \left(\i\hbar\right)^2 \sum_{pqrs} \int_{t_0}^t \d \cbar t \, \mathcal{U}_{ir}(t,\cbar t)\Bigg[ \left( \frac{\d}{\d t} \varepsilon^{-1,>}_{lpjq}(t,\cbar t) \right) \mathcal{G}^{\tn{F},>}_{rqsp}(\cbar t)
   \nonumber \\ &\, \qquad \qquad \qquad 
   - \left( \frac{\d}{\d t} \varepsilon^{-1,<}_{lpjq}(t,\cbar t) \right) \mathcal{G}^{\tn{F},<}_{rqsp}(\cbar t)\Bigg] \mathcal{U}_{sk}(\cbar t, t) \nonumber \\
   &\,= \frac{1}{\i\hbar} \sum_{pq} \Big[ \mathfrak{h}^{\varepsilon,\tn{HF}}_{pjql}(t) + \mathfrak{h}^{\varepsilon,\tn{corr}}_{pjql}(t) \Big] \mathcal{G}_{iqkp}(t)
 \, , \nonumber
\end{align}
whereas the third contribution to \refeqn{eq:eom_parts_epsilon}, which stems from the derivative of the propagators, is 
\begin{align}
    &\left[\frac{\d}{\d t} \mathcal{G}_{ijkl} (t)\right]_{\mathcal{U}}\nonumber\\
   &\,= \pm \left(\i\hbar\right)^2 \sum_{pqrs} \int_{t_0}^t \d \cbar t \, \left( \frac{\d}{\d t} \mathcal{U}_{ir}(t,\cbar t) \right) \Big[ \varepsilon^{-1,>}_{lpjq}(t,\cbar t) \mathcal{G}^{\tn{F},>}_{rqsp}(\cbar t)
   \nonumber \\
    &\, \qquad \qquad \qquad \qquad
   - \varepsilon^{-1,<}_{lpjq}(t,\cbar t) \mathcal{G}^{\tn{F},<}_{rqsp}(\cbar t)\Big] \mathcal{U}_{sk}(\cbar t, t) \nonumber \\
   &\,\quad \pm \left(\i\hbar\right)^2 \sum_{pqrs} \int_{t_0}^t \d \cbar t \, \mathcal{U}_{ir}(t,\cbar t)\Big[ \varepsilon^{-1,>}_{lpjq}(t,\cbar t) \mathcal{G}^{\tn{F},>}_{rqsp}(\cbar t)
   \nonumber \\ &\, \qquad \qquad \qquad 
   - \varepsilon^{-1,<}_{lpjq}(t,\cbar t) \mathcal{G}^{\tn{F},<}_{rqsp}(\cbar t)\Big] \left( \frac{\d}{\d t} \mathcal{U}_{sk}(\cbar t, t) \right) \nonumber\\
   &\,= \frac{1}{\i \hbar} \sum_p 
   \Big[
   h^\textnormal{HF}_{ip} (t) \mathcal{G}_{pjkl}(t) - 
\mathcal{G}_{ijpl}(t) h^\textnormal{HF}_{pk} (t)\Big]\nonumber\\
&\,= - \frac{1}{\i \hbar} \sum_{pq}  \mathcal{G}_{qjpl}(t) \Big[ \mathfrak{h}^{\varepsilon,\tn{HF}}_{qkpi}(t) \Big]^*\, ,
\end{align}
Finally, the three contributions to the derivative of $\mathcal{G}$ are combined to reveal
\begin{align}
 \i\hbar&\frac{\d}{\d t} \mathcal{G}_{ijkl} (t) = \Psi_{ijkl}(t)
    \label{eq:g2-eq-gw}
\\
  & + \sum_{pq} \bigg\{ \mathfrak{h}^{\varepsilon}_{qjpl}(t) \Big[ \mathcal{G}_{qkpi}(t) \Big]^* - \mathcal{G}_{qjpl}(t) \Big[\mathfrak{h}^{\varepsilon}_{qkpi}(t)\Big]^* \bigg\} \, ,
\nonumber 
\end{align}
where $\mathfrak{h}^{\varepsilon}(t)$ was defined in \refeqn{eq:h-epsilon-def}.
With this we have obtained the equations of the G1--G2 scheme for the $GW$ approximation. 
For $\mathfrak{h}^{\varepsilon,\tn{corr}}(t) \equiv 0$, we recover the equations from the SOA, cf. \refeqn{eq:result_eom_soa}, since the remaining Hamiltonian contribution can be expressed as a commutator.
Equation \eqref{eq:g2-eq-gw} is the most compact formulation that visualizes the intrinsic structure of $\mathcal{G}$ in the $GW$ approximation. 

For practical use, it is convenient to separate the correlation contributions from the mean-field terms via the introduction of an additional quantity:
\begin{align}
 \i\hbar&\frac{\d}{\d t} \mathcal{G}_{ijkl} (t) - \left[h^{(2),\tn{HF}}, \mathcal{G}\right]_{ijkl} (t) \nonumber \\
 &= \Psi_{ijkl}(t)
  + \Pi_{ijkl}(t) - \Big[\Pi_{lkji}(t)\Big]^*\, , \label{eq:g2-eq-gw-pi}
\end{align}
where polarization effects are included in
\begin{align}
 \Pi_{ijkl}(t) = \sum_{pq} \mathfrak{h}^{\varepsilon,\tn{corr}}_{qjpl}(t) \mathcal{G}_{ipkq}(t) \, . \label{eq:gwpi}
\end{align}
Equation \eqref{eq:g2-eq-gw} agrees with the polarization approximation of density-matrix theory, cf. Refs.~\cite{hohenester97,bonitz_qkt}. In the Markov limit this leads to the quantum generalization of the Balescu--Lenard kinetic equation \cite{balescu_60,lenard_60,klimontovich_75}.

Here, we have employed the standard definition of $GW$ in NEGF theory, which is widely used in literature (see, e.g., Refs.~\cite{stefanucci_diagrammatic_2014,von_friesen_successes_2009,schluenzen_jpcm_19}), in which the screened interaction [\eqref{eq:W-def}] does not include exchange terms. The generalization to also describe exchange processes is, however, straightforwardly carried out. For the G1--G2 scheme, this is achieved by simply replacing $\Psi_{ijkl}(t)$ by $\Psi^\pm_{ijkl}(t)$ in \refeqns{eq:g2-eq-gw}{eq:g2-eq-gw-pi}. 

Again we have succeeded to eliminate all time integrations which means that \refeqn{eq:g2-eq-gw} can be solved with an effort that is first order in $N_\tn{t}$. Note that the conventional HF-GKBA scheme with $GW$ selfenergy scales as $N_\tn{t}^3$ indicating a huge advantage of the G1--G2 formulation \cite{schluenzen_19_prl}. More computational details will be given below, in Sec.~\ref{s:numerics}.

\subsection{$GW$-G1--G2 equations for the Hubbard model}\label{ss:gw-hubbard}
For the Hubbard system [cf. \refeqn{eq:h-hubbard}] we again use the interaction matrix \eqref{eq:wmatrix-hubbard}. %
With that, the equations of motion \eqref{eq:g2-eq-gw-pi} become, %
\begin{align}
 \i\hbar\frac{\d}{\d t} & \mathcal{G}^{\uparrow\downarrow\uparrow\downarrow}_{ijkl} (t) - \left[h^{(2),\tn{HF}}_{\uparrow\downarrow}, \mathcal{G}^{\uparrow\downarrow\uparrow\downarrow}\right]_{ijkl} (t) \label{eq:g2-gw-hubbard1} \\
 &= %
  \Psi^{\uparrow\downarrow\uparrow\downarrow}_{ijkl}(t)
   + \Pi^{\uparrow\downarrow\uparrow\downarrow}_{ijkl}(t) - \Big[ \Pi^{\uparrow\downarrow\uparrow\downarrow}_{lkji}(t)\Big]^*\quad \textnormal{and} \nonumber  \\
 \i\hbar\frac{\d}{\d t} & \mathcal{G}^{\uparrow\uparrow\uparrow\uparrow}_{ijkl} (t) - \left[h^{(2),\tn{HF}}_{\uparrow\uparrow}, \mathcal{G}^{\uparrow\uparrow\uparrow\uparrow}\right]_{ijkl} (t) \nonumber \\
 &=  \Pi^{\uparrow\uparrow\uparrow\uparrow}_{ijkl}(t) - \Big[ \Pi^{\uparrow\uparrow\uparrow\uparrow}_{lkji}(t)\Big]^*\, ,
\label{eq:g2-gw-hubbard2}
\end{align}
where we introduced the polarization terms,
\begin{align}
 &\Pi^{\uparrow\downarrow\uparrow\downarrow}_{ijkl}(t) = - (\i\hbar)^2 U(t)\times \label{eq:pihubbardtrans}
 \\&\,
  \sum_{p} \Big[ G^{>,\downarrow}_{jp} (t) G^{<,\downarrow}_{pl} (t) - G^{<,\downarrow}_{jp} (t) G^{>,\downarrow}_{pl} (t) \Big] \mathcal{G}^{\uparrow\uparrow\uparrow\uparrow}_{ipkp}(t) \, , \nonumber\\
&\Pi^{\uparrow\uparrow\uparrow\uparrow}_{ijkl}(t) = - (\i\hbar)^2 U(t)\times \label{eq:pihubbardcis}
 \\&\,
  \sum_{p} \Big[ G^{>,\uparrow}_{jp} (t) G^{<,\uparrow}_{pl} (t) - G^{<,\uparrow}_{jp} (t) G^{>,\uparrow}_{pl} (t) \Big] \mathcal{G}^{\uparrow\downarrow\uparrow\downarrow}_{ipkp}(t) \, . \nonumber
\end{align}
Notice that there are two separate spin combinations (four when considering $\uparrow\,\leftrightarrow\,\downarrow$) for the two-particle Green function that enter \refeqns{eq:g2-gw-hubbard1}{eq:g2-gw-hubbard2}. Due to the cross-coupling in the two polarization terms, they cannot be solved independently~\cite{joost_pss_18,schluenzen_jpcm_19}. 
Numerical results for the $GW$-G1--G2 scheme are presented in Sec.~\ref{s:numerics}.

\subsection{$GW$-G1--G2 equations for jellium}\label{ss:gw-jellium}
For the uniform electron gas [cf. \refeqn{eq:h-jellium})] we again use the interaction matrix \eqref{eq:jellium-w-matrix},
and define
\begin{align}
    \Pi_{\bm{p},\bm{\cbar p},\bm{q}}^{\alpha \beta}(t) &:= \Pi_{\bm{p}-\bm{q},\bm{\cbar p}+\bm{q},\bm{p},\bm{\cbar p}}^{\alpha \beta \alpha \beta}(t)\, .
\end{align}
With that, the equation \eqref{eq:g2-eq-gw-pi} for the time-diagonal two-particle Green function [recall the definition \eqref{eq:g2-def-jellium}] becomes,
\begin{align}
 \i\hbar&\frac{\d}{\d t} \mathcal{G}_{\bm{p} \bm{\cbar p} \bm{q}}^{ \alpha \beta }(t) - \mathcal{G}_{\bm{p} \bm{\cbar p} \bm{q}}^{ \alpha \beta }(t) 
 \left(h^{\tn{HF},\alpha}_{\bm{p} -\bm{q}} + h^{\tn{HF},\beta}_{\bm{\cbar{p}} +\bm{q}} - h^{\tn{HF},\alpha}_{\bm{p}} - h^{\tn{HF},\beta}_{\bm{\cbar{p}}} \right) \nonumber \\
 &= \underbrace{\frac{1}{\left(\i\hbar\right)^2} v_{\left|\bm{q}\right|}(t) \Phi_{\bm{p} \bm{\cbar p} \bm{q}}^{ \alpha \beta }(t)}_{\eqqcolon\Psi_{\bm{p} \bm{\cbar p} \bm{q}}^{ \alpha \beta }(t)} + \Pi_{\bm{p},\bm{\cbar p},\bm{q}}^{\alpha \beta}(t) - \Big[ \Pi_{\bm{\cbar p}+\bm{q},\bm{p}-\bm{q},\bm{q}}^{\beta\alpha}(t)\Big]^*\, , 
\end{align}
with the momentum representation of the polarization term, given by
\begin{align}
 \Pi_{\bm{p},\bm{\cbar p},\bm{q}}^{\alpha \beta}(t) =& \pm (\i\hbar)^2\Big[  G^{>,\beta}_{\bm{\cbar p}+\bm{q}}(t) G^{<,\beta}_{\bm{\cbar p}}(t) - G^{<,\beta}_{\bm{\cbar p}+\bm{q}}(t) G^{>,\beta}_{\bm{\cbar p}}(t)\Big] \times \nonumber \\ & \quad \times v_{\left|\bm{q}\right|}(t) \sum_{\bm{k},\sigma} \mathcal{G}_{\bm{p}\bm{k}\bm{q}}^{\alpha \sigma}(t)\,. \label{eq:jellium-gwpi}
\end{align}
As we will discuss in Sec.~\ref{s:numerics}, the $GW$ equations for jellium can be solved particularly efficiently.

\section{$T$-matrix selfenergies}\label{s:t-matrix}
We next turn to the case of strong coupling where the second-Born approximation is not applicable. It is well known that the entire Born series can be summed up, giving rise to the $T$-matrix (or binary-collision or ladder) approximation. Here we first consider the case of a static pair interaction. The extension to a dynamically screened $T$ matrix will be considered in Sec.~\ref{s:dyn-scr-ladder}. We start by considering, in Sec.~\ref{ss:tpp}, the $T$ matrix in the particle--particle channel after which we analyze, in Sec.~\ref{ss:tph}, the $T$ matrix in the particle--hole channel.

\subsection{$T$ matrix in the particle--particle channel}\label{ss:tpp}

For the particle--particle $T$ matrix, the selfenergy has the form \cite{kadanoff-baym-book,schluenzen_cpp16},
\begin{align}
 \Sigma_{ij}^\gtrless(t,t') = \i \hbar \sum_{kl} T^{\tn{pp},\gtrless}_{ikjl}(t,t') G^\lessgtr_{lk}(t',t)\, .
 \label{eq:sigma-t}
\end{align}
Here, the $T$ matrix is expressed as
\begin{align}
 T_{ijkl}^{\tn{pp},\gtrless}(t,t') = \sum_{pq} w_{ijpq}(t) \Omega^{{\tn{pp},\gtrless}}_{pqkl}(t,t')\, ,
 \label{eq:tmatrix}
\end{align}
which allows us to rewrite the selfenergy \eqref{eq:sigma-t}:
\begin{align}
 \Sigma_{ij}^\gtrless(t,t') = \i \hbar \sum_{klpq} w_{ikpq}(t) \Omega^{\tn{pp},\gtrless}_{pqjl}(t,t') G^\lessgtr_{lk}(t',t)\, .
 \label{eq:sigma-t2} \quad
\end{align}
In \refeqns{eq:tmatrix}{eq:sigma-t2} the quantity $\Omega^\tn{pp}$ is the nonequilibrium generalization of the M\o ller operator from scattering theory~\cite{taylor-scattering,kremp-etal.97ap}.
The collision integral \eqref{eq:definition_D} of the time-diagonal equation then becomes,
\begin{align}
 I_{ij} (t) 
&= \i\hbar \sum_{klpqr} w_{ipqr}(t) \int_{t_0}^t \d \cbar t \, \Big[ \Omega^{\tn{pp},>}_{qrkl}(t,\cbar t) \mathcal{G}^{\tn{H},<}_{kljp}(\cbar t, t) \nonumber \\ 
& \qquad \qquad \qquad \qquad \qquad - \Omega^{\tn{pp},<}_{qrkl}(t,\cbar t) \mathcal{G}^{\tn{H},>}_{kljp}(\cbar t, t)\Big] \nonumber \\
&= \pm \i \hbar \sum_{klp} w_{iklp}(t) \mathcal{G}_{kpjl} (t) \, ,
 \end{align}
which results in the following expression for the time-diagonal element of the two-particle Green function,
\begin{align}
 \mathcal{G}_{ijkl} (t)=& \pm \sum_{pq} \int_{t_0}^t \d \cbar t \, \Big[ \Omega^{\tn{pp},>}_{ijpq}(t,\cbar t) \mathcal{G}^{\tn{H},<}_{pqkl}(\cbar t, t) \nonumber \\
 & \qquad \qquad - \Omega^{\tn{pp},<}_{ijpq}(t,\cbar t) \mathcal{G}^{\tn{H},>}_{pqkl}(\cbar t, t)\Big] \, .
 \end{align} 
 By construction, the $T$ matrix obeys the following symmetry [cf. \refeqn{eq:wsymm_b}],
 \begin{align} \label{eq:Tsymm}
  T^{\tn{pp},\gtrless}_{ijkl}(t,t') &= -\left[T^{\tn{pp},\gtrless}_{klij}(t',t)\right]^*\, .
 \end{align}

 The $T$ matrix sums up the particle--particle collisions via the
 recursive equation (nonequilibrium Lippmann--Schwinger equation; compared to the standard definition of the $T$ matrix, here the singular part has been subtracted \cite{schluenzen_cpp16,schluenzen_jpcm_19}),
\begin{widetext}
 \begin{align}
  T&^{\tn{pp},\gtrless}_{ijkl}(t,t') = \pm \i \hbar \sum_{pqrs} w_{ijpq}(t) \mathcal{G}^{\tn{H},\gtrless}_{pqrs}(t, t') w^\pm_{rskl}(t') \\
  &  \quad+ \i\hbar \sum_{pqrs} w_{ijpq}(t) \Bigg\{ \int_{t_0}^t \d \cbar t\, \Big[ \mathcal{G}^{\tn{H},>}_{pqrs}(t, \cbar t) - \mathcal{G}^{\tn{H},<}_{pqrs}(t, \cbar t) \Big] T^{\tn{pp},\gtrless}_{rskl}(\cbar t, t') + \int_{t_0}^{t'} \d \cbar t\, \mathcal{G}^{\tn{H},\gtrless}_{pqrs}(t, \cbar t) \Big[ T^{\tn{pp},<}_{rskl}(\cbar t, t') - T^{\tn{pp},>}_{rskl}(\cbar t, t')\Big] \Bigg\}\, .\nonumber
 \end{align}
 Following this and using the symmetries of \refeqns{eq:wsymm_b}{eq:Tsymm} the relation for the M\o ller operator is readily derived,
 \begin{align}
  \Omega&^{\tn{pp},\gtrless}_{ijkl}(t,t') = \pm \i \hbar \sum_{pq} \mathcal{G}^{\tn{H},\gtrless}_{ijpq}(t, t') w^\pm_{pqkl}(t') \\
  & + \i\hbar \sum_{pqrs} \Bigg\{ \int_{t_0}^t \d \cbar t\, \Big[ \mathcal{G}^{\tn{H},>}_{ijpq}(t, \cbar t) - \mathcal{G}^{\tn{H},>}_{ijpq}(t, \cbar t) \Big] w_{pqrs} (\cbar t) \Omega^{\tn{pp},\gtrless}_{rskl}(\cbar t, t') + \int_{t_0}^{t'} \d \cbar t\, \mathcal{G}^{\tn{H},\gtrless}_{ijpq}(t, \cbar t) w_{pqrs} (\cbar t) \Big[ \Omega^{\tn{pp},<}_{rskl}(\cbar t, t') - \Omega^{\tn{pp},>}_{rskl}(\cbar t, t')\Big] \Bigg\}\, \nonumber\\
  & \qquad \qquad  = \pm \i \hbar \sum_{pq} \mathcal{G}^{\tn{H},\gtrless}_{ijpq}(t, t') w^\pm_{pqkl}(t') \\
  & + \i\hbar \sum_{pqrs} \Bigg\{ \int_{t_0}^t \d \cbar t\, \Omega^{\tn{pp},\gtrless}_{rspq}(t',\cbar t) \Big[ \mathcal{G}^{\tn{H},<}_{pqij}(\cbar t, t) - \mathcal{G}^{\tn{H},>}_{pqij}(\cbar t, t) \Big] + \int_{t_0}^{t'} \d \cbar t\, \Big[ \Omega^{\tn{pp},>}_{rspq}(t', \cbar t) - \Omega^{\tn{pp},<}_{rspq}(t', \cbar t) \Big] \mathcal{G}^{\tn{H},\gtrless}_{pqij}(\cbar t, t) \Bigg\}^* w_{rskl} (t') \, .\nonumber
 \end{align}

The time-diagonal equation for $\Omega^{\tn{pp}}$ can be further simplified,
   \begin{align}\label{eq:omega-pp-diag}
  \Omega^{\tn{pp},\gtrless}_{ijkl}(t,t) 
  =& \pm \i \hbar \sum_{pq} \mathcal{G}^{\tn{H},\gtrless}_{ijpq}(t) w^\pm_{pqkl}(t) + \i\hbar \sum_{pqrs} \Bigg[\int_{t_0}^t \d \cbar t\, \Big( \Omega^{\tn{pp},>}_{pqrs}(t,\cbar t)  \mathcal{G}^{\tn{H},<}_{rsij}(\cbar t, t) - \Omega^{\tn{pp},<}_{pqrs}(t,\cbar t) \mathcal{G}^{\tn{H},>}_{rsij}(\cbar t, t) \Big)\Bigg]^* w_{pqkl} (t) \nonumber \\
  =& \pm \i \hbar \sum_{pq} \mathcal{G}^{\tn{H},\gtrless}_{ijpq}(t) w^\pm_{pqkl}(t) \pm \i \hbar \sum_{pq} \Big[\mathcal{G}_{pqij}(t)\Big]^* w_{pqkl}(t)\, .
  \end{align}
  \end{widetext}

\subsubsection{$T^\tn{pp}$ approximation within the HF-GKBA}
We now apply the HF-GKBA [cf. \eqrefr{eq:GKBA}{eq:prop_b}] and find the following expressions for $\mathcal{G}^\GKBA$,
 \begin{align}
 \mathcal{G}^\GKBA_{ijkl} (t)=& \pm \left(\i \hbar\right)^2 \sum_{pqrs} \int_{t_0}^t \d \cbar t \, \Big[ \Omega^{\tn{pp},>}_{ijpq}(t,\cbar t) \mathcal{G}^{\tn{H},<}_{pqrs}(\cbar t)\nonumber\\ &\quad - \Omega^{\tn{pp},<}_{ijpq}(t,\cbar t) \mathcal{G}^{\tn{H},>}_{pqrs}(\cbar t)\Big] \mathcal{U}^{(2)}_{rskl}(\cbar t, t) \, , \label{eq:gtwoppgkba}
 \end{align} 
 as well as for $\Omega^\tn{pp}_\GKBA$,
  \begin{align}
  \Omega^{\tn{pp},\gtrless}_{ijkl}&(t\ge t') %
  \label{eq:tau_func} \\
  =& \pm \left(\i \hbar\right)^3  \sum_{pqrs} \mathcal{U}^{(2)}_{ijrs}(t, t') \mathcal{G}^{\tn{H},\gtrless}_{rspq}(t')  w^\pm_{pqkl}(t') \nonumber \\
  & + \left(\i \hbar\right)^3 \sum_{pqrsuv} \Bigg[\int_{t_0}^t \d \cbar t\, \mathcal{U}^{(2)}_{ijrs}(t,\cbar t) \times \nonumber \\ &\quad \times \Big(\mathcal{G}^{\tn{H},>}_{rspq}(\cbar t) - \mathcal{G}^{\tn{H},<}_{rspq}(\cbar t)\Big) w_{pquv} (\cbar t) \Omega^{\tn{pp},\gtrless}_{uvkl}(\cbar t,t') \nonumber \\
  &\nonumber \quad + \int_{t_0}^{t'} \d \cbar t\, \mathcal{U}^{(2)}_{ijrs}(t, \cbar t) \mathcal{G}^{\tn{H},\gtrless}_{rspq}(\cbar t) w_{pquv} (\cbar t) \times \nonumber \\
  &\nonumber \quad \times \Big(\Omega^{\tn{pp},<}_{uvkl}(\cbar t, t') - \Omega^{\tn{pp},>}_{uvkl}(\cbar t, t')\Big) \Bigg] \, ,\nonumber
 \end{align}
 where $\mathcal{U}$ is given by \refeqns{eq:eom-ua}{eq:eom-ub}.
With \refeqn{eq:tau_func}
we easily find an expression for the time derivative of the time-off-diagonal values of $\Omega^{\tn{pp}}$,
\begin{align}\label{eq:omega-pp}
 &\frac{\d}{\d t} \Omega^{\tn{pp},\gtrless}_{ijkl}(t\ge t') \\ 
 &\qquad= \frac{1}{\i\hbar} \sum_{pq} \left(\mathfrak{h}^{\Omega^\tn{pp}, \tn{HF}}_{ijpq}(t) + \mathfrak{h}^{\Omega^\tn{pp}, \tn{corr}}_{ijpq}(t)\right) \Omega^{\tn{pp},\gtrless}_{pqkl}(t\ge t') \, . 
\nonumber
\end{align}
As for the case of $GW$ selfenergies, here we introduced two quasi-Hamiltonians, 
\begin{align}
 \mathfrak{h}^{\Omega^\tn{pp}, \tn{HF}}_{ijkl}(t) &= h^\textnormal{(2),HF}_{ijkl}(t)\, , \\
  \mathfrak{h}^{\Omega^\tn{pp}, \tn{corr}}_{ijkl}(t) &= \left(\i \hbar\right)^2\sum_{pq} \Big[ \mathcal{G}^{\tn{H},>}_{ijpq}(t) - \mathcal{G}^{\tn{H},<}_{ijpq}(t) \Big] w_{pqkl}(t)\, . \label{eq:HTPPcorr} \phantom{.....}
\end{align}
Combining these Hamiltonians again into a single one,
\begin{align}
    \mathfrak{h}^{\Omega^\tn{pp}}_{ijkl}(t) = \mathfrak{h}^{\Omega^\tn{pp}, \tn{HF}}_{ijkl}(t)+\mathfrak{h}^{\Omega^\tn{pp}, \tn{corr}}_{ijkl}(t)\,,
    \label{eq:h-om-pp-def}
\end{align}
the equation \eqref{eq:omega-pp} for the M\o ller operator is transformed into a time-dependent two-particle Schrödinger equation,
\begin{align}
 & \i\hbar \frac{\d}{\d t} \Omega^{\tn{pp},\gtrless}_{ijkl}(t\ge t') 
 = \sum_{pq}  \mathfrak{h}^{\Omega^\tn{pp}}_{ijpq}(t) \Omega^{\tn{pp},\gtrless}_{pqkl}(t\ge t') \, . \label{eq:omega-pp-tdse}
\end{align}
This equation is analogous to the Schrödinger equation for the inverse dielectric function, \refeqn{eq:tdse-eps}, the main difference being the modified Hamiltonian \eqref{eq:h-om-pp-def}.
\subsubsection{$T^\tn{pp}$-G1--G2 equations for a general basis}
To derive the G1--G2 scheme for the particle--particle $T$ matrix, we have to take the derivative of $\mathcal{G}$, yielding,
  \begin{align}\label{eq:eom_parts_pp}
   &\frac{\d}{\d t} \mathcal{G}_{ijkl} (t) \\ &= \left[\frac{\d}{\d t} \mathcal{G}_{ijkl} (t)\right]_{\int^{}_{}} + \left[\frac{\d}{\d t} \mathcal{G}_{ijkl} (t)\right]_{\Omega^\tn{pp}} + \left[\frac{\d}{\d t} \mathcal{G}_{ijkl} (t)\right]_{\mathcal{U}^{(2)}}\nonumber\,,
   \end{align}
   The derivative of the integration boundaries results in, 
    \begin{align}\label{eq:eom_parts_int_pp}
   &\left[\frac{\d}{\d t} \mathcal{G}_{ijkl} (t)\right]_{\int^{}_{}} \\
   &\quad=\pm\sum_{pq} \Big[ \Omega^{\tn{pp},>}_{ijpq}(t,t) \mathcal{G}^{\tn{H},<}_{pqkl}(t) - \Omega^{\tn{pp},<}_{ijpq}(t, t) \mathcal{G}^{\tn{H},>}_{pqkl}(t)\Big]  \nonumber\\
   &\quad=\i\hbar \sum_{pqrs} w^\pm_{rspq}(t) \Big[ \mathcal{G}^{\tn{H},>}_{ijrs}(t) \mathcal{G}^{\tn{H},<}_{pqkl}(t) - \mathcal{G}^{\tn{H},<}_{ijrs}(t) \mathcal{G}^{\tn{H},>}_{pqkl}(t)\Big]\nonumber\\
   &\qquad + \i\hbar \sum_{pqrs} \Big[\mathcal{G}_{rsij}(t)\Big]^* w_{rspq}(t) \Big[ \mathcal{G}^{\tn{H},<}_{pqkl}(t) - \mathcal{G}^{\tn{H},>}_{pqkl}(t)\Big]\nonumber\\
   &\quad = \frac{1}{\i\hbar} \Psi^\pm_{ijkl}(t)
  - \frac{1}{\i\hbar} \sum_{pq} \Big[\mathfrak{h}^{\Omega^\tn{pp}, \tn{corr}}_{klpq}(t) \mathcal{G}_{pqij}(t)\Big]^*\,,\nonumber
   \end{align}
   while the time derivative of the M\o ller operator yields,
    \begin{align}\label{eq:eom_parts_omega_pp}
   &\left[\frac{\d}{\d t} \mathcal{G}_{ijkl} (t)\right]_{\Omega^\tn{pp}} \\
   &\quad=\pm \left(\i \hbar\right)^2 \sum_{pqrs} \int_{t_0}^t \d \cbar t \, \Bigg[ \left( \frac{\d}{\d t} \Omega^{\tn{pp},>}_{ijpq}(t,\cbar t) \right) \mathcal{G}^{\tn{H},<}_{pqrs}(\cbar t) 
   \nonumber \\ &\, \qquad
   - \left( \frac{\d}{\d t} \Omega^{\tn{pp},<}_{ijpq}(t,\cbar t) \right) \mathcal{G}^{\tn{H},>}_{pqrs}(\cbar t) \Bigg] \mathcal{U}^{(2)}_{rskl}(\cbar t, t) \nonumber \\
   &\quad = \frac{1}{\i\hbar} \sum_{pq} \left(\mathfrak{h}^{\Omega^\tn{pp}, \tn{HF}}_{ijpq}(t) + \mathfrak{h}^{\Omega^\tn{pp}, \tn{corr}}_{ijpq}(t)\right) \mathcal{G}_{pqkl}(t)\nonumber \,.
   \end{align}
   The last contribution originates from the derivative of the two-particle propagator,
  \begin{align}\label{eq:eom_parts_U_pp}
   &\left[\frac{\d}{\d t} \mathcal{G}_{ijkl} (t)\right]_{\mathcal{U}^{(2)}}\\
   &\quad=\pm \left(\i \hbar\right)^2 \sum_{pqrs} \int_{t_0}^t \d \cbar t \, \Big[ \Omega^{\tn{pp},>}_{ijpq}(t,\cbar t) \mathcal{G}^{\tn{H},<}_{pqrs}(\cbar t) 
   \nonumber \\ &\, \qquad
   - \Omega^{\tn{pp},<}_{ijpq}(t,\cbar t) \mathcal{G}^{\tn{H},>}_{pqrs}(\cbar t) \Big] \left( \frac{\d}{\d t} \mathcal{U}^{(2)}_{rskl}(\cbar t, t) \right) \nonumber\\
   &\quad= \frac{1}{\i\hbar} \sum_{pq} \mathcal{G}_{ijpq}(t) \mathfrak{h}^{\Omega^\tn{pp}, \tn{HF}}_{pqkl}(t) \nonumber \,, 
   \end{align}

Combining the three contributions to the derivative of $\mathcal{G}$ reveals
\begin{align}
 \i\hbar&\frac{\d}{\d t} \mathcal{G}_{ijkl} (t) =  
\Psi^\pm_{ijkl}(t)
  \nonumber \\
  & + \sum_{pq} \bigg\{ \mathfrak{h}^{\Omega^\tn{pp}}_{ijpq}(t) \Big[ \mathcal{G}_{klpq}(t) \Big]^* - \mathcal{G}_{ijpq}(t) \Big[\mathfrak{h}^{\Omega^\tn{pp}}_{klpq}(t)\Big]^* \bigg\} \, ,
  \label{eq:g2-eq-tpp} \phantom{.....}
\end{align}
where $\mathfrak{h}^{\Omega^\tn{pp}}(t)$ was introduced in \refeqn{eq:h-om-pp-def}.
This is the central equation for the G1--G2 scheme in $T$-matrix approximation for the particle--particle channel \cite{bonitz_qkt,kremp-etal.97ap}. Compared to the equation of motion for $\mathcal{G}$ in second Born approximation, \refeqn{eq:result_eom_soa}, this equation contains, in addition, 
the particle--particle ladder terms which are generated by the quasi-Hamiltonian of \refeqn{eq:HTPPcorr}. Again, for practical use, it is convenient to separate the correlation contributions from the mean-field terms via the introduction of an additional quantity:
\begin{align}
 \i\hbar&\frac{\d}{\d t} \mathcal{G}_{ijkl} (t) - \left[h^{(2),\tn{HF}}, \mathcal{G}\right]_{ijkl} (t) \label{eq:g2-eq-tpp-lambda} \\
 &= \Psi^\pm_{ijkl}(t)
  + \Lambda^\tn{pp}_{ijkl}(t) - \Big[\Lambda^\tn{pp}_{klij}(t)\Big]^*\, ,\nonumber
\end{align}
where the particle--particle ladder term is defined by
\begin{align}
 \Lambda^\tn{pp}_{ijkl}(t) = \sum_{pq} \mathfrak{h}^{\Omega^\tn{pp},\tn{corr}}_{ijpq}(t) \mathcal{G}_{pqkl}(t) \, . \label{eq:lambdapp}
\end{align}
Without the $\Lambda$-terms we exactly recover the equation of motion for $\mathcal{G}$ in second-order Born approximation. Inclusion of the $\Lambda$-terms, on the other hand, allows one to take into account multiple scattering and large-angle scattering effects that are important for strongly correlated systems. These terms correspond to the summation of the infinite Born series.

\subsubsection{$T^\tn{pp}$-G1--G2 equations for the Hubbard model}
We now apply this result to the Hubbard Hamiltonian and find,
\begin{align}
 \i\hbar\frac{\d}{\d t} & \mathcal{G}^{\uparrow\downarrow\uparrow\downarrow}_{ijkl} (t) - \left[h^{(2),\tn{HF}}_{\uparrow\downarrow}, \mathcal{G}^{\uparrow\downarrow\uparrow\downarrow}\right]_{ijkl} (t) \\
 &= \Psi^{\uparrow\downarrow\uparrow\downarrow}_{ijkl}(t) + \Lambda^{\tn{pp},\uparrow\downarrow\uparrow\downarrow}_{ijkl}(t) - \Big[ \Lambda^{\tn{pp},\uparrow\downarrow\uparrow\downarrow}_{klij}(t)\Big]^*\,,\nonumber
\end{align}
where we introduced the particle--particle ladder term
\begin{align} \label{eq:lambdapp_hubb}
 &\Lambda^{\tn{pp},\uparrow\downarrow\uparrow\downarrow}_{ijkl}(t) = (\i\hbar)^2 U(t)\times
 \\&\,
  \sum_{p} \Big[ G^{>,\uparrow}_{ip} (t) G^{>,\downarrow}_{jp} (t) - G^{<,\uparrow}_{ip} (t) G^{<,\downarrow}_{jp} (t) \Big] \mathcal{G}^{\uparrow\downarrow\uparrow\downarrow}_{ppkl}(t) \, . \nonumber
\end{align}
In the present case there exists only one distinct spin combination (two when considering $\uparrow\,\leftrightarrow\,\downarrow$) of the particle pair that enters the single-particle EOM [cf. \refeqns{eq:eom_g1_hubbard}{eq:colint_hubbard}] which simplifies the equations.
Numerical results for the $T^\tn{pp}$-G1--G2 scheme are presented in Sec.~\ref{s:numerics}.

\subsubsection{$T^\tn{pp}$-G1--G2 equations for jellium}\label{sss:tpp-jellium}
Turning now to the uniform electron gas, \refeqn{eq:h-jellium}, we again use the interaction matrix \eqref{eq:jellium-w-matrix},
and define
\begin{align}
    \Lambda_{\bm{p},\bm{\cbar p},\bm{q}}^{\tn{pp},\alpha \beta}(t) &:= \Lambda_{\bm{p}-\bm{q},\bm{\cbar p}+\bm{q},\bm{p},\bm{\cbar p}}^{\tn{pp},\alpha \beta \alpha \beta}(t)\, .
\end{align}
With that, the equation of motion for the time-diagonal two-particle Green function becomes,
\begin{align}
 \i\hbar&\frac{\d}{\d t} \mathcal{G}_{\bm{p} \bm{\cbar p} \bm{q}}^{ \alpha \beta }(t) - \mathcal{G}_{\bm{p} \bm{\cbar p} \bm{q}}^{ \alpha \beta }(t) 
 \left(h^{\tn{HF},\alpha}_{\bm{p} -\bm{q}} + h^{\tn{HF},\beta}_{\bm{\cbar{p}} +\bm{q}} - h^{\tn{HF},\alpha}_{\bm{p}} - h^{\tn{HF},\beta}_{\bm{\cbar{p}}} \right) \nonumber \\
 &= \Psi_{\bm{p} \bm{\cbar p} \bm{q}}^{ \pm, \alpha \beta }(t) + \Lambda_{\bm{p},\bm{\cbar p},\bm{q}}^{\tn{pp},\alpha \beta}(t) - \Big[ \Lambda_{\bm{p}-\bm{q},\bm{\cbar p}+\bm{q},-\bm{q}}^{\tn{pp},\alpha\beta}(t)\Big]^*\, , 
\end{align}
where the momentum representation of the particle--particle ladder term is given by
\begin{align}
 \Lambda_{\bm{p},\bm{\cbar p},\bm{q}}^{\tn{pp},\alpha \beta}(t) =&  (\i\hbar)^2\Big[  G^{>,\alpha}_{\bm{p}-\bm{q}}(t) G^{>,\beta}_{\bm{\cbar p}+\bm{q}}(t) - G^{<,\alpha}_{\bm{p}-\bm{q}}(t) G^{<,\beta}_{\bm{\cbar p}+\bm{q}}(t)\Big] \times \nonumber \\ & \quad \times \sum_{\bm{k}} v_{\left|\bm{k}-\bm{q}\right|}(t) \mathcal{G}_{\bm{p}\bm{\cbar p}\bm{k}}^{\alpha \beta}(t)\,. \label{eq:jellium-lambdapp}
\end{align}

\subsection{Particle--hole $T$ matrix}\label{ss:tph}
For the $T$ matrix in the particle--hole channel \cite{schluenzen_jpcm_19}, the derivations of the single-time equations are performed in similar fashion as for the particle--particle $T$ matrix in Sec.~\ref{ss:tpp}. The detailed derivation is given in Appendix~\ref{app:tph}. Here, we summarize the main findings.

\subsubsection{$T^\tn{ph}$-G1--G2 equations for a general basis}
As for the $GW$ and the TPP approximations, two quasi-Hamiltonians are  introduced, 
\begin{align}
 \mathfrak{h}^{\Omega^\tn{ph}, \tn{HF}}_{ijkl}(t) &= \delta_{jl}h^\tn{HF}_{ik}-\delta_{ik}h^\tn{HF}_{jl}\, , \\
  \mathfrak{h}^{\Omega^\tn{ph}, \tn{corr}}_{ijkl}(t) &= \left(\i \hbar\right)^2\sum_{pq} \Big[ \mathcal{G}^{\tn{F},>}_{iqlp}(t) - \mathcal{G}^{\tn{F},<}_{iqlp}(t) \Big] w_{pjkq}(t)\, , \label{eq:HTPHcorr} \phantom{.....}
  \end{align}
  and combined into a single quantity,
  \begin{align}
  \mathfrak{h}^{\Omega^\tn{ph}}_{ijkl}(t)
  &=\mathfrak{h}^{\Omega^\tn{ph}, \tn{HF}}_{ijkl}(t) + \mathfrak{h}^{\Omega^\tn{ph}, \tn{corr}}_{ijkl}(t)\,.
  \label{eq:homph-def}
\end{align}
The corresponding  M\o ller operator of the particle--hole $T$ matrix again obeys a time-dependent two-particle Schrödinger equation,
\begin{align}
 &\i\hbar\frac{\d}{\d t} \Omega^{\tn{ph},\gtrless}_{ijkl}(t\ge t')
 =  \sum_{pq}  \mathfrak{h}^{\Omega^\tn{ph}}_{ipql}(t)  \Omega^{\tn{ph},\gtrless}_{qjkp}(t\ge t')
 \, . 
 \label{eq:omph-tdse}
\end{align}
The time derivative of $\mathcal{G}$ in TPH approximation follows as
\begin{align}
 \i\hbar&\frac{\d}{\d t} \mathcal{G}_{ijkl} (t) =  \Psi^\pm_{ijkl}(t)
  \label{eq:g2-eq-tph}
  \\
  & + \sum_{pq} \bigg\{ \mathfrak{h}^{\Omega^\tn{ph}}_{ipql}(t) \Big[ \mathcal{G}_{kpqj}(t) \Big]^* - \mathcal{G}_{ipql}(t) \Big[\mathfrak{h}^{\Omega^\tn{ph}}_{kpqj}(t)\Big]^* \bigg\} \, .
  \nonumber 
\end{align}
Again, for practical use, it is convenient to separate the correlation contributions from the mean-field terms via the introduction of 
an additional quantity:
\begin{align}
 \i\hbar&\frac{\d}{\d t} \mathcal{G}_{ijkl} (t) - \left[h^{(2),\tn{HF}}, \mathcal{G}\right]_{ijkl} (t) \label{eq:g2-eq-tph-lambda} \\
 &= \Psi^\pm_{ijkl}(t)
  + \Lambda^\tn{ph}_{ijkl}(t) - \Big[\Lambda^\tn{ph}_{klij}(t)\Big]^*\, ,\nonumber
\end{align}
where the particle--hole ladder term is defined by
\begin{align}
 \Lambda^\tn{ph}_{ijkl}(t) = \sum_{pq} \mathfrak{h}^{\Omega^\tn{ph},\tn{corr}}_{ipql}(t) \mathcal{G}_{qjkp}(t) \, . \label{eq:lambdaph}
\end{align}
As in the case of the particle--particle $T$ matrix, Sec.~\ref{ss:tpp}, neglect of the $\Lambda$-terms exactly recovers the equation of motion for $\mathcal{G}$ in second-order Born approximation. Inclusion of theses terms, on the other hand, accounts for the entire Born series.

\subsubsection{$T^\tn{ph}$-G1--G2 equations for the Hubbard basis}
\label{sss:tph-hubbard}
For the Hubbard system (for the definitions, see Sec.~\ref{ss:hubbard-soa}), we find,
\begin{align}
 \i\hbar&\frac{\d}{\d t} \mathcal{G}^{\uparrow\downarrow\uparrow\downarrow}_{ijkl} (t) - \left[h^{(2),\tn{HF}}_{\uparrow\downarrow}, \mathcal{G}^{\uparrow\downarrow\uparrow\downarrow}\right]_{ijkl} (t) \\
 &= \Psi^{\uparrow\downarrow\uparrow\downarrow}_{ijkl}(t)
  +\Lambda^{\tn{ph},\uparrow\downarrow\uparrow\downarrow}_{ijkl}(t) - \Big[ \Lambda^{\tn{ph},\uparrow\downarrow\uparrow\downarrow}_{klij}(t)\Big]^*\,,\nonumber
\end{align}
where we introduced the particle--hole ladder term for the Hubbard system
\begin{align} \label{eq:lambdaph_hubb}
 &\Lambda^{\tn{ph},\uparrow\downarrow\uparrow\downarrow}_{ijkl}(t) = (\i\hbar)^2 U(t)\times
 \\&\,
  \sum_{p} \Big[ G^{>,\uparrow}_{ip} (t) G^{<,\downarrow}_{pl} (t) - G^{<,\uparrow}_{ip} (t) G^{>,\downarrow}_{pl} (t) \Big] \mathcal{G}^{\uparrow\downarrow\uparrow\downarrow}_{pjkp}(t) \, . \nonumber
\end{align}
Similar to the behavior in the TPP case, only one spin combination (two when considering $\uparrow\,\leftrightarrow\,\downarrow$) contributes to the single-particle EOM in \refeqns{eq:eom_g1_hubbard}{eq:colint_hubbard}. 
The $T^\tn{ph}$-G1--G2 scheme for the Hubbard model is numerically tested in Sec.~\ref{s:numerics}.

\subsubsection{$T^\tn{ph}$-G1--G2 equations for jellium}\label{sss:tph-jellium}
For the uniform electron gas, \refeqn{eq:h-jellium}, we again use the interaction matrix \eqref{eq:jellium-w-matrix},
and define
\begin{align}
    \Lambda_{\bm{p},\bm{\cbar p},\bm{q}}^{\tn{ph},\alpha \beta}(t) &:= \Lambda_{\bm{p}-\bm{q},\bm{\cbar p}+\bm{q},\bm{p},\bm{\cbar p}}^{\tn{ph},\alpha \beta \alpha \beta}(t)\, .
\end{align}
With that, the equation of motion for the time-diagonal two-particle Green function becomes,
\begin{align}
 \i\hbar&\frac{\d}{\d t} \mathcal{G}_{\bm{p} \bm{\cbar p} \bm{q}}^{ \alpha \beta }(t) - \mathcal{G}_{\bm{p} \bm{\cbar p} \bm{q}}^{ \alpha \beta }(t) 
 \left(h^{\tn{HF},\alpha}_{\bm{p} -\bm{q}} + h^{\tn{HF},\beta}_{\bm{\cbar{p}} +\bm{q}} - h^{\tn{HF},\alpha}_{\bm{p}} - h^{\tn{HF},\beta}_{\bm{\cbar{p}}} \right) \nonumber \\
 &= \Psi_{\bm{p} \bm{\cbar p} \bm{q}}^{ \pm, \alpha \beta }(t) + \Lambda_{\bm{p},\bm{\cbar p},\bm{q}}^{\tn{ph},\alpha \beta}(t) - \Big[ \Lambda_{\bm{p}-\bm{q},\bm{\cbar p}+\bm{q},-\bm{q}}^{\tn{ph},\alpha\beta}(t)\Big]^*\, , 
\end{align}
with the momentum representation of the particle--hole ladder term, given by
\begin{align} 
 \Lambda_{\bm{p},\bm{\cbar p},\bm{q}}^{\tn{ph},\alpha \beta}(t) =&  (\i\hbar)^2\Big[  G^{>,\alpha}_{\bm{p}-\bm{q}}(t) G^{<,\beta}_{\bm{\cbar p}}(t) - G^{<,\alpha}_{\bm{p}-\bm{q}}(t) G^{>,\beta}_{\bm{\cbar p}}(t)\Big] \times \nonumber \\ & \quad \times \sum_{\bm{k}} v_{\left|\bm{k}\right|}(t) \mathcal{G}_{\bm{p},\bm{\cbar p}-\bm{k},\bm{q}+\bm{k}}^{\alpha \beta}(t)\,.
 \label{eq:jellium-lambdaph}
\end{align}

\section{Dynamically-Screened-Ladder Approximation}\label{s:dyn-scr-ladder}
So far we have considered three important selfenergy approximations: the second-Born approximation, $GW$ and the particle--particle and particle--hole $T$ matrices. While $GW$ describes dynamical screening, for weakly coupled systems, the $T$-matrix selfenergy accounts for strong coupling but neglects dynamic screening effects. Therefore, the question arises how to combine strong coupling and dynamical screening into a single model in a computationally feasible way. An approximate to realize this within NEGF theory is the fluctuating-exchange approximation (FLEX) that combines $T$ matrix and $GW$ contributions according to $\Sigma=\Sigma_{\rm TPP}+\Sigma_{\rm TPH}+\Sigma_{GW}-2\Sigma_{\rm SOA}$, where the last term is needed to avoid double counting, for more details, see Ref.~\cite{schluenzen_jpcm_19}. A fully selfconsistent treatment of dynamical-screening and strong-coupling effects is provided by the dynamically-screened-ladder approximation that has been studied in the context of the bound-state problem in a plasma medium in equilibrium \cite{zimmermann78}. For more details, see  Ref.~\cite{kraeft-green-book}. 

The G1--G2 scheme allows for a straightforward way to combine the $GW$ (including exchange) and both $T$-matrix approximations in a selfconsistent way for arbitrary nonequilibrium situations. This is achieved by including in the EOM of the time-diagonal two-particle Green function the terms with all effective Hamiltonians that were derived for 
$GW$, the particle--particle and the particle--hole $T$ matrix, respectively, cf. \eqrefss{eq:h-epsilon-def}{eq:h-om-pp-def}{eq:homph-def}. 
Then, the EOM for $\mathcal{G}$, in a general basis becomes,
\begin{align}
 \i\hbar&\frac{\d}{\d t} \mathcal{G}_{ijkl} (t) - \left[h^{(2),\tn{HF}}, \mathcal{G}\right]_{ijkl} (t) \label{eq:g2-dsl-equation} = \Psi^\pm_{ijkl}(t)
\\
  & + \sum_{pq} \bigg\{ \mathfrak{h}^{\varepsilon,\tn{corr}}_{qjpl}(t) \Big[ \mathcal{G}_{qkpi}(t) \Big]^* - \mathcal{G}_{qjpl}(t) \Big[\mathfrak{h}^{\varepsilon,\tn{corr}}_{qkpi}(t)\Big]^* \bigg\} \nonumber
  \\
  & + \sum_{pq} \bigg\{ \mathfrak{h}^{\Omega^\tn{pp},\tn{corr}}_{ijpq}(t) \Big[ \mathcal{G}_{klpq}(t) \Big]^* - \mathcal{G}_{ijpq}(t) \Big[\mathfrak{h}^{\Omega^\tn{pp},\tn{corr}}_{klpq}(t)\Big]^* \bigg\}\nonumber\\
  & + \sum_{kl} \bigg\{ \mathfrak{h}^{\Omega^\tn{ph},\tn{corr}}_{ipql}(t) \Big[ \mathcal{G}_{kpqj}(t) \Big]^* - \mathcal{G}_{ipql}(t) \Big[\mathfrak{h}^{\Omega^\tn{ph},\tn{corr}}_{kpqj}(t)\Big]^* \bigg\}\,.
\nonumber
\end{align}
Alternatively, we can rewrite this equation by using the polarization ($\Pi$) and ladder ($\Lambda$) terms that were defined by \eqrefss{eq:gwpi}{eq:lambdapp}{eq:lambdaph},
\begin{align}
 \i\hbar&\frac{\d}{\d t} \mathcal{G}_{ijkl} (t) - \left[h^{(2),\tn{HF}}, \mathcal{G}\right]_{ijkl} (t) = \Psi^\pm_{ijkl}(t)
     \label{eq:g2-dsl-equation-pi-lambda}
\\
  & + \Pi_{ijkl}(t) - \Big[\Pi_{lkji}(t)\Big]^* + \Lambda_{ijkl}(t) - \Big[\Lambda_{klij}(t)\Big]^*\, ,
  \nonumber
\end{align}
where we combined both ladder terms into
\begin{align}
 \Lambda_{ijkl}(t) &= \Lambda^\tn{pp}_{ijkl}(t) + \Lambda^\tn{ph}_{ijkl}(t)\, .
\end{align}
Obviously, \refeqn{eq:g2-dsl-equation-pi-lambda} is a generalization of all previous cases: it additively includes 
the contributions of the second-order Born selfenergy (second line), polarization terms that account for dynamical screening and strong coupling terms. The SOA term that appears in each of the different approximations is included only once, so no double counting occurs.
Since all contributions are treated on the same footing, this equation amounts to a simultaneous full account of dynamical screening and strong binary correlations. Alternatively, this approximation can be obtained from reduced-density-operator theory by neglecting three-particle and higher correlations \cite{bonitz_qkt}; an early discussion was presented by Wang and Cassing \cite{wang-cassing-85}.

It is easily verified that the entire \refeqn{eq:g2-dsl-equation} requires a CPU-time that has the same linear scaling with $N_\tn{t}$ as all the special cases that were studied before. On the other hand, the polarization and ladder terms determine the scaling with the basis size $N_\tn{b}$. This is summarized in Tab.~\ref{tab:scaling} and discussed in more detail in Sec.~\ref{s:numerics}.

\section{Verification of the numerical scaling}\label{s:numerics}
As was shown in the previous sections, the G1--G2 scheme reduces the time-diagonal Keldysh-Kadanoff-Baym equation within the HF-GKBA to a memory-less, time-local form. This means, the theoretical scaling is first order in the propagation duration. This dramatic acceleration is achieved by propagating, in addition to the single-particle Green function, also the time-diagonal two-particle Green function $\mathcal{G}$. This function has, in general, four basis indices and, thus, a dimensionality of $N_\tn{b}^4$, where $N_\tn{b}$ is the single-particle basis dimension. The total scaling of the G1--G2 scheme with $N_\tn{b}$ depends on the selfenergy and on the type of basis. In the following, we investigate this scaling more in detail, extending the analysis of Ref.~\cite{schluenzen_19_prl}.

\subsection{Second-order Born selfenergy}\label{ss:numerics_soa}
We start by analyzing the $N_\tn{b}$-scaling of the SOA-HF-GKBA equation for $\mathcal{G}$, \refeqn{eq:result_eom_soa}, which we rewrite in a different form
\begin{align}
 &\i\hbar\frac{\d}{\d t} \mathcal{G}_{ijkl} (t) - \left[h^{(2),\tn{HF}}, \mathcal{G}\right]_{ijkl} (t) \label{eq:soa-skalierung} \\
 &= \left(\i\hbar\right)^2\sum_{p} 
  G^>_{ip}(t) \sum_q G^>_{jq}(t) \sum_r G^<_{rk}(t) \sum_s w^\pm_{pqrs}(t) G^<_{sl}(t) \nonumber \\
  &-\left(\i\hbar\right)^2\sum_{p} 
  G^<_{ip}(t) \sum_q G^<_{jq}(t) \sum_r G^>_{rk}(t) \sum_s w^\pm_{pqrs}(t) G^>_{sl}(t) \, . \nonumber
\end{align}
The r.h.s. of this equation contains four sums of dimensionality $N_\tn{b}$ which are all independent of each other. They are evaluated by successive execution of the occurring tensor contractions. This means the total scaling of the CPU time, in this case, is of order $N_\tn{b}^5$.

For the Hubbard basis a first look at \eqrefrno{eq:g2-hubbard}{eq:phi-hubbard}{eq:twopart_hamiltonian_hubbard} suggests an $N_\tn{b}^5$-scaling, due to the commutator term in \refeqn{eq:g2-hubbard} and the summation in the $\Phi$ term of \refeqn{eq:phi-hubbard}. However, in the Hubbard model the scaling can be further reduced. 
Note that the Hartree--Fock Hamiltonian, $h^\tn{HF}(t)$, is a tridiagonal matrix and, thus, the commutator can be computed with $N_\tn{b}^4$ effort:
\begin{align}
    &\left[h^{(2),\tn{HF}}_{\uparrow\downarrow}, \mathcal{G}^{\uparrow \downarrow \uparrow\downarrow}\right]_{ijkl} (t) \nonumber \\
    &\quad= \sum_p \Big[h^{\tn{HF},\uparrow}_{ip}(t) \mathcal{G}^{\uparrow \downarrow \uparrow \downarrow}_{pjkl}(t) + h^{\tn{HF},\downarrow}_{jp}(t) \mathcal{G}^{\uparrow \downarrow \uparrow \downarrow}_{ipkl}(t) \nonumber \\
    &\quad \qquad- \mathcal{G}^{\uparrow \downarrow \uparrow \downarrow}_{ijpl}(t) h^{\tn{HF},\uparrow}_{pk}(t) - \mathcal{G}^{\uparrow \downarrow \uparrow \downarrow}_{ijkp}(t) h^{\tn{HF},\downarrow}_{pl}(t)\Big] \nonumber \\
    &\quad = - \i\hbar U(t) \mathcal{G}^{\uparrow \downarrow \uparrow \downarrow}_{ijkl}(t) \Big[G^{\downarrow}_{ii}(t) + G^{\uparrow}_{jj}(t) - G^{\downarrow}_{kk}(t) - G^{\uparrow}_{ll}(t)\Big] \nonumber \\
    &\quad \qquad-J\sum_p \Big[\delta_{\left<i,p\right>} \mathcal{G}^{\uparrow \downarrow \uparrow \downarrow}_{pjkl}(t) + \delta_{\left<j,p\right>} \mathcal{G}^{\uparrow \downarrow \uparrow \downarrow}_{ipkl}(t) \nonumber \\
    &\quad \qquad \qquad- \mathcal{G}^{\uparrow \downarrow \uparrow \downarrow}_{ijpl}(t) \delta_{\left<p,k\right>} - \mathcal{G}^{\uparrow \downarrow \uparrow \downarrow}_{ijkp}(t) \delta_{\left<p,l\right>}\Big] \, .
\end{align}
On the other hand, the $\Phi$ term can be simplified by using the identity of \refeqn{eq:g-glsymm}:
\begin{align}
&\Phi^{\uparrow\downarrow\uparrow\downarrow}_{ijkl}(t) \nonumber \\
&\quad = \left(\i\hbar\right)^4 \sum_{p}
  \bigg\{ \Big[G^{<,\uparrow}_{ip}(t) + \frac{1}{\i\hbar} \delta_{ip}\Big] \Big[G^{<,\downarrow}_{jp}(t) + \frac{1}{\i\hbar} \delta_{jp}\Big] \times \nonumber \\ 
  &\qquad \quad  \times G^{<,\uparrow}_{pk}(t) G^{<,\downarrow}_{pl}(t) -G^{<,\uparrow}_{ip}(t) G^{<,\downarrow}_{jp}(t) \times \nonumber \\
  &\quad \qquad \times \Big[G^{<,\uparrow}_{pk}(t) + \frac{1}{\i\hbar} \delta_{pk}\Big] \Big[G^{<,\downarrow}_{pl}(t) + \frac{1}{\i\hbar} \delta_{pl}\Big] \bigg\} \nonumber \\
    &\quad = \left(\i\hbar\right)^2 \left(\delta_{ij} - \delta_{kl} \right) G^{<,\uparrow}_{ik}(t) G^{<,\downarrow}_{jl}(t)
    \label{eq:soa-cubic-scheme}\\
    &\quad \quad + \left(\i\hbar\right)^3 \left[ G^{<,\uparrow}_{ij}(t) G^{<,\uparrow}_{jk}(t) - G^{<,\uparrow}_{lk}(t) G^{<,\uparrow}_{il}(t) \right] G^{<,\downarrow}_{jl}(t) \nonumber\\
    & \quad \quad + \left(\i\hbar\right)^3 \left[ G^{<,\downarrow}_{ji}(t) G^{<,\downarrow}_{il}(t) - G^{<,\downarrow}_{kl}(t) G^{<,\downarrow}_{jk}(t) \right] G^{<,\uparrow}_{ik}(t)\nonumber\,.
\end{align}
Here, the leading contribution to the difference, ${G^<G^<G^<G^<-G^<G^<G^<G^<}$, cancels (contribution with four functions $G^<$) which reduces the complexity. For the Hubbard basis, this reduces the 
numerical effort
of the G1--G2 scheme to a $N^4_\tn{b}$-scaling compared to the $N^5_\tn{b}$-scaling in the straightforward implementation \cite{schluenzen_19_prl}. 
In total, an acceleration is achieved for the SOA-G1--G2 scheme, compared to the ordinary HF-GKBA if $N_\tn{t}\gtrsim N_\tn{b}$, as summarized in Tab.~\ref{tab:scaling}.

The reformulation above that eliminates products of four $G^<$ functions can be made for any basis choice. However, for the general basis 
this does not result in an improved $N_\tn{b}$-scaling.
For the jellium basis the \eqrefr{eq:jellium-g1}{eq:jellium-g2} reveal a particularly favorable scaling with the basis size with $N_\tn{b}^3$ for which the above reformulation does not provide further improvement.

\subsection{$GW$ selfenergy}\label{ss:numerics_gw}

The additional terms of the $GW$ approximation can change the $N_\tn{b}$ scaling compared to the SOA case discussed in the previous section. For the general basis, the leading-order terms for the scaling with the basis size are found in \refeqns{eq:gwhcorr}{eq:gwpi} which reveal a $N_\tn{b}^6$-scaling. For this case no further reductions are possible, cf. Tab.~\ref{tab:scaling}.

For the Hubbard basis the polarization terms [\refeqns{eq:pihubbardtrans}{eq:pihubbardcis}] can be reformulated by again using \refeqn{eq:g-glsymm} to get
\begin{align}
    &\Pi^{\uparrow\downarrow\uparrow\downarrow}_{ijkl}(t) = - \i\hbar U(t)
   G^{<,\downarrow}_{jl} (t) \Big[\mathcal{G}^{\uparrow\uparrow\uparrow\uparrow}_{ijkj}(t) - \mathcal{G}^{\uparrow\uparrow\uparrow\uparrow}_{ilkl}(t)\Big] \, , \\
   &\Pi^{\uparrow\uparrow\uparrow\uparrow}_{ijkl}(t) = - \i\hbar U(t)
   G^{<,\uparrow}_{jl} (t) \Big[\mathcal{G}^{\uparrow\downarrow\uparrow\downarrow}_{ijkj}(t) - \mathcal{G}^{\uparrow\downarrow\uparrow\downarrow}_{ilkl}(t)\Big] \, .
\end{align}
From this, it is obvious that, compared to the second-order Born approximation, no further complexity is added for $GW$ in the Hubbard case, and the scaling with the basis size remains $N_\tn{b}^4$.

To explore the $N_\tn{b}$-scaling for the jellium basis we recall the polarization term, \refeqn{eq:jellium-gwpi},
\begin{align}
 \Pi_{\bm{p},\bm{\cbar p},\bm{q}}^{\alpha \beta}(t) =& \pm (\i\hbar)^2\Big[  G^{>,\beta}_{\bm{\cbar p}+\bm{q}}(t) G^{<,\beta}_{\bm{\cbar p}}(t) - G^{<,\beta}_{\bm{\cbar p}+\bm{q}}(t) G^{>,\beta}_{\bm{\cbar p}}(t)\Big] \times \nonumber \\ & \quad \times v_{\left|\bm{q}\right|}(t) \sum_{\bm{k},\sigma} \mathcal{G}_{\bm{p}\bm{k}\bm{q}}^{\alpha \sigma}(t)\,.
 \nonumber
\end{align}
As one can see, the tensor contraction over $\bm{k}$ can be executed independently of $\bm{\cbar p}$. Thus, the full scaling of the $GW$--G1--G2 scheme  for a jellium basis remains of order $N_\tn{b}^3$, as in the case of the standard HF-GKBA. 

\subsection{$T$-matrix selfenergies}

The $T$-matrix equations [Sec.~\ref{s:t-matrix}] behave very similar to the $GW$ equations. For a general basis set with a four-index interaction tensor both, TPP and TPH scale as $N_\tn{b}^6$ which can be directly seen from \refeqns{eq:HTPPcorr}{eq:lambdapp}, as well as \refeqns{eq:HTPHcorr}{eq:lambdaph}.

For the Hubbard basis we can now use \refeqn{eq:g-glsymm} to eliminate contributions that are of second order in $G^<$ from the ladder terms in \refeqn{eq:lambdapp_hubb},
\begin{align}
 &\Lambda^{\tn{pp},\uparrow\downarrow\uparrow\downarrow}_{ijkl}(t) = \delta_{ij} U(t) \mathcal{G}^{\uparrow\downarrow\uparrow\downarrow}_{ijkl}(t) \\
 &\qquad + \i\hbar U(t) \Big[ G^{<,\downarrow}_{ji} (t) \mathcal{G}^{\uparrow\downarrow\uparrow\downarrow}_{iikl}(t) + G^{<,\uparrow}_{ij} (t) \mathcal{G}^{\uparrow\downarrow\uparrow\downarrow}_{jjkl}(t) \Big]  \, , \nonumber
\end{align}
as well as in \refeqn{eq:lambdaph_hubb},
\begin{align}
 &\Lambda^{\tn{ph},\uparrow\downarrow\uparrow\downarrow}_{ijkl}(t)\\
 &\qquad= \i\hbar U(t) \Big[G^{<,\downarrow}_{il} (t) \mathcal{G}^{\uparrow\downarrow\uparrow\downarrow}_{ijki}(t) - G^{<,\uparrow}_{il} (t) \mathcal{G}^{\uparrow\downarrow\uparrow\downarrow}_{ljkl}(t)\Big] \, . \nonumber
\end{align}
For both cases one can see that the remaining scaling order of the equations is $N_\tn{b}^4$ since all internal summations have been eliminated. 

In the jellium basis the $T$ matrices show a different scaling behavior compared to $GW$. To see this, we reproduce the two ladder terms of \refeqns{eq:jellium-lambdapp}{eq:jellium-lambdaph},
\begin{align}
 \Lambda_{\bm{p},\bm{\cbar p},\bm{q}}^{\tn{pp},\alpha \beta}(t) =&  (\i\hbar)^2\Big[  G^{>,\alpha}_{\bm{p}-\bm{q}}(t) G^{>,\beta}_{\bm{\cbar p}+\bm{q}}(t) - G^{<,\alpha}_{\bm{p}-\bm{q}}(t) G^{<,\beta}_{\bm{\cbar p}+\bm{q}}(t)\Big] \times \nonumber \\ & \quad \times \sum_{\bm{k}} v_{\left|\bm{k}-\bm{q}\right|}(t) \mathcal{G}_{\bm{p}\bm{\cbar p}\bm{k}}^{\alpha \beta}(t)\,, 
\end{align}
\begin{align} 
 \Lambda_{\bm{p},\bm{\cbar p},\bm{q}}^{\tn{ph},\alpha \beta}(t) =&  (\i\hbar)^2\Big[  G^{>,\alpha}_{\bm{p}-\bm{q}}(t) G^{<,\beta}_{\bm{\cbar p}}(t) - G^{<,\alpha}_{\bm{p}-\bm{q}}(t) G^{>,\beta}_{\bm{\cbar p}}(t)\Big] \times \nonumber \\ & \quad \times \sum_{\bm{k}} v_{\left|\bm{k}\right|}(t) \mathcal{G}_{\bm{p},\bm{\cbar p}-\bm{k},\bm{q}+\bm{k}}^{\alpha \beta}(t)\,.
\end{align}
Evidently, in both cases the tensor contraction of $\bm{k}$ depends on all other momenta $\bm{p},\bm{\cbar p}, \bm{q}$. Thus, the final scaling with the basis size becomes of order $N_\tn{b}^4$.
A summary of the numerical scaling with the propagation duration and the basis size is presented in \tref{tab:scaling}.

At the same time, any practical implementation of the G1--G2 scheme could, in principle, carry a large overhead that prevents to achieve the theoretical scaling with the simulation duration and the basis dimension within a relevant parameter range. We, therefore, have implemented the G1--G2 scheme for each of the selfenergies discussed in this paper and present representative numerical results in Sec.~\ref{ss:numerik-hubbard}. 
\def\arraystretch{1.5}
\begin{table}[t]
\begin{tabular}{c|c|c|c|c}
\multicolumn{2}{c}{}&\multicolumn{2}{c}{HF-GKBA}&\multirow[b]{ 2}{*}{\shortstack{speedup\\ ratio}}\\ \cline{3-4}
 $\Sigma$ & Basis & standard & G1--G2 & \\
 \hline
  & general & $\mathcal{O}\left(N_\tn{b}^5 N_\tn{t}^2\right)$ &  $\mathcal{O}\left(N_\tn{b}^5 N_\tn{t}^1\right)$ & $\mathcal{O}\left(N_\tn{t}\right)$\\
 SOA & Hubbard & $\mathcal{O}\left(N_\tn{b}^3 N_\tn{t}^2\right)$ & $\mathcal{O}\left(N_\tn{b}^4 N_\tn{t}^1\right)$ & $\mathcal{O}\left(N_\tn{t}/N_\tn{b}\right)$\\
  & jellium 
  & $\mathcal{O}\left(N_\tn{b}^3 N_\tn{t}^2\right)$ & $\mathcal{O}\left(N_\tn{b}^3 N_\tn{t}^1\right)$ & $\mathcal{O}\left(N_\tn{t}\right)$\\
   \hline
  & general & $\mathcal{O}\left(N_\tn{b}^6 N_\tn{t}^3\right)$ & $\mathcal{O}\left(N_\tn{b}^6 N_\tn{t}^1\right)$ & $\mathcal{O}\left(N^2_\tn{t}\right)$\\
 $GW$ & Hubbard & $\mathcal{O}\left(N_\tn{b}^3 N_\tn{t}^3\right)$ & $\mathcal{O}\left(N_\tn{b}^4 N_\tn{t}^1\right)$ & $\mathcal{O}\left(N^2_\tn{t}/N_\tn{b}\right)$\\
  & jellium & $\mathcal{O}\left(N_\tn{b}^3 N_\tn{t}^3\right)$ & $\mathcal{O}\left(N_\tn{b}^3 N_\tn{t}^1\right)$ & $\mathcal{O}\left(N^2_\tn{t}\right)$\\
  \hline
  & general & $\mathcal{O}\left(N_\tn{b}^6 N_\tn{t}^3\right)$ & $\mathcal{O}\left(N_\tn{b}^6 N_\tn{t}^1\right)$ & $\mathcal{O}\left(N^2_\tn{t}\right)$\\
  TPP & Hubbard& $\mathcal{O}\left(N_\tn{b}^3 N_\tn{t}^3\right)$ & $\mathcal{O}\left(N_\tn{b}^4 N_\tn{t}^1\right)$ & $\mathcal{O}\left(N^2_\tn{t}/N_\tn{b}\right)$\\
  & jellium & $\mathcal{O}\left(N_\tn{b}^3 N_\tn{t}^3\right)$ & $\mathcal{O}\left(N_\tn{b}^4 N_\tn{t}^1\right)$ &$\mathcal{O}\left(N^2_\tn{t}/N_\tn{b}\right)$\\
  \hline
  & general & $\mathcal{O}\left(N_\tn{b}^6 N_\tn{t}^3\right)$ & $\mathcal{O}\left(N_\tn{b}^6 N_\tn{t}^1\right)$ & $\mathcal{O}\left(N^2_\tn{t}\right)$\\
 TPH & Hubbard & $\mathcal{O}\left(N_\tn{b}^3 N_\tn{t}^3\right)$ & $\mathcal{O}\left(N_\tn{b}^4 N_\tn{t}^1\right)$ &$\mathcal{O}\left(N^2_\tn{t}/N_\tn{b}\right)$\\
  & jellium & $\mathcal{O}\left(N_\tn{b}^3 N_\tn{t}^3\right)$ & $\mathcal{O}\left(N_\tn{b}^4 N_\tn{t}^1\right)$ & $\mathcal{O}\left(N^2_\tn{t}/N_\tn{b}\right)$\\
  \hline
  & general & %
  --
  & $\mathcal{O}\left(N_\tn{b}^6 N_\tn{t}^1\right)$ & %
  --
  \\
 DSL & Hubbard & %
 --
 & $\mathcal{O}\left(N_\tn{b}^4 N_\tn{t}^1\right)$ & %
 --
 \\
   & jellium & %
   --
   & $\mathcal{O}\left(N_\tn{b}^4 N_\tn{t}^1\right)$ & %
   --
\end{tabular}
\caption{\label{tab:scaling} Scaling of the CPU time with the number of time steps $N_\tn{t}$ and basis dimension $N_\tn{b}$ of the traditional non-Markovian HF-GKBA and the present time-local scheme (G1--G2), for three relevant basis sets and the selfenergy approximations considered in this paper: the second-Born approximation (SOA), $GW$ approximation ($GW$), the particle--particle (TPP) and particle--hole (TPH) $T$ matrices, and the dynamically-screened-ladder approximation (DSL). Last column: CPU speedup ratio of the G1--G2 scheme compared to standard HF-GKBA. For DSL, currently no standard HF-GKBA version exists. Note that full two-time NEGF simulations always have cubic scaling with $N_\tn{t}$.
}
\end{table}

\subsection{Numerical results for the Hubbard basis}\label{ss:numerik-hubbard}
As we have shown above (cf. Tab.~\ref{tab:scaling}), the Hubbard basis is the most unfavorable case for the G1--G2 scheme. Therefore, we choose this case for numerical demonstrations.
In Ref.~\cite{schluenzen_19_prl} we presented the first numerical tests of this scheme and demonstrated that, for finite Hubbard clusters the predicted linear scaling is indeed, achieved for SOA and $GW$ selfenergies, already for rather small values $N_\tn{t}$. 
\begin{figure}[h]
\includegraphics[]{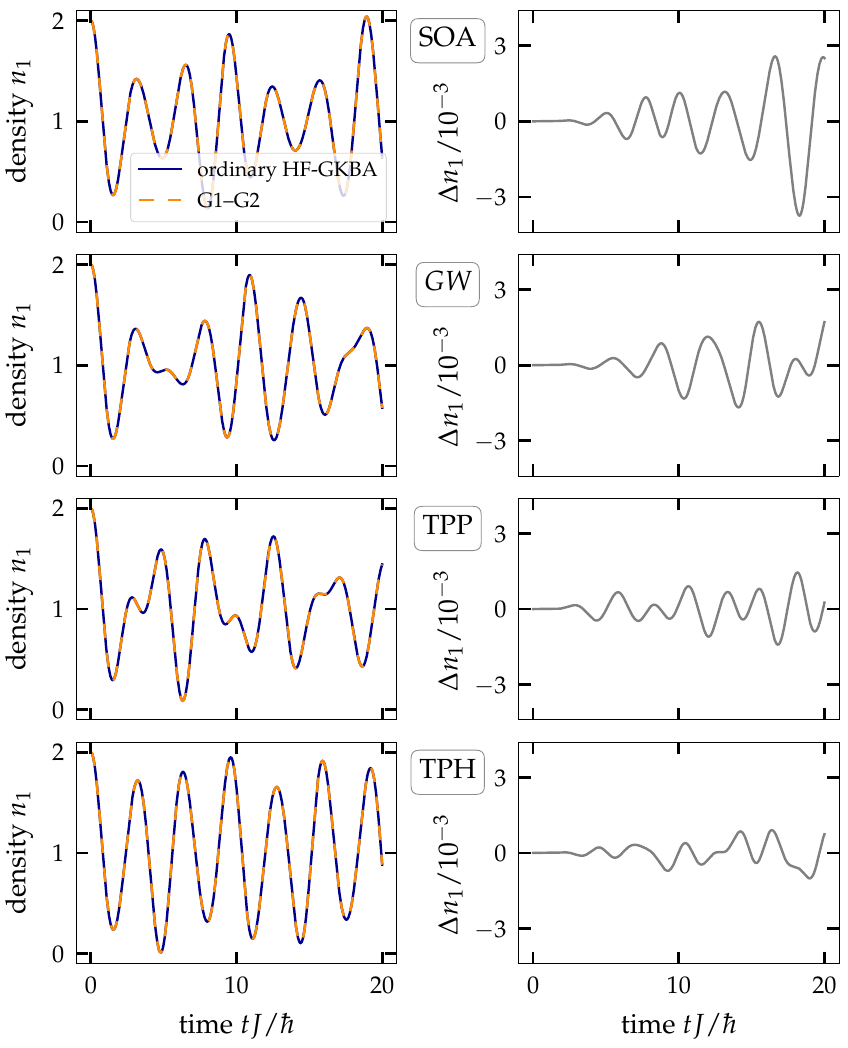}
\caption{Comparison of the ordinary HF-GKBA and the G1--G2 schemes for a Hubbard dimer with $U=J$ at half filling. The initial state was uncorrelated. Rows correspond to SOA, $GW$, TPP and TPH selfenergies. Right column shows the deviation $\Delta n_1 (t)= n_1^\mathrm{G1-G2}(t) - n_1^\mathrm{ordinary}(t)$ of the densities of both schemes on site 1.}
\label{fig:densities}
\end{figure}

Here we extend these simulations to the $T$-matrix selfenergies and the DSL approximation. Furthermore, we explicitly verify the $N_\tn{b}$-scaling. As a first test, we verify that the derived formulas of the G1--G2 scheme are equivalent to the original (non-Markovian) HF-GKBA formulation. As a test case we consider, in Fig.~\ref{fig:densities} the time evolution in a Hubbard dimer for SOA, $GW$, TPP and TPH selfenergies. The agreement is excellent, and the deviations are mostly due to the original HF-GKBA, as discussed in Ref.~\cite{schluenzen_19_prl}.
\begin{figure}[h]
\includegraphics[]{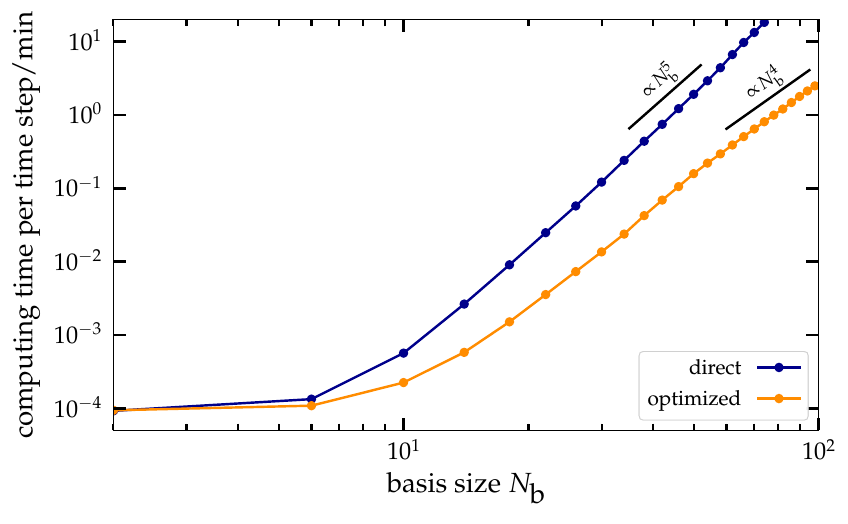}
\caption{CPU time scaling of the SOA-G1--G2 scheme with the basis size $N_\tn{b}$ comparing the direct, \refeqn{eq:phi-hubbard} \cite{schluenzen_19_prl}, and the optimized implementation, \refeqn{eq:soa-cubic-scheme}. Results are for a 1D Hubbard chain. %
}
 \label{fig:nb-scaling}
\end{figure}

Next, we verify the scaling with the basis dimension $N_\tn{b}$ for the SOA selfenergy. In Fig.~\ref{fig:nb-scaling} we show results for a large number of Hubbard chains of varying length, $N_\tn{b}=2\dots 100$.
We clearly confirm the $N^5_\tn{b}$-scaling for the standard implementation of the G1--G2 scheme that uses \refeqn{eq:phi-hubbard} \cite{schluenzen_19_prl}. This asymptotic behavior is reached already for $N_\tn{b}\gtrsim 20$. The second curve is for the same setup but uses the optimization, \refeqn{eq:soa-cubic-scheme}. Again, the predicted improved scaling according to $N^4_\tn{b}$ is clearly identified, at least for $N_\tn{b}\gtrsim 50$. 
This confirms the expected speedup of the SOA-G1--G2 scheme compared to the standard HF-GKBA, if $N_\tn{t}\gtrsim N_\tn{b}$. Thus, even for the most unfavorable case of a Hubbard basis [cf. Tab.~\ref{tab:scaling}] the scaling advantage should be reached already for small simulation durations.
\begin{figure}[h]
\includegraphics[]{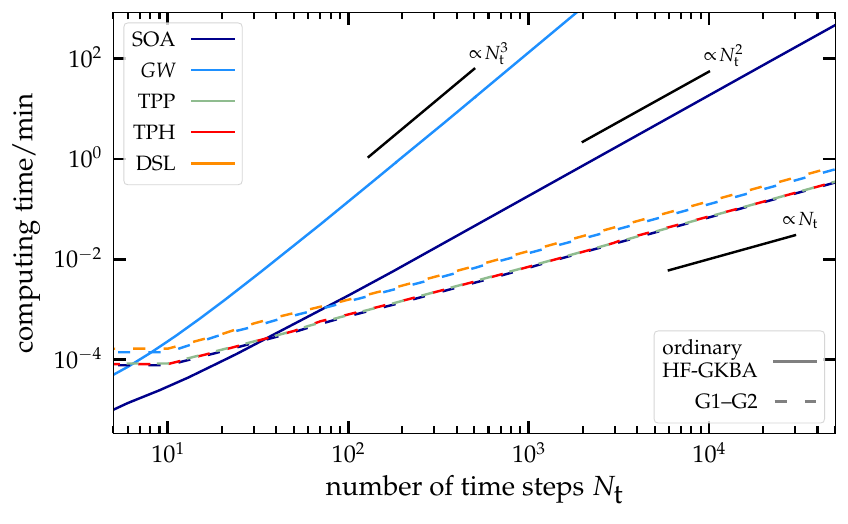}
\caption{CPU time scaling with the simulation duration $N_\tn{t}$, comparing the standard HF-GKBA to the G1--G2 scheme. G1--G2-data are shown for five selfenergy approximations (indicated in the legend) all of which clearly exhibit linear scaling. In contrast, the standard HF-GKBA scales as $N_\tn{t}^2$, for SOA, and $N^3_\tn{t}$, for $GW$. Results are for
 a 10-site Hubbard chain.
 }
 \label{fig:nt-scaling}
\end{figure}

To explore the scaling with $N_\tn{t}$ in more detail we have performed a series of simulations for all selfenergy approximations, comparing the standard HF-GKBA to the G1--G2 scheme. The results are shown in Fig.~\ref{fig:nt-scaling} and confirm the quadratic (cubic) scaling of the CPU time with $N_\tn{t}$, for the standard HF-GKBA with SOA ($GW$) selfenergy. Similar cubic scaling is observed for the two $T$-matrix approximations (not shown) whereas simulations with DSL approximation are not possible, at the moment. Let us now turn to the G1--G2 results (dashed lines). Each of the curves exhibits the predicted linear scaling, already for $N_\tn{t}\gtrsim 20$. Interestingly, in the G1--G2 scheme, the CPU time required for the rather involved $T$-matrix approximations is only slightly above the time required for the comparatively simple SOA case. Equally remarkable is the observation that the $GW$ and DSL approximations, which, in Hubbard, rely on cross-coupling spin components, are rather close to the former selfenergies.

Note that, for the present small system (10-site Hubbard chain) ``break even'' of the G1--G2 scheme is reached for all selfenergies compared to the ordinary SOA-HF-GKBA (dark blue curve) well below $N_\tn{t}=100$ whereas the original $GW$-HF-GKBA (light blue) is unfavorable, practically from the start. For larger times, the ordinary $GW$-HF-GKBA quickly turns out unfeasible (e.g., for $N_\tn{t}\sim 10^3$ it requires $10^4$ times longer simulations than $GW$-G1--G2), and the same applies to the $T$-matrix selfenergies. Thus, we conclude that, it is not just a quantitative gain in CPU time that the G1--G2 scheme delivers but, in many cases, highly accurate simulations (beyond the simple SOA selfenergy) become possible at all that are (currently) impossible otherwise.
\begin{figure}[h]
\includegraphics[]{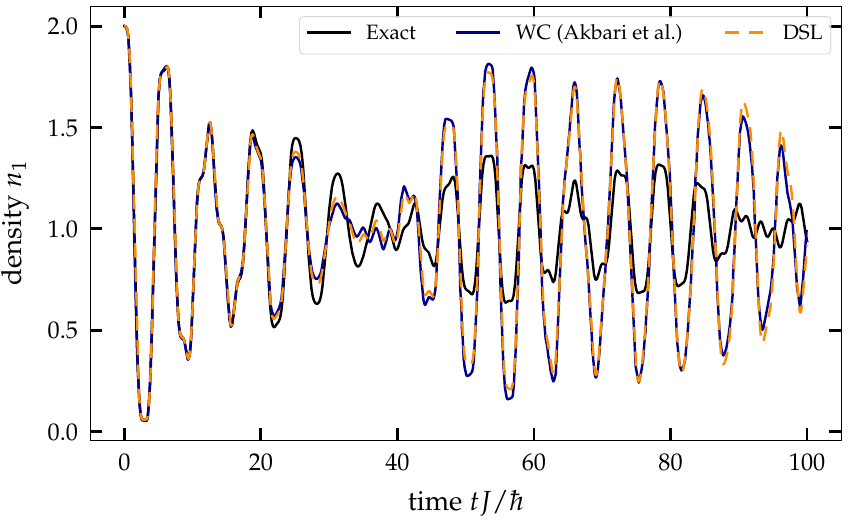}
\caption{Density evolution comparing the DSL-G1--G2 scheme to the results of Ref. \cite{akbari_prb_12} and exact-diagonalization simulations. The system is a four-site Hubbard chain at $U=0.1J$ and half filling; the simulations started from a noninteracting (uncorrelated) initial state.}
 \label{fig:dsl-densities}
\end{figure}

In particular, at increased coupling, $U/J \gtrsim 2$, SOA selfenergies are known to be inaccurate (for an analysis see Ref.~\cite{schluenzen_jpcm_19}) and for reliable simulations, more advanced approximations are crucial. In that context the DSL approximation is particularly attractive because it contains the dominant correlation effects selfconsistently. Until now such simulations have only occasionally been reported, for very small systems and short propagation times. An example of a four-site Hubbard chain is shown in Fig.~\ref{fig:dsl-densities}. We observe excellent agreement of our DSL-G1--G2 scheme to the Wang--Cassing approximation simulations of Akbari \textit{et al.} \cite{akbari_prb_12} confirming the equivalence of the two approximations. The results show excellent quantitative agreement with exact diagonalization data (black curve), however, for times $tJ/\hbar \gtrsim 30$ deviations are growing.

\section{Discussion and Outlook}\label{s:discussion}
In this paper we analyzed the properties of nonequilibrium Green functions in the frame of the generalized Kadanoff--Baym ansatz with Hartree--Fock propagators (HF-GKBA). Due to the non-Markovian structure of the collision integral, HF-GKBA simulations have an unfavorable quadratic (cubic) scaling with the number of time steps, for second-order Born (more complicated) selfenergies. 
At the same time, it has been reported earlier that this memory integral can be formally eliminated in favor of coupled time-local differential equations for the single-particle and two-particle density matrix \cite{bonitz_qkt,hermanns_jpcs13}. An equivalent formulation in the framework of nonequilibrium Green functions has been established in Ref.~\cite{schluenzen_19_prl}---the G1--G2 scheme.
The formal equivalence between both approaches is important because it means that the G1--G2 scheme retains all attractive properties of the HF-GKBA: it is total-energy conserving and time-reversible \cite{bonitz_cpp18}. Furthermore, all selfenergies from NEGF theory that have been derived, e.g. using diagrammatic techniques, can be transformed into a time-local form, by applying the HF-GKBA. 

On the other hand, the former analyses concentrated, e.g., mainly on spatially homogeneous systems (jellium)~\cite{bonitz_qkt} and did not include computational aspects such as the CPU time requirement. The scaling with the propagation time and basis size have only recently been analyzed in detail in conjunction with the G1--G2 scheme \cite{schluenzen_19_prl}, and it was confirmed that the $N_\tn{t}^1$-scaling can be achieved in practice. Here, we substantially extended these results, including additional high-level selfenergies such as the particle--particle and particle--hole $T$-matrix selfenergies and the screened-ladder approximation. In each case $N_\tn{t}^1$-scaling of the CPU time could be confirmed giving rise to a remarkable $N_\tn{t}^2$-scaling advantage compared to the standard HF-GKBA scheme (Fig.~\ref{fig:nt-scaling}) which was found to be independent of the single-particle basis used for the simulations. Furthermore, we re-analyzed the CPU-time scaling with the basis dimension $N_\tn{b}$ and observed that the G1--G2 scheme has an overhead, compared to standard HF-GKBA, that is, \textit{at most}, first order in $N_\tn{b}$, cf. Tab.~\ref{tab:scaling}. Even for the most unfavorable basis---the Hubbard basis---the G1--G2 scheme has only a $N_\tn{b}^1$ overhead (down from a $N_\tn{b}^2$ overhead reported in Ref.~\cite{schluenzen_19_prl}) which could be achieved by a reformulation of the scattering term in the G2--equation, cf. Sec.~\ref{ss:numerics_soa}. Thus, we expect that the G1--G2 scheme outperforms the standard HF-GKBA approach, in all cases of practical relevance, which can be seen from the CPU-time scaling ratio summarized in the right column of Tab.~\ref{tab:scaling}.

With the G1--G2 scheme NEGF simulations (within the HF-GKBA) have been brought to the same 
CPU time scaling as many other time-dependent approaches, including semiclassical molecular dynamics, hydrodynamics, Boltzmann-type kinetic equations, TDDFT (adiabatic approximation), and the time-dependent Schrödinger equation. Most importantly, now long simulations are feasible that were previously prohibited by the memory structure (resulting in the $N^2_\tn{t}$ or $N^3_\tn{t}$ discussed above) without compromising the quality of the treatment of electronic correlations. We also showed that the inclusion of initial correlations in the G1--G2 scheme is trivial, and their propagation again requires a CPU time effort that is of order $N_\tn{t}$. 
Also the precomputation of the correlated initial state, e.g. via imaginary time stepping or adiabatic switching, see, e.g., Ref.~\cite{bonitz_pss_18}, can be carried out separately and does not effect the propagation scaling. 

While we presented numerical results only for the Hubbard model, even larger gains, compared to the standard HF-GKBA, are predicted for jellium (e.g. electron gas, dense quantum plasmas, electron--hole plasmas etc.) and for more general basis sets where the interaction tensor has four indices (e.g. electron dynamics in atoms and molecules). At the same time, the removal of the memory integral as the main CPU time bottleneck was achieved by computing the dynamics of an additional quantity---the time-diagonal two-particle Green function $\mathcal{G}_{ijkl}$. Thus, the new bottleneck in the G1--G2 scheme is the memory cost to store this four-dimensional tensor (only the current values are required), but this can be mitigated by suitable parallelization concepts.

By mapping NEGF simulations to a time-local scheme for single-time quantities, it should be expected that close connections exist with reduced-density-operator theory (RDO) \cite{bonitz_qkt,hermanns_jpcs13,schluenzen_19_prl}. The latter has been an independent many-body approach that has been successfully applied in many areas, including semiconductor optics, e.g. Refs.~\cite{lindberg_prb88,axt_zphysb_94}, dense plasmas \cite{bonitz-etal.96pla}, correlated electrons \cite{hohenester97,akbari_prb_12,lacroix_prb14}, nuclear matter \cite{schuck_epja_16}, and cold atoms \cite{schmelcher}. Our results indicate the correspondence between important selfenergy approximations of NEGF theory to closure relations of RDO and confirm  and extend earlier results on the particle--particle $T$ matrix \cite{kremp-etal.97ap} and the $GW$ approximation \cite{hohenester97}. We also investigated the 
simultaneous treatment  of strong coupling and dynamical screening effects by combining ladder and polarization terms in the equation for $\mathcal{G}$. This lead us to the dynamically-screened-ladder approximation (DSL), in Sec.~\ref{s:dyn-scr-ladder}. This approximation includes all two-particle interaction contributions and is, thus, equivalent to an approximation considered by Wang and Cassing before \cite{wang-cassing-85}. The equivalence of the two approximations was confirmed by the excellent agreement with the numerical results of Akbari \textit{et al.}~\cite{akbari_prb_12} for a small Hubbard cluster, cf. Fig.~\ref{fig:dsl-densities}.

Despite the high quality of the DSL, we also observed that it is in quantitative agreement with exact diagonalization (CI) data (black curve in Fig.~\ref{fig:dsl-densities}) only during the initial relaxation phase (for times $tJ/\hbar \lesssim 30$) \cite{akbari_prb_12}. So, clearly more systematic comparisons to CI results, for a broader range of coupling strengths and filling fractions, are desirable to understand the applicability limits of the DSL. While CI simulations are limited to very small particle numbers (basis size $N_\tn{b}$) the G1--G2 scheme in DSL and simpler approximations can treat much larger systems.
To go beyond those parameters where the DSL approximation is valid, further improved approximations are in high demand. This will require to partially include three-particle correlations. Examples are the Kirkwood superposition approximation of classical statistical physics \cite{kirkwood_35} (for recent applications see  Refs.~\cite{singer_doi:10.1063/1.1776552,fortov_kirwood_08}), the approximation by Nakatsuji and Yasuda 
\cite{nakatsuji_prl_96,lackner_pra_17}, and
selfenergy corrections to the BBGKY hierarchy
\cite{bonitz_qkt}.
Another route to improvements starts from nonequilibrium Green functions theory where one approach is to apply the GKBA but replace the Hartree--Fock propagators by correlated propagators \cite{bonitz_pss_18}. Another concept is to replace the GKBA entirely, by an improved reconstruction ansatz. In both cases, the procedure outlined in the present paper will allow one to derive the corresponding improved G1--G2 scheme.
Since the applicability limits of the GKBA are still not fully explored, full two-time NEGF simulations will remain indispensible for tests and benchmarks, see, e.g., Ref.~\cite{kwong-etal.98pss}.

In conclusion, let us come back to the remarkable capability of the G1--G2 scheme to efficiently perform long-time simulations of correlated electron dynamics. With this it should be feasible to reach thermodynamic equilibrium (or a quasi-stationary or pre-thermalized state) of the electrons. At the same time, slower processes, such as the equilibration with heavier particles (e.g. with the lattice in solids or with ions in dense plasmas) will make it desirable to 
develop a multiscale approach. This can be based on approximate solutions of the G1--G2 equations, e.g. by using retardation expansions \cite{bonitz_qkt} or the correlation-time approximation \cite{bonitz96pla}, eventually approaching the Markovian Boltzmann equation or local thermodynamic equilibrium. In that case a connection of the kinetic simulations to quantum hydrodynamic models, see, e.g., Refs.~\cite{zhandos_pop18,bonitz_pop_19}, could be a promising approach.

\section*{Acknowledgements}
We thank K. Balzer and C. Makait for valuable comments. This work was supported 
by grant shp00015 for CPU time at 
the Norddeutscher Verbund f\"ur Hoch- und H\"ochstleistungsrechnen (HLRN).

J.J. and N.S. contributed equally to this work.

\appendix

\section{Properties of the time-evolution operator}\label{app:Uprop}
In the following, we derive important properties of the one- and two-particle propagators.
\subsection{Symmetry relations}\label{app:Usymm}
The single-particle time-evolution operator $\mathcal{U}$ fulfills the symmetry
\begin{align}
 \Big[\mathcal{U}_{ji}(t',t)\Big]^* &= \Big[G_{ji}^\mathrm{R}(t',t) - G_{ji}^\mathrm{A}(t',t)\Big]^*\nonumber\\
 &= -\mathcal{U}_{ij}(t,t') \, , \label{eq:ret_adv_sym}
\end{align}
where $\left[G_{ji}^\mathrm{A/R}(t,t') \right]^* = G_{ij}^\mathrm{R/A}(t',t)$ has been used.
Likewise, the two-particle propagator obeys,
\begin{align}
 \left[\mathcal{U}_{klij}^{(2)}(t,t')\right]^* &=  \Big[\mathcal{U}_{ki}(t,t') \Big]^* \Big[\mathcal{U}_{lj}(t,t')\Big]^*\nonumber \\
 &= \mathcal{U}_{ijkl}^{(2)}(t',t) \, ,
 \label{eq:ret_adv_sym2}
\end{align}
where \refeqn{eq:ret_adv_sym} has been used.

\subsection{Group property}\label{app:Ugroup}
Utilizing \refeqns{group_ret}{group_adv}, now the group property for the propagator $\mathcal{U}$ is derived for all relevant time orderings. Starting with
\begin{align}
& \i \hbar \sum_k  \mathcal{U}_{ik}(t,\cbar{t}) \mathcal{U}_{kj}(\cbar{t},t') \\ \nonumber
&= \i \hbar \sum_k \left[G_{ik}^\mathrm{R}(t,\cbar{t}) - G_{ik}^\mathrm{A}(t,\cbar{t})\right] \left[G_{kj}^\mathrm{R}(\cbar{t},t') - G_{kj}^\mathrm{A}(\cbar{t},t')\right]\, ,
\end{align}
five different cases have to be considered. For $t=\cbar{t}=t'$ one gets
\begin{align}
\sum_k \mathcal{U}_{ik}(t,t) \mathcal{U}_{kj}(t,t) 
    = 
    \sum_k  \frac{\delta_{ik}\delta_{kj}}{\left(\i\hbar\right)^2}  = \frac{\delta_{ij} }{(\i\hbar)^2} 
    = \frac{1}{\i\hbar}\mathcal{U}_{ij}(t,t) \,.
\nonumber
\end{align}
For $t=\cbar{t}$ one gets
\begin{align}
    \sum_k \mathcal{U}_{ik}(t,t) \mathcal{U}_{kj}(t,t') = 
    \sum_k  \frac{1}{\i\hbar} \delta_{ik}\mathcal{U}_{kj}(t,t') =\frac{1}{\i\hbar} \mathcal{U}_{ij}(t,t')\,,\nonumber
\end{align}
as well as for $\cbar{t}=t'$,
\begin{align}
    \sum_k \mathcal{U}_{ik}(t,t') \mathcal{U}_{kj}(t',t') &= 
    \sum_k  \mathcal{U}_{ik}(t,t') \frac{1}{\i\hbar} \delta_{kj} \nonumber %
    = \frac{1}{\i\hbar}\mathcal{U}_{ij}(t,t')\,.
\end{align}
For $t>\cbar{t}>t'$, the propagators reduce to $\mathcal{U}_{ij}(t,t') = G^\tn{R}_{ij}(t,t')$, for which \refeqn{group_ret} is directly applicable. For the analogous case, $t<\cbar{t}<t'$, one obtains $\mathcal{U}_{ij}(t,t') = -G^\tn{A}_{ij}(t,t')$ which, together with \refeqn{group_adv}, leads to,
\begin{align}
    \i \hbar \sum_k \mathcal{U}_{ik}(t,\cbar{t}) \mathcal{U}_{kj}(\cbar{t},t') =  \mathcal{U}_{ij}(t,t')\,,
\nonumber
\end{align}
for all $t,t'$.
A direct consequence of this group property is [cf. \refeqn{prop_two}],
\begin{align}
\nonumber
 \mathcal{U}^{(2)}_{ijkl}(t,t') = \left(\i\hbar\right)^2 \sum_{pq} \mathcal{U}^{(2)}_{ijpq}(t,\cbar{t}) \mathcal{U}^{(2)}_{pqkl}(\cbar{t},t')\,,
\end{align}
for the two-particle propagator.

\subsection{Equations of motion}\label{app:Ueom}
Using the EOM for the retarded/advanced Green functions, \refeqn{eq:eom_twotime}, the EOMs for the modified propagator immediately follows, where we separately consider the time evolution along the first and second time arguments and along the time diagonal:
\begin{align}
\i\hbar \frac{\d}{\d t} \mathcal{U}_{ij}(t,t') &=  \sum_k h^\tn{HF}_{ik}(t) G^\tn{R}_{kj} (t,t') + \delta_{ij} \delta(t,t') 
\nonumber\\ \nonumber
& \quad - \sum_k h^\tn{HF}_{ik}(t) G^\tn{A}_{kj} (t,t') - \delta_{ij} \delta(t,t')  \nonumber \\
&= \sum_k h^\tn{HF}_{ik}(t) \mathcal{U}_{kj} (t,t') \, , \label{eq:eom-ua} \\
\i\hbar \frac{\d}{\d t} \mathcal{U}_{ij}(t',t) &= - \sum_k G^\tn{R}_{ik} (t',t) h^\tn{HF}_{kj}(t) - \delta_{ij} \delta(t,t') 
\nonumber\\ \nonumber 
& \quad + \sum_k G^\tn{A}_{ik} (t',t) h^\tn{HF}_{kj}(t) + \delta_{ij} \delta(t,t')  \nonumber \\
&= - \sum_k \mathcal{U}_{ik} (t',t) h^\tn{HF}_{kj}(t) \, , \label{eq:eom-ub}\\
\i\hbar \frac{\d}{\d t} \mathcal{U}_{ij}(t=t)
&= \left[h^\tn{HF}(t), G^\tn{R} (t,t)\right]_{ij} - \left[h^\tn{HF}(t), G^\tn{A} (t,t)\right]_{ij} \nonumber \\ 
&= \left[h^\tn{HF}(t), \mathcal{U} (t,t)\right]_{ij} \, .
\end{align}
Obviously, $\mathcal{U}$ has no time-singular term, but obeys a Schr\"odinger-type equation of motion.
For the two-particle propagator follows,
\begin{align}
 \frac{\d}{\d t} \left[ \mathcal{U}^{(2)}_{ijkl} (t,\cbar t) \right] &= 
 \frac{\d}{\d t} \left[ \mathcal{U}_{ik} (t,\cbar t) \right] \mathcal{U}_{jl} (t,\cbar t) \\ 
 & \quad + \mathcal{U}_{ik} (t,\cbar t)  \frac{\d}{\d t} \left[ \mathcal{U}_{jl} (t,\cbar t) \right] 
 \\
 &= \left[ \frac{1}{\i\hbar} \sum_p h^\tn{HF}_{ip} (t) \mathcal{U}_{pk} (t,\cbar t) \right] \mathcal{U}_{jl}(t,\cbar t) 
 \nonumber
 \\
 & \quad + \mathcal{U}_{ik} (t,\cbar t) \left[  \frac{1}{\i\hbar} \sum_p h^\tn{HF}_{jp} (t) \mathcal{U}_{pl} (t,\cbar t) \right]\, \nonumber\\
 &=  \frac{1}{\i\hbar} \sum_p h^\tn{HF}_{ip} (t) \mathcal{U}^{(2)}_{pjkl} (t,\cbar t) \\ 
 & \quad + \frac{1}{\i\hbar} \sum_p h^\tn{HF}_{jp} (t) \mathcal{U}^{(2)}_{ipkl} (t,\cbar t) \, ,  \label{prod_ret_pre} 
\end{align}
To simplify the notation, we use the two-particle Hartree--Fock Hamiltonian [cf. \refeqn{twopart_hamiltonian}] 
so that 
\begin{align}
 \sum_{pq} h^{(2),\tn{HF}}_{ijpq}(t) \mathcal{U}^{(2)}_{pqkl} =& \sum_p h^\tn{HF}_{ip} (t) \mathcal{U}^{(2)}_{pjkl} (t,\cbar t)\\ \nonumber &+ \sum_p h^\tn{HF}_{jp} (t) \mathcal{U}^{(2)}_{ipkl} (t,\cbar t)\,,
\end{align}
and \refeqn{prod_ret_pre} can be rewritten as,
\begin{align}
 \frac{\d}{\d t} \left[ \mathcal{U}^{(2)}_{ijkl} (t,\cbar t) \right] \label{prod_ret_app}
 &= \frac{1}{\i\hbar} \sum_{pq} h^{(2),\tn{HF}}_{ijpq} (t) \mathcal{U}^{(2)}_{pqkl} (t,\cbar t)\, .
\end{align}
In the same way the derivative with respect to the second time argument is found,
\begin{align}
 \frac{\d}{\d t} \left[  \mathcal{U}^{(2)}_{ijkl} (\cbar t,t) \right] &= 
 \frac{\d}{\d t} \left[ \mathcal{U}_{ik} (\cbar t,t) \right] \mathcal{U}_{jl} (\cbar t,t) 
 \nonumber\\ \nonumber & \quad + \mathcal{U}_{ik} (\cbar t,t)  \frac{\d}{\d t} \left[ \mathcal{U}_{jl} (\cbar t,t) \right] \\
 &= \left[ -\frac{1}{\i\hbar} \sum_p \mathcal{U}_{ip} (\cbar t,t) h^\tn{HF}_{pk} (t) \right] \mathcal{U}_{jl}(\cbar t,t) \nonumber\\
 & \quad+ \mathcal{U}_{ik} (\cbar t,t) \left[ -\frac{1}{\i\hbar} \sum_p  \mathcal{U}_{jp} (\cbar t,t) h^\tn{HF}_{pl} (t)\right]\, \nonumber\\
  &= -\frac{1}{\i\hbar} \sum_{pq} \mathcal{U}^{(2)}_{ijpq}(\cbar t,t) h^{(2),\tn{HF}}_{pqkl} (t)
\nonumber
\end{align}

\section{Particle--hole $T$ matrix}\label{app:tph}
For the $T$ matrix in the particle--hole channel \cite{schluenzen_jpcm_19}, the  derivation of the G1--G2 scheme is performed in similar fashion as for the particle--particle $T$ matrix in Sec.~\ref{ss:tpp}.
The selfenergy has the form,
\begin{align}
 \Sigma_{ij}^\gtrless(t,t') = \i \hbar \sum_{kl} T^{\tn{ph},\gtrless}_{ikjl}(t,t') G^\gtrless_{lk}(t,t')\, ,
 \label{eq:sigma-tph}
\end{align}
where now the particle--hole $T$ matrix is expressed as
\begin{align}
 T_{ijkl}^{\tn{ph},\gtrless}(t,t') = \sum_{pq} w_{ipql}(t) \Omega^{{\tn{ph},\gtrless}}_{qjkp}(t,t')\, ,
 \label{eq:tmatrixph}
\end{align}
which allows us to rewrite the selfenergy \eqref{eq:sigma-tph}:
\begin{align}
 \Sigma_{ij}^\gtrless(t,t') = \i \hbar \sum_{klpq} w_{ipql}(t) \Omega^{\tn{ph},\gtrless}_{qkjp}(t,t') G^\gtrless_{lk}(t,t')\, .
 \label{eq:sigma-tph2}
\end{align}
In \refeqns{eq:tmatrixph}{eq:sigma-tph2},  $\Omega^\tn{ph}$ denotes the nonequilibrium generalization of the M\o ller operator in the particle--hole channel. %
The collision integral \eqref{eq:definition_D} of the time-diagonal equation then becomes,
\begin{align}
 I_{ij} (t) 
&= \i\hbar \sum_{klpqr} w_{ipqr}(t) \int_{t_0}^t \d \cbar t \, \Big[ \Omega^{\tn{ph},>}_{qlkp}(t,\cbar t) \mathcal{G}^{\tn{F},<}_{krlj}(\cbar t,t) \nonumber \\ 
& \qquad \qquad \qquad \qquad \qquad - \Omega^{\tn{ph},<}_{qlkp}(t,\cbar t) \mathcal{G}^{\tn{F},>}_{krlj}(\cbar t,t)\Big] \nonumber \\\nonumber
&= \pm \i \hbar \sum_{klp} w_{iklp}(t) \mathcal{G}_{lpjk} (t) \, ,
 \end{align}
which results in the following expression for the time-diagonal element of the two-particle Green function,
\begin{align}
 \mathcal{G}_{ijkl} (t)=& \pm \sum_{pq} \int_{t_0}^t \d \cbar t \, \Big[ \Omega^{\tn{ph},>}_{iqpl}(t,\cbar t) \mathcal{G}^{\tn{F},<}_{pjqk}(\cbar t,t) \nonumber \\
 & \qquad \qquad - \Omega^{\tn{ph},<}_{iqpl}(t,\cbar t) \mathcal{G}^{\tn{F},>}_{pjqk}(\cbar t,t)\Big] \, .
\label{eq:g2-tph}
 \end{align} 
 By construction, the particle--hole $T$ matrix obeys the following symmetry [cf. \refeqn{eq:wsymm_a}],
 \begin{align}
  T^{\tn{ph},\gtrless}_{ijkl}(t,t') &= T^{\tn{ph},\lessgtr}_{jilk}(t',t)\, .
 \end{align}
The particle--hole $T$ matrix sums up the particle--hole collisions via the recursive equation (again the singular part has been subtracted compared to its standard definition~\cite{schluenzen_jpcm_19})
 \begin{widetext}
 \begin{align}
     T_{ijkl}^{\tn{ph},\gtrless}&(t,t') = \pm \i \hbar \sum_{pqrs} w_{iqpl}(t) G^{\tn{F},\gtrless}_{psqr}(t,t') w^\pm_{rjks}(t') \\
     & + \i\hbar \sum_{pqrs} w_{iqpl}(t) \bigg[ \int_{t_0}^t \d \cbar t\, \Big( G^{\tn{F},>}_{psqr} (t,\cbar t) - G^{\tn{F},<}_{psqr} (t,\cbar t) \Big) T^{\tn{ph},\gtrless}_{rjks}(\cbar t, t') +\int_{t_0}^{t'} \d \cbar t \, G^{\tn{F},\gtrless}_{psqr} (t,\cbar t) \Big( T^{\tn{ph},<}_{rjks}(\cbar t, t') - T^{\tn{ph},>}_{rjks}(\cbar t, t')\Big)\bigg] \, ,\nonumber
 \end{align}
 whereas the M\o ller operator obeys
  \begin{align}
     &\Omega_{ijkl}^{\tn{ph},\gtrless}(t,t') = \pm \i \hbar \sum_{pq} G^{\tn{F},\gtrless}_{iplq}(t,t') w^\pm_{qjkp}(t') \\
     & + \i\hbar \sum_{pqrs} \bigg[ \int_{t_0}^t \d \cbar t\, \Big( G^{\tn{F},>}_{iplq} (t,\cbar t) - G^{\tn{F},<}_{iplq} (t,\cbar t) \Big) w_{qrsp}(\cbar t) \Omega^{\tn{ph},\gtrless}_{sjkr}(\cbar t, t') +\int_{t_0}^{t'} \d \cbar t \, G^{\tn{F},\gtrless}_{iplq} (t,\cbar t) w_{qrsp}(\cbar t) \Big( \Omega^{\tn{ph},<}_{sjkr}(\cbar t, t') - \Omega^{\tn{ph},>}_{sjkr}(\cbar t, t')\Big)\bigg]\nonumber\\
     &\Omega_{ijkl}^{\tn{ph},\gtrless}(t,t') = \pm \i \hbar \sum_{pq} G^{\tn{F},\gtrless}_{iplq}(t,t') w^\pm_{qjkp}(t') \nonumber \\
     & + \i\hbar \sum_{pqrs} \bigg[ \int_{t_0}^t \d \cbar t\, \Big( G^{\tn{F},<}_{piql} (\cbar t,t) - G^{\tn{F},>}_{piql} (\cbar t,t) \Big) \Omega^{\tn{ph},\lessgtr}_{rqps}(t', \cbar t)  +\int_{t_0}^{t'} \d \cbar t \, G^{\tn{F},\lessgtr}_{piql} (\cbar t,t) \Big( \Omega^{\tn{ph},>}_{rqps}(t', \cbar t) - \Omega^{\tn{ph},<}_{rqps}(t', \cbar t)\Big) \bigg] w_{sjkr}(t')\, .\nonumber
 \end{align}
 The time-diagonal equation for $\Omega^{\tn{ph}}$ can be further simplified,
   \begin{align}
  \Omega^{\tn{ph},\gtrless}_{ijkl}(t,t) 
  =& \pm \i \hbar \sum_{pq} G^{\tn{F},\gtrless}_{iplq}(t) w^\pm_{qjkp}(t) + \i\hbar \sum_{pqrs} \int_{t_0}^t \d \cbar t\, \Big( \mathcal{G}^{\tn{F},<}_{piql}(\cbar t,t) \Omega^{\tn{ph},>}_{rqps}(t,\cbar t)   - \mathcal{G}^{\tn{F},>}_{piql}(\cbar t,t) \Omega^{\tn{ph},<}_{rqps}(t,\cbar t) \Big) w_{sjkr} (t) \nonumber \\
  =& \pm \i \hbar \sum_{pq} G^{\tn{F},\gtrless}_{iplq}(t) w^\pm_{qjkp}(t) \pm \i \hbar \sum_{pq} \mathcal{G}_{ipql}(t) w_{qjkp}(t)\, .
  \end{align}
 \end{widetext}
 
\subsection{$T^\tn{ph}$ approximation within the HF-GKBA}
Applying the HF-GKBA 
to \refeqn{eq:g2-tph} yields
 \begin{align}
 \mathcal{G}^\GKBA_{ijkl}& (t)= \pm \left(\i \hbar\right)^2 \sum_{pqrs} \int_{t_0}^t \d \cbar t \, \mathcal{U}_{jr}(t,\cbar t) \mathcal{U}_{sk}(\cbar t, t) \times \nonumber \\ \times& \Big[ \Omega^{\tn{ph},>}_{iqpl}(t,\cbar t) %
 \mathcal{G}^{\tn{F},<}_{prqs}(\cbar t) -\Omega^{\tn{ph},<}_{iqpl}(t,\cbar t) %
 \mathcal{G}^{\tn{F},>}_{prqs}(\cbar t)\Big] \, , \nonumber
 \end{align} 
and, for 
 the M\o ller operator,
   \begin{align}\nonumber
     \Omega_{ijkl}^{\tn{ph},\gtrless}&(t\ge t') \\
     =& \pm \left(\i \hbar\right)^3 \sum_{pqrs} \mathcal{U}_{ir}(t,t')  G^{\tn{F},\gtrless}_{rpsq}(t') \mathcal{U}_{sl}(t',t) w^\pm_{qjkp}(t') \nonumber \\
     & + \left(\i \hbar\right)^3 \sum_{pqrsuv} \bigg[ \int_{t_0}^t \d \cbar t\, w_{qrsp}(\cbar t) \mathcal{U}_{iu}(t,\cbar t) \times \nonumber \\
     & \quad \times \Big( G^{\tn{F},>}_{upvq} (t,\cbar t) - G^{\tn{F},<}_{upvq} (t,\cbar t) \Big) \mathcal{U}_{vl}(\cbar t,t) \Omega^{\tn{ph},\gtrless}_{sjkr}(\cbar t, t') \nonumber \\
     & \quad + \int_{t_0}^{t'} \d \cbar t \, \mathcal{U}_{iu}(t,\cbar t) G^{\tn{F},\gtrless}_{upvq} (t,\cbar t) \mathcal{U}_{vl}(\cbar t,t)  w_{qrsp}(\cbar t) \times\nonumber \\ & \quad \times \Big( \Omega^{\tn{ph},<}_{sjkr}(\cbar t, t') - \Omega^{\tn{ph},>}_{sjkr}(\cbar t, t')\Big)\bigg]\, ,
     \label{eqap:omph}
 \end{align}
 where $\mathcal{U}$ is given by \refeqns{eq:eom-ua}{eq:eom-ub}.
With \refeqn{eqap:omph} we obtain the time derivative,
\begin{align}
 &\frac{\d}{\d t} \Omega^{\tn{ph},\gtrless}_{ijkl}(t\ge t') \\ 
 &= \frac{1}{\i\hbar} \sum_p
 \left\{ h^\textnormal{HF}_{ip}(t) \Omega^{\tn{ph},\gtrless}_{pjkl}(t\ge t') 
 -\Omega^{\tn{ph},\gtrless}_{ijkp}(t\ge t') h^\textnormal{HF}_{pl}(t) \right\}
 \nonumber \\
 &\quad \pm \i\hbar \sum_{pqrs} \Big[ \mathcal{G}^{\tn{F},>}_{iplq}(t) - \mathcal{G}^{\tn{F},<}_{iplq}(t) \Big] w_{qrsp}(t)  \Omega^{\tn{ph},\gtrless}_{sjkr} (t \ge t')\nonumber\\
 &= \frac{1}{\i\hbar} \sum_{pq} \Big[ \mathfrak{h}^{\Omega^\tn{ph}, \tn{HF}}_{ipql}(t) + \mathfrak{h}^{\Omega^\tn{ph}, \tn{corr}}_{ipql}(t) \Big] \Omega^{\tn{ph},\gtrless}_{qjkp}(t\ge t')
 \, , \nonumber
\end{align}
where we introduced the Hamiltonians
\begin{align}
 \mathfrak{h}^{\Omega^\tn{ph}, \tn{HF}}_{ijkl}(t) &= \delta_{jl}h^\tn{HF}_{ik}-\delta_{ik}h^\tn{HF}_{jl}\, , \nonumber\\
  \mathfrak{h}^{\Omega^\tn{ph}, \tn{corr}}_{ijkl}(t) &= \left(\i \hbar\right)^2\sum_{pq} \Big[ \mathcal{G}^{\tn{F},>}_{iplq}(t) - \mathcal{G}^{\tn{F},<}_{iplq}(t) \Big] w_{qjkp}(t)\, , %
  \nonumber
\end{align}
that can be combined to
\begin{align}
\mathfrak{h}^{\Omega^\tn{ph}}_{ijkl}(t)
  &=\mathfrak{h}^{\Omega^\tn{ph}, \tn{HF}}_{ijkl}(t) + \mathfrak{h}^{\Omega^\tn{ph}, \tn{corr}}_{ijkl}(t)\,,
   \nonumber
\end{align}
and the M\o ller operator obeys a  Schrödinger equation,
\begin{align}
 &\i\hbar\frac{\d}{\d t} \Omega^{\tn{ph},\gtrless}_{ijkl}(t\ge t')
 =  \sum_{pq}  \mathfrak{h}^{\Omega^\tn{ph}}_{ipql}(t)  \Omega^{\tn{ph},\gtrless}_{qjkp}(t\ge t')
 \, . \label{eq:omph-tdse2}
\end{align}
\subsection{$T^\tn{ph}$-G1--G2 equations for a general basis}
  Next, we compute the time derivative of $\mathcal{G}$,
  \begin{align}\label{eq:eom_parts_ph}
   &\frac{\d}{\d t} \mathcal{G}_{ijkl} (t) \\ &= \left[\frac{\d}{\d t} \mathcal{G}_{ijkl} (t)\right]_{\int^{}_{}} + \left[\frac{\d}{\d t} \mathcal{G}_{ijkl} (t)\right]_{\Omega^\tn{ph}} + \left[\frac{\d}{\d t} \mathcal{G}_{ijkl} (t)\right]_{\mathcal{U}}\nonumber\,,
   \end{align}
   and obtain for the first part,
    \begin{align}\label{eq:eom_parts_int_ph}
   &\left[\frac{\d}{\d t} \mathcal{G}_{ijkl} (t)\right]_{\int^{}_{}} \\
   &\quad=\pm\sum_{pq} \Big[ \Omega^{\tn{ph},>}_{iqpl}(t, t) \mathcal{G}^{\tn{F},<}_{pjqk}(t, t) - \Omega^{\tn{ph},<}_{iqpl}(t, t) \mathcal{G}^{\tn{F},>}_{pjqk}(t, t)\Big]  \nonumber\\
   &\quad=\i\hbar \sum_{pqrs} w^\pm_{rqps}(t) \Big[ \mathcal{G}^{\tn{F},>}_{islr}(t) \mathcal{G}^{\tn{F},<}_{pjqk}(t) - \mathcal{G}^{\tn{F},<}_{islr}(t) \mathcal{G}^{\tn{F},>}_{pjqk}(t)\Big]\nonumber\\
   &\qquad + \i\hbar \sum_{pqrs} \mathcal{G}_{isrl}(t) w_{rqps}(t) \Big[ \mathcal{G}^{\tn{F},<}_{pjqk}(t) - \mathcal{G}^{\tn{F},>}_{pjqk}(t)\Big]\nonumber\\
   &\quad = \frac{1}{\i\hbar} \Psi^\pm_{ijkl}(t)
  - \frac{1}{\i\hbar} \sum_{pq} \Big[\mathfrak{h}^{\Omega^\tn{ph}, \tn{corr}}_{kqpj}(t)\Big]^* \mathcal{G}_{iqpl}(t)\,,\nonumber
   \end{align}
and, for the second part,
    \begin{align}\label{eq:eom_parts_omega_ph}
   &\left[\frac{\d}{\d t} \mathcal{G}_{ijkl} (t)\right]_{\Omega^\tn{ph}} \\
   &\quad=\pm \left(\i \hbar\right)^2 \sum_{pqrs} \int_{t_0}^t \d \cbar t \, \mathcal{U}_{jr}(t,\cbar t) \Bigg[ \left( \frac{\d}{\d t} \Omega^{\tn{ph},>}_{iqpl}(t,\cbar t) \right) \mathcal{G}^{\tn{F},<}_{prqs}(\cbar t) \nonumber \\ &\, \qquad
   - \left( \frac{\d}{\d t} \Omega^{\tn{ph},<}_{iqpl}(t,\cbar t) \right) \mathcal{G}^{\tn{F},>}_{prqs}(\cbar t) \Bigg] \mathcal{U}_{sk}(\cbar t, t) \nonumber \\
   &\quad = \frac{1}{\i\hbar} \sum_{pq} \left[\mathfrak{h}^{\Omega^\tn{ph}, \tn{HF}}_{ipql}(t) + \mathfrak{h}^{\Omega^\tn{ph}, \tn{corr}}_{ipql}(t)\right] \mathcal{G}_{qjkp}(t)\nonumber \,,
   \end{align}
and, for the third part,   
  \begin{align}\label{eq:eom_parts_U_ph}
   &\left[\frac{\d}{\d t} \mathcal{G}_{ijkl} (t)\right]_{\mathcal{U}}\\
   &\quad=\pm \left(\i \hbar\right)^2 \sum_{pqrs} \int_{t_0}^t \d \cbar t \, \left( \frac{\d}{\d t}\mathcal{U}_{jr}(t,\cbar t)\right) \Bigg[ \Omega^{\tn{ph},>}_{iqpl}(t,\cbar t)  \mathcal{G}^{\tn{F},<}_{prqs}(\cbar t) \nonumber \\ &\, \qquad
   - \Omega^{\tn{ph},<}_{iqpl}(t,\cbar t) \mathcal{G}^{\tn{F},>}_{prqs}(\cbar t) \Bigg] \mathcal{U}_{sk}(\cbar t, t) \nonumber \\
    &\qquad \pm \left(\i \hbar\right)^2 \sum_{pqrs} \int_{t_0}^t \d \cbar t \, \mathcal{U}_{jr}(t,\cbar t) \Bigg[ \Omega^{\tn{ph},>}_{iqpl}(t,\cbar t)  \mathcal{G}^{\tn{F},<}_{prqs}(\cbar t) \nonumber \\ &\, \qquad
   - \Omega^{\tn{ph},<}_{iqpl}(t,\cbar t) \mathcal{G}^{\tn{F},>}_{prqs}(\cbar t) \Bigg] \left( \frac{\d}{\d t} \mathcal{U}_{sk}(\cbar t, t) \right)  \nonumber \\
   &\quad= \frac{1}{\i\hbar} \sum_{pq} \mathcal{G}_{ipql}(t) \mathfrak{h}^{\Omega^\tn{ph}, \tn{HF}}_{jqpk}(t) \nonumber \,, 
   \end{align}
Combining the three contributions yields the derivative,
\begin{align}
 \i\hbar&\frac{\d}{\d t} \mathcal{G}_{ijkl} (t) =  \Psi^\pm_{ijkl}(t)
\nonumber \\
  & + \sum_{kl} \bigg\{ \mathfrak{h}^{\Omega^\tn{ph}}_{ipql}(t) \Big[ \mathcal{G}_{kpqj}(t) \Big]^* - \mathcal{G}_{ipql}(t) \Big[\mathfrak{h}^{\Omega^\tn{ph}}_{kpqj}(t)\Big]^* \bigg\} \, ,
  \nonumber
\end{align}
which is the result presented in the main part of the paper.

\section{Integral solution $\mathcal{G}(t)$ and initial correlations for higher-order selfenergies} \label{app:IC}
While initial correlations are trivially added to the differential G1--G2 scheme as initial condition, as we demonstrated in Sec.~\ref{ss:inicor}, for the integral representation of $\mathcal{G}$, this problem is more involved. We, therefore, outline, in this appendix, the solution for higher-order selfenergies by extending our SOA result, \refeqn{eq:g2-ic_compact}.
Since the derivations are carried out analogously to Sec.~\ref{ss:inicor} and Sec.~\ref{ss:integral_sol}, respectively, we only give the resulting equations.
Performing the time derivative of the integral expressions recovers the differential equations for the respective selfenergy, cf. \eqrefss{eq:g2-eq-gw}{eq:g2-eq-tpp}{eq:g2-eq-tph}.
\subsection{$GW$ Selfenergy}
In the case of the $GW$ selfenergy \refeqn{eq:g2-ic_compact} becomes
\begin{align}\label{eq:Gtwo-epsilon}
 &\mathcal{G}^\GKBA_{ijkl}(t) = \left(\i\hbar\right)^4 \sum_{pqrs}\int_{t_0}^t \mathrm{d}\cbar{t}\, \mathcal{U}_{lqjs}^{(2),\varepsilon}(t,\cbar{t}) \times
 \\ & \quad\nonumber \times
   \Big[ \delta(t_0,\cbar{t}) \mathcal{G}^0_{pqrs} + 
  \frac{1}{\i\hbar}\Psi_{pqrs}(\cbar{t}) \Big]
  \left[\mathcal{U}_{irkp}^{(2),\varepsilon}(t,\cbar{t})\right]^*\,,
\end{align}
where
\begin{align}
\label{eq:U_prop_gw}
 \mathcal{U}&_{ijkl}^{(2),\varepsilon}(t,t') = \mathcal{U}_{kj}(t,t')\mathcal{U}_{li}(t',t) \\ \nonumber&+ \i\hbar \sum_{pqrs} \int_{t'}^t \d\cbar{t}\, \mathcal{U}_{kp}(t,\cbar{t})\mathcal{U}_{qi}(\cbar{t},t) \mathfrak{h}^{\varepsilon, \tn{corr}}_{rpsq}(\cbar{t}) \mathcal{U}_{rjsl}^{(2),\varepsilon}(\cbar{t},t')\,.
\end{align}
The equation of motion for these modified propagators can also be brought to a differential form:
\begin{align}
 & \i\hbar\frac{\d}{\d t} \mathcal{U}^{(2),\varepsilon}_{ijkl}(t\ge t') 
 =  \sum_{pq}  \mathfrak{h}^{\varepsilon}_{pkqi}(t) \mathcal{U}^{(2),\varepsilon}_{pjql}(t\ge t')
 \, . 
\label{eq:tdse-Ueps}
\end{align}
As one observes, $\mathcal{U}_{ijkl}^{(2),\varepsilon}$ obeys the same equation as $\varepsilon^{-1,\gtrless}_{ijkl}$ itself [cf. \refeqn{eq:tdse-eps}]. They are, however, not identical, since the time-diagonal values differ [cf. \refeqns{eq:unity_prop}{eq:eps_diag}].

\subsection{$T$ matrix in the particle--particle channel}\label{ss:tpp-inicor}
For the particle--particle $T$-matrix approximation similar equations can be derived. The equivalent of \refeqn{eq:g2-ic_compact} takes the form,
\begin{align}\label{eq:Gtwo-omega-pp}
 &\mathcal{G}^\GKBA_{ijkl}(t) = \left(\i\hbar\right)^4 \sum_{pqrs}\int_{t_0}^t \mathrm{d}\cbar{t}\, \mathcal{U}_{ijpq}^{(2),\Omega^\tn{pp}}(t,\cbar{t})  \times
 \\ & \quad\nonumber \times
 \Big[ \delta(t_0,\cbar{t}) \mathcal{G}^0_{pqrs} + \frac{1}{\i\hbar} \Psi^\pm_{pqrs}(\cbar{t}) \Big]
  \left[\mathcal{U}_{klrs}^{(2),\Omega^\tn{pp}}(t,\cbar{t})\right]^*\,,
\end{align}
where
\begin{align}
\label{eq:U_prop_pp}
 \mathcal{U}&_{ijkl}^{(2),\Omega^\tn{pp}}(t,t') = \mathcal{U}_{ijkl}^{(2)}(t,t') \\ \nonumber&+ \i\hbar \sum_{pqrs} \int_{t'}^t \d\cbar{t}\, \mathcal{U}_{ijpq}^{(2)}(t,\cbar{t}) \mathfrak{h}^{\Omega^\tn{pp}, \tn{corr}}_{pqrs}(\cbar{t}) \mathcal{U}_{rskl}^{(2),\Omega^\tn{pp}}(\cbar{t},t')\,.
\end{align}
The corresponding differential equation for the two-particle propagator mirrors the respective equation for $\Omega^\tn{pp}$ [cf. \refeqn{eq:omega-pp-tdse}],
\begin{align}
 & \i\hbar \frac{\d}{\d t} \mathcal{U}_{ijkl}^{(2),\Omega^\tn{pp}}(t\ge t')
 = \sum_{pq}  \mathfrak{h}^{\Omega^\tn{pp}}_{ijpq}(t) \mathcal{U}_{pqkl}^{(2),\Omega^\tn{pp}}(t\ge t') \, . \label{eq:tdse-Uomegapp} \qquad
\end{align}
As for $GW$, the time-diagonal values of both quantities do, however, not coincide.
\subsection{$T$ matrix in the particle--hole channel}\label{ss:tph-inicor}
Finally, in the particle--hole $T$-matrix approximation \refeqn{eq:g2-ic_compact} is replaced by
\begin{align}\label{eq:Gtwo-omega-ph}
 &\mathcal{G}^\GKBA_{ijkl}(t) = \i\hbar \sum_{pqrs}\int_{t_0}^t \mathrm{d}\cbar{t}\, \mathcal{U}_{iyul}^{(2),\Omega^\tn{ph}}(t,\cbar{t})  \times
 \\ & \quad\nonumber \times
  \Big[ \delta(t_0,\cbar{t}) \mathcal{G}^0_{pqrs} + \frac{1}{\i\hbar} \Psi^\pm_{pqrs}(\cbar{t}) \Big]
  \left[\mathcal{U}_{kvxj}^{(2),\Omega^\tn{ph}}(t,\cbar{t})\right]^*\,,
\end{align}
with
\begin{align}
\label{eq:U_prop_ph}
 \mathcal{U}&_{ijkl}^{(2),\Omega^\tn{ph}}(t,t') = \mathcal{U}_{ik}(t,t')\mathcal{U}_{jl}(t',t) \\ \nonumber&+ \i\hbar \sum_{pqrs} \int_{t'}^t \d\cbar{t}\, \mathcal{U}_{iq}(t,\cbar{t})\mathcal{U}_{pl}(\cbar{t},t) \mathfrak{h}^{\Omega^\tn{ph}, \tn{corr}}_{qrsp}(\cbar{t}) \mathcal{U}_{sjkr}^{(2),\Omega^\tn{ph}}(\cbar{t},t')\,.
\end{align}
The last equation can again be transformed into its differential form,
\begin{align}
 &\i\hbar\frac{\d}{\d t} \mathcal{U}^{(2),\Omega^\tn{ph}}_{ijkl}(t\ge t')
 =  \sum_{pq}  \mathfrak{h}^{\Omega^\tn{ph}}_{ipql}(t)  \mathcal{U}^{(2),\Omega^\tn{ph}}_{qjkp}(t\ge t')
 \, , 
\end{align}
which matches \refeqn{eq:omph-tdse} for $\Omega^\tn{ph}$ in analogy to \refeqns{eq:tdse-Ueps}{eq:tdse-Uomegapp}.

\bibliography{bibliography,dfg_pngf,mb-ref}

\end{document}